\documentclass[twocolumn,twocolappendix]{aastex631}
\usepackage{amsmath}
\usepackage{rotating,tabularx}

\newcommand{\kms}{\mathrm{km\,s^{-1}}}
\newcommand{\lya}{$\rm Ly\alpha$}
\newcommand{\lyb}{Ly$\beta$}

\newcommand{\K}{{\,{\rm K}}}
\newcommand{\cm}{{\rm cm}}
\newcommand{\dl}{$\delta_{\rm Ly\alpha}$}
\newcommand{\deltalya}{\delta_{\rm Ly\alpha}}
\newcommand{\zabs}{$z_{\rm abs}$}
\newcommand{\zabsorb}{z_{\rm abs}}
\newcommand{\nH}{n_{\rm H}}
\newcommand{\LAF}{\lya\ forest}
\newcommand{\picca}{\texttt{picca}}

\newcommand{\CII}{\ion{C}{2}}
\newcommand{\CIII}{\ion{C}{3}}
\newcommand{\CIV}{\ion{C}{4}}
\newcommand{\NV}{\ion{N}{5}}

\newcommand{\OI}{\ion{O}{1}}
\newcommand{\OVI}{\ion{O}{6}}
\newcommand{\MgII}{\ion{Mg}{2}}
\newcommand{\AlII}{\ion{Al}{2}}
\newcommand{\AlIII}{\ion{Al}{3}}
\newcommand{\SiII}{\ion{Si}{2}}
\newcommand{\SiIII}{\ion{Si}{3}}
\newcommand{\SiIV}{\ion{Si}{4}}
\newcommand{\FeII}{\ion{Fe}{2}}

\newcommand{\comquest}[1]{}
\newcommand{\mmp}[1]{}
\shorttitle{Metal Lines Associated with the Lyman-alpha Forest}
\begin{document}
\title{Metal Lines Associated with the Lyman-alpha Forest from eBOSS Data}
\email{liyang@shao.ac.cn, zhengzheng@astro.utah.edu}

\author{Li Yang}
\affiliation{
Shanghai Astronomical Observatory, Chinese Academy of Sciences \\
80 Nandan Road, Shanghai 200030, People's Republic of China
}
\affiliation{
School of Astronomy and Space Sciences, University of Chinese Academy of Sciences \\
No.19A Yuquan Road, Beijing 100049, People’s Republic of China \\
}
\affiliation{
Department of Physics and Astronomy, University of Utah \\
115 S 1400 E, Salt Lake City, UT 84112, USA
}

\author{Zheng Zheng}
\affiliation{
Department of Physics and Astronomy, University of Utah \\
115 S 1400 E, Salt Lake City, UT 84112, USA
}

\author{H\'elion~du~Mas~des~Bourboux}
\affiliation{
Department of Physics and Astronomy, University of Utah \\
115 S 1400 E, Salt Lake City, UT 84112, USA
}

\author{Kyle Dawson}
\affiliation{
Department of Physics and Astronomy, University of Utah \\
115 S 1400 E, Salt Lake City, UT 84112, USA
}

\author{Matthew M. Pieri}
\affiliation{
Aix Marseille Universit\'e, CNRS, CNES, Laboratoire d’Astrophysique de Marseille \\
Marseille 13388, France
}

\author{Graziano Rossi}
\affiliation{
Department of Physics and Astronomy, Sejong University,\\
209, Neungdong-ro, Gwangjin-gu, Seoul, South Korea
}

\author{Donald P. Schneider}
\affiliation{
Institute for Gravitation and the Cosmos, Pennsylvania State University\\
University Park, PA 16802, USA
}
\affiliation{
Department of Astronomy and Astrophysics, Eberly College of Science, The Pennsylvania State University\\ 
525 Davey Laboratory, University Park, PA 16802, USA
}

\author{Axel de la Macorra}
\affiliation{
Instituto de F\'isica, Universidad Nacional Aut\'onoma de M\'exico, \\ 
Circuito de la Investigaci\'on Cient\'ifica Ciudad Universitaria, 04510  CDMX, M\'exico
}

 \date{\today}

\begin{abstract}
We investigate the metal species associated with the \lya\ forest in the eBOSS quasar spectra. Metal absorption lines are revealed in the stacked spectra from cross-correlating the selected \lya\ absorbers in the forest and the flux fluctuation field. Up to 13 metal species are identified associated with relatively strong \lya\ absorbers (those with flux fluctuation $-1.0<\deltalya<-0.6$ and with neutral hydrogen column density of $\sim 10^{15-16}\cm^{-2}$) over absorber redshift range of $2<\zabsorb<4$. The column densities of these species decrease toward higher redshift and for weaker \lya\ absorbers. From modelling the column densities of various species, we find that the column density pattern suggests contributions from multiple gas components both in the circumgalactic medium (CGM) and in the intergalactic medium (IGM). While the low-ionization species (e.g., \ion{C}{2}, \ion{Si}{2}, and \ion{Mg}{2}) can be explained by high-density, cool gas  ($T\sim 10^4$ K) from the CGM, the high-ionization species may reside in low-density or high-temperature gas in the IGM. The measurements provide inputs to model metal contamination in the \LAF\ baryon acoustic oscillations measurement. Comparison with metal absorptions in high-resolution quasar spectra and with hydrodynamic galaxy formation simulations can further elucidate the physical conditions of these \lya\ absorbers.
\end{abstract}

\section{Introduction}
\label{section::Introduction}
\setcounter{footnote}{0}

The Lyman-$\alpha$ (\lya) forest, namely the ensemble of absorption lines in quasar's continua at wavelengths below quasar's \lya\ emission caused by the absorption of intervening neutral hydrogen (e.g., \citealt{Cen94,Bi97,Rauch98}), has become a powerful cosmological and astrophysical probe.

As a tracer of the underlying matter density field, the \LAF\ is a powerful cosmological probe that has been successfully used to constrain cosmological parameters. In particular, the \LAF\ data from the spectroscopic observation of quasars in the the Sloan Digital Sky Survey (SDSS) has been used to measure the Baryon Acoustic Oscillations (BAO), adding unique high-redshift ($z\sim 3$) data points to constrain cosmology \citep[e.g.,][]{Slosar13,Bautista17,Helion2020}.

The \LAF\ also encodes valuable information about the circumgalactic and intergalactic media (CGM and IGM), including their density and temperature distribution. While the IGM and CGM are the source of material for star formation and galaxy formation, their properties are also affected by gas outflow from galaxies as a result of feedback processes. The outflow can bring heavy elements into the CGM and IGM. The metal content is connected to the stellar population, abundance pattern, and star formation history. The metal species at various ionization levels can constrain the ionizing ultraviolet (UV) background. The \LAF\ observation provides a means to reveal metals. As metal absorption in the \LAF\ is an important systematic component to be modelled in \LAF\ BAO measurement \citep{Bautista15}, studying metals in the \LAF\  also aids the cosmological application of the \LAF.

High-resolution spectroscopic observations of quasars with long exposure times are usually the choice to study metal lines in the \LAF\ or associated with the \LAF\ 
{
\citep[e.g.,][]{Ellison2000,Schaye2003,Simcoe2004,Lehner2019,Lehner2021}. For example, \citet{Lehner2021} study $2.2 \lesssim z \lesssim 3.6$ \ion{H}{1}-selected absorbers with neutral hydrogen column density in the range of $14.6 \leq \log (N_{\rm HI}/{\rm cm}^{-2}) \leq 20$. The metallicity is found to have a broad distribution (e.g., an interquartile range from $-3.6$ to $-1.8$
 dex with a median of $-2.4$ dex for $\log N_{\rm HI}=$ 14.5--16.2 absorbers),  and the median metallicity increases and the interquartile range decreases with increasing neutral hydrogen column density.
}
High-resolution spectroscopic observations of quasars usually probe only the \LAF\ along the sight line of a single quasar or those of a small number of quasars. In contrast, the large spectroscopic survey of quasars from the various phases of the SDSS survey is able to probe hundreds of thousands of quasar sight lines with moderate resolution and short exposure times. Although the signal-to-noise ratio and the spectral resolution from the SDSS observation typically do not allow the identification of metal absorption lines associated with the \LAF\ in any individual quasar spectrum, the large number of quasar spectra can be used to perform a statistical study of metal lines.

\citet{Pieri2010} and \citet{Pieri2014} develop a technique to study metal absorption lines using composite spectrum of \lya\ forest absorbers identified in SDSS spectra of quasars. 
{
\lya\ absorbers in the \lya\ forest of each quasar are selected based on the transmitted flux fraction. For each selected absorber, the quasar's spectrum is shifted to the restframe of the absorber. The shifted quasar spectra for all the selected absorbers are then stacked to produce a composite spectrum. Features uncorrelated with the absorbers are highly suppressed in the composite spectrum, and the coherently stacked metal absorption lines associated with the selected absorbers can be detected with high significance.
}
Applying the technique to the data from the SDSS \citep{Pieri2010} and from the Baryon Oscillation Spectroscopic Survey (BOSS) of SDSS III \citep{Pieri2014} has revealed a series of metal absorption lines in the composite spectrum to probe the physical conditions and metal enrichment of the CGM and IGM at $z\sim$ 2--3.

In this paper, following the same basic idea in \citet{Pieri2010}, we study the metal lines associated with the \LAF\footnote{In our notation, \lya\ forest is the region blueward of quasar \lya\ emission line, and  we do not specifically limit the column density of the absorption systems (absorbers). As we will see, the absorbers we select can have contributions from high-column-density systems. More accurately, we study metal lines associated with \lya\ absorbers in the \lya\ forest region.}  by applying a cross-correlation method to obtain the stacked spectra using data from the extended BOSS (eBOSS; \citealt{Dawson2016}) survey of the SDSS IV. We study the dependence of the metal absorption lines on the \lya\ fluctuation field and their redshift evolution over $2<z<4$.

In Section~\ref{sec:data}, we describe the data and the method to compute the stacked spectra of \lya\ absorbers. We then present the main results in Section~\ref{sec:results}, including the column densities of 13 metal species as a function of \lya\ absorber strength and redshift. A simple model is discussed to understand the results. Finally, we summarize and discuss our findings in Section~\ref{sec:summary}. We leave some details on column density measurements in Appendix~\ref{sec:appendix_N_metals}.

\section{Data Samples and Reduction Methods}
\label{sec:data}

The \lya\ forest data in this work are based on the quasar observation from the Sloan Digital Sky Survey (SDSS; \citealt{York2000}), gathered during SDSS-III by the Baryon Oscillation Spectroscopic Survey (BOSS; \citealt{Eisenstein2011,Dawson2013}), and during SDSS-IV by the extended BOSS (eBOSS; \citealt{Dawson2016,Blanton2017}), with a small fraction observed during SDSS-I and II \citep{Schneider2010}.

All quasar spectra used for the analysis here were obtained using the eBOSS spectrographs mounted on the 2.5 m Sloan Foundation telescope (\citealt{Gunn2006,Smee2013,Myers2015}) at the Apache Point Observatory, publicly available in the sixteenth data release (DR16; \citealt{Ahumada2020}). The spectral resolution varies with wavelength (3600-10400\AA), from $R = 1800$ to 2200, and we adopt $R=2000$ in our analysis, and this approximation has no effect on our results.

The DR16 quasar catalog (DR16Q; \citealt{Lyke2020}) is adopted to extract quasar spectra from  \texttt{SpAll-v5\_13\_0.fits}\footnote{\url{https://data.sdss.org/datamodel/files/BOSS_SPECTRO_REDUX/RUN2D/spAll.html}}, processed with version \texttt{v5\_13\_0} of the eBOSS spectroscopic pipeline. These are ``co-added'' spectra constructed from 
typically four exposures of 15 min and resampled at wavelength pixels of width
$\Delta \log\lambda \sim 10^{-4}$ ($ c \Delta \lambda / \lambda \sim 69\, \kms$). Broad absorption line quasars (BALs) with BALPROB $> 0.9$ \citep{Lyke2020} are removed from our analysis. We end up with 725,635 quasars.

Following \citet{Helion2020}, we discard pixels flagged as problematic in the flux calibration or sky subtraction by the pipeline and and correct for small residual flux calibration errors from the eBOSS pipeline. We then mask out spectral intervals, in observed wavelength, where the variance increases sharply owing to unmodeled emission lines from the sky, as well the wavelength range of \ion{Ca}{2} H\&K absorption of the Milky Way. Damped Lyman-Alpha systems (DLAs) are identified (\citealt{Chabanier2021,Lyke2020,Parks2018}), and all pixels where the absorption due to the DLA is higher than $20\%$ are not used\footnote{
We perform a test of removing pixels with DLA-caused absorption higher than 5\% and find that this more conservative cut has little effect on our final stacked results. Also, by default we mask \lya\ and \lyb\ lines associated with DLAs. We do a further test of masking the metal lines from DLAs. We find that, except weakening the shadow lines in the stacked spectrum, it has no effect on the measurements of metals in the \lya\ absorbers we select, as the DLA metal lines do not contaminate them.
}. 
Finally, a spectral region is used only if it contains at least 50 pixels to avoid overfitting the mean transmitted flux. 

\subsection{\lya\ Absorbers and the Forests}

\begin{figure*}
\centering
\includegraphics[width=0.45\textwidth]{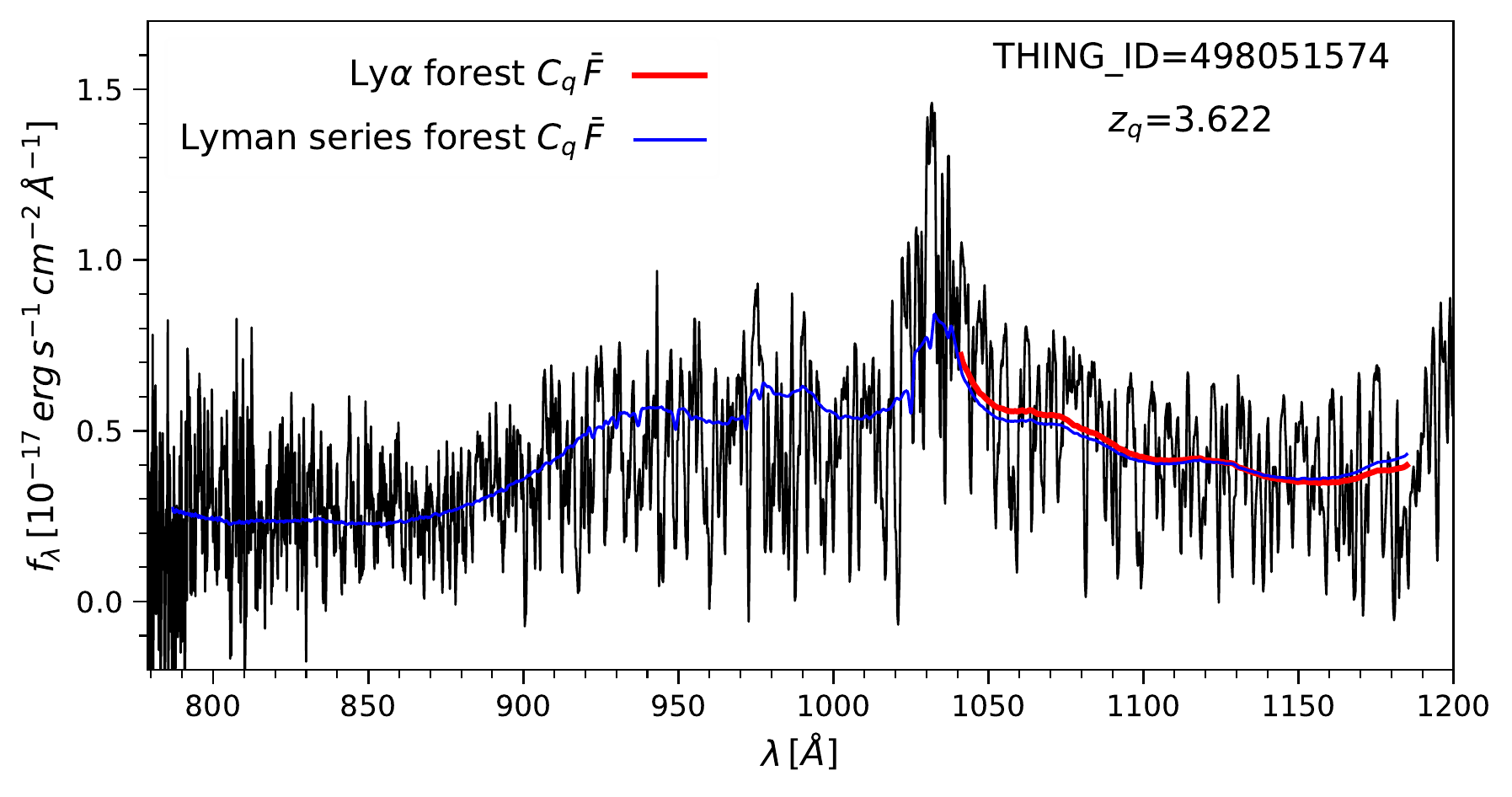} 
\includegraphics[width=0.45\textwidth]{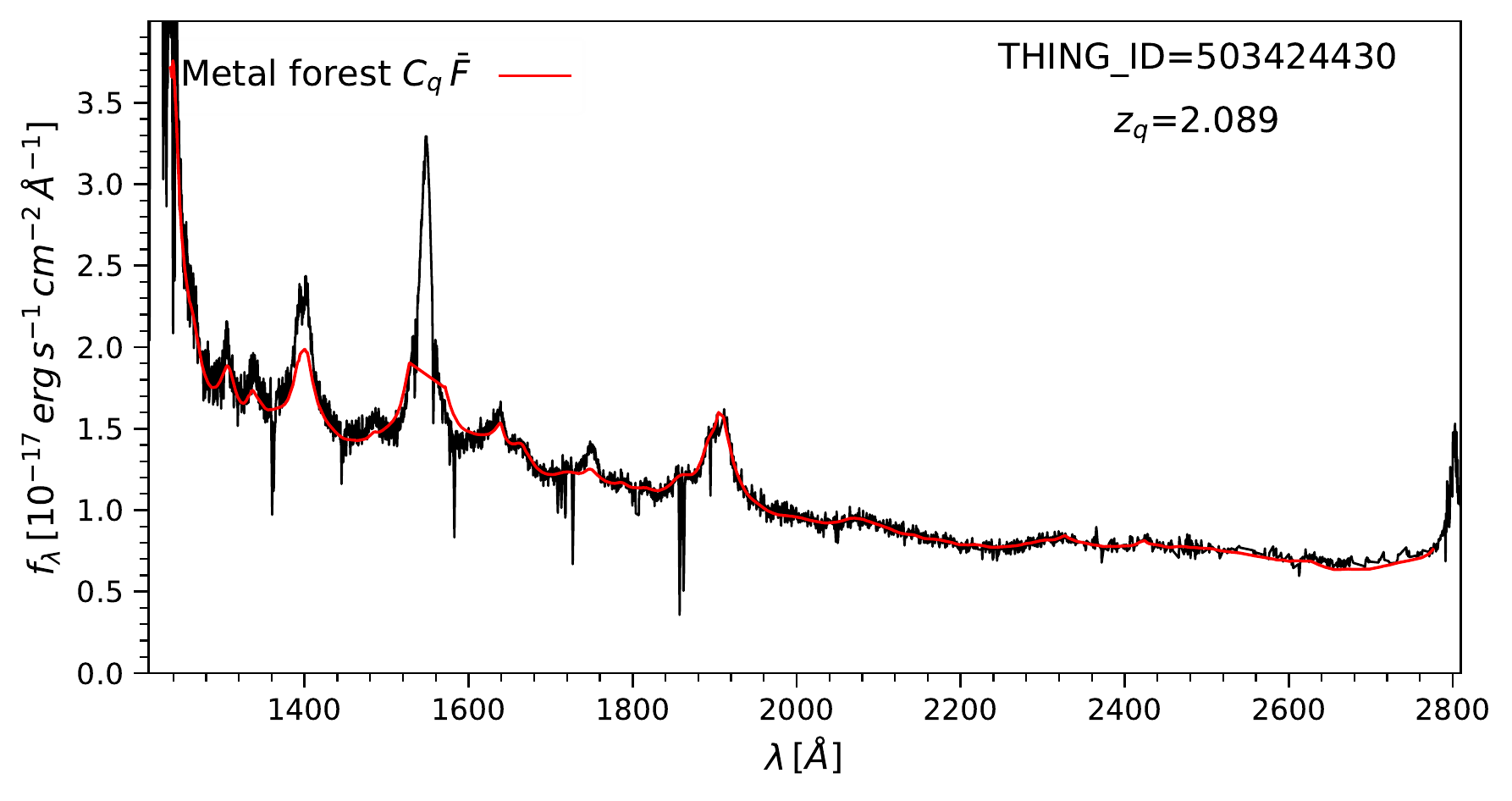}
\caption{Illustration of different forests and mean flux fits. The spectra are in quasar's rest-frame. The colored curves are the fits for the mean expected flux [$C_q\bar{F}$ in eq.~(\ref{eqn:delta})] in the \lya\ forest (red in the left panel), in the Lyman series forest (blue in the left panel), and in the metal forest (red in the right panel). See the text. Note that in the right panel the wavelength range within $\pm 15 $\AA\ of the quasar's \ion{C}{4} emission is excluded for the $C_q\bar{F}$ fit, as the emission line center are too sharp to produce a good fit. The IDs of the two quasars in the DR16Q catalog are labelled in each panel.
}
\label{fig:illustration_CFbar}
\end{figure*}

In this work, \lya\ absorption systems (\lya\ absorbers) are selected based on the fluctuation field of the transmitted flux in the \lya\ forest. With a selected \lya\ absorber, the fluctuation fields in the  Lyman series range and in the range longer than the \lya\ wavelength are stacked for our study. We call these two ranges Lyman series forest and metal absorption forest, respectively. To measure the flux fluctuation field of a given forest, we make use of the publicly available Python ``Package for IGM Cosmological-Correlations Analyses'' \texttt{picca}\footnote{\url{https://github.com/igmhub/picca}} \citep{picca}. 

The measurement of the flux fluctuation field for the \lya\ forest, the Lyman series forest, and the corresponding metal forest follows an approach similar to the one established for the \lya\ forest BAO analysis, as presented in \citet{Helion2020}. In short, for each quasar $q$, the fluctuation field $\delta_q(\lambda)$ in the transmitted flux for each forest is computed as
\begin{equation}
\delta_q(\lambda)=  \frac{f_q(\lambda)}{C_q(\lambda)\bar{F}(\lambda)}-1,
\label{eqn:delta}
\end{equation}
where $f_q(\lambda)$ is the observed flux density and $C_q(\lambda)$ is the continuum (the flux density that would be observed in the absence of absorption).The quantity $\bar{F}(\lambda)$ is the mean transmission. The product $C_q(\lambda)\bar{F}(\lambda)$ is the mean expected flux for this quasar. It is modelled as a universal function in quasar's rest-frame (with a uniform forest spectral template), corrected by a linear function of $\log\lambda$. The model parameters are determined by maximizing the likelihood between the model flux and the observed flux in the forest region for each quasar (see \citealt{Helion2020} for detail).

{
As noted in \citet{Bautista17} and \citealt{Helion2020}, the above fitting method introduces a small systematic bias in $\deltalya$. The difference in the correlation functions between the true $\deltalya$ field and the one derived from the above fitting method based on mock data (Fig.11 of \citealt{Bautista17}) allows us to estimate this bias, and we find that the derived $\deltalya$ is about 0.02 lower than the true value, too small to have any significant effect on our results. Besides the systematic bias, the mean transmitted flux fitting for each individual quasar also introduces a statistical fluctuation in $\deltalya$, which is included in the uncertainty of $\deltalya$ \citep[e.g.,][]{Helion2020}. Such a fluctuation would result in a spread in the true \dl\ values for a given selected \dl. The high signal-to-noise ratio (above ten; see below) we impose to select \lya\ absorbers limits such an effect to be small. In retrospect, this is supported by the clear trend we find in the dependence of metal column density on \dl\ (to be presented later). Finally, it is possible that the mean transmitted flux fitting is not accurate at other wavelengths (e.g., around a metal line). As mentioned in Section~\ref{sec:stacked}, the stacked spectrum is normalized by a local linear fit around each metal line, which largely removes the effect caused by the inaccuracy in the above fitting.
}

With the measured \lya\ fluctuation field \dl\ in the \lya\ forest, we construct a catalog of \lya\ absorbers. We select \lya\ absorbers in the redshift range $2.0<\zabsorb<4.0$, where \zabs\ is the redshift computed from the \lya\ absorption. The lower redshift limit is a consequence of the \lya\ forest exiting the observed wavelength coverage for quasars with $z\lesssim 2$. The upper redshift limit is adopted as there are only a small number of $z>4.0$ quasars in DR16Q. For the purpose of our analysis, the \lya~forest is defined to be in the wavelength range of $\lambda_{\rm RF}=$1041--1185\AA\ in quasar's rest-frame {(similar to that used by \citealt{McDonald2006})}, where the proximity effects of the quasar's \lya\ and \lyb\ emission line profiles are small \citep{Vid2013}. 

In this work, we concentrate on overdense regions and define \lya\ absorber pixels as local minima among three adjacent pixels in the \lya\ forest. We select absorbers in four bins of flux fluctuation,  $\deltalya=-0.95$, $-0.85$, $-0.75$, and $-0.65$, respectively, with bin width of 0.1. These are the ranges for absorbers with noticeable metal detection, which is the focus of this work. The analysis of hydrogen distribution properties for absorbers in other ranges of \dl, including those corresponding to underdense regions, will be presented in a separate paper. We require that for each absorber pixel the signal-to-noise ratio in \dl\ is above ten\footnote{The signal-to-noise ratio in \dl\ is calculated as ${\rm SNR}_\delta=|\delta_q|/\tilde{\sigma}_{{\rm pip},q}$, where $\tilde\sigma_{{\rm pip},q}=\sigma_{{\rm pip},q}/(C_q \bar{F})$, $\sigma^2_{{\rm pip},q}$ is the pipeline estimate of the flux variance, and $\delta_q$ and $C_q \bar{F}$ are the same as in equation~(\ref{eqn:delta}).}. Such a cut preserves as many absorbers as possible for the study of metal absorption and at the same time it keeps meaningful divisions of the four absorber samples. To investigate the redshift evolution of the metals in \lya\ absorbers, we further divide the absorbers into 10 redshift bins, with $\zabsorb=2.1\pm 0.1$, $2.3\pm 0.1$, ..., and $3.9\pm 0.1$. In Table~\ref{table:num_forests}, the numbers of \lya\ absorbers in different \dl\ and redshift bins are provided. 

Like the \lya\ forest, we can define the \lyb\ forest region to be in the wavelength range of $\lambda_{\rm RF}=$978--1014\AA\ in quasar's rest-frame, and similarly for higher order Lyman series. In practice, we denote the wavelength range of $\lambda_{\rm RF}=$787--1185\AA\ in quasar's rest-frame as a single forest, namely the Lyman series forest, even though there are metal lines in this range. {
For metal lines redward of quasar's \lya\ emission, each can have its own forest region. For example, in \citet{Helion2019}, they adopt quasar rest-fame wavelength range of 1260--1375\AA\ as the \ion{Si}{4} (1394/1403\AA\ doublet) forest and 1420--1520\AA\ as the \ion{C}{4} (1548/1551\AA\ doublet) forest. In this work, we define a single metal forest as the range of $\lambda_{\rm RF}=$1236--2776\AA\ in quasar's rest-frame. The range corresponds to $\sim 5000 \, {\rm km \, s^{-1}}$ redward of the quasar's \lya\ emission and $\sim 2600 \, {\rm km \, s^{-1}}$ blueward of the quasar's \MgII\ emission, and we also exclude the pixels within $\pm 15$\AA\  (in quasar's rest-frame) around the quasar's \CIV\ emission, reducing the effect of the quasar's emission lines on the $C_q\bar{F}$ fit.
} Figure~\ref{fig:illustration_CFbar} illustrates the different forests and the $C_q\bar{F}$ fits in equation~(\ref{eqn:delta}) for each forest. 

With quasars in the redshift range of $2.08<z_q<4.82$, applying the absorber redshift cuts of $2<\zabsorb<4$ and strength cuts of $-1.0<\deltalya<-0.6$, we end up with 47,118 \lya\ forests ($\lambda_{\rm RF}$=1041--1185\AA), 46,804 Lyman series forests ($\lambda_{\rm RF}$=787--1185\AA), and 47,076 metal forests ($\lambda_{\rm RF}$=1236--2776\AA). \lya\ absorbers selected from the \lya\ forests will be correlated with flux fluctuations in the Lyman series forests and metal forests to study the metal content associated with those \lya\ absorbers.

\begin{table*}
    \caption{Number of \lya\ absorbers in the sample as a function of \lya\ flux fluctuation \dl\ and absorber redshift \zabs
    \label{table:num_forests}
    }
\centering
    \begin{tabular}{crrrrrrrrrr}
        \hline
        \dl\ & & & & & \zabs\ & & & & & \\
          &
[2.0,2.2] &
[2.2,2.4] &
[2.4,2.6] &
[2.6,2.8] &
[2.8,3.0] &
[3.0,3.2] &
[3.2,3.4] &
[3.4,3.6] &
[3.6,3.8] &
[3.8,4.0] \\
        \hline \hline
        $[-1.0,-0.9]$ & $2,420$ & $4,085$ & $4,628$ & $4,296$ & $3,880$ & $2,891$ & $1,935$ & $925$ & $424$ & $137$ \\ [0.5ex] 
$[-0.9,-0.8]$ & $2,162$ & $3,864$ & $4,555$ & $4,133$ & $3,458$ & $2,510$ & $1,614$ & $794$ & $403$ & $114$ \\ [0.5ex] 
$[-0.8,-0.7]$ & $1,886$ & $3,713$ & $3,997$ & $3,563$ & $2,988$ & $2,144$ & $1,372$ & $607$ & $265$ & $83$ \\ [0.5ex] 
$[-0.7,-0.6]$ & $1,416$ & $3,027$ & $3,424$ & $2,966$ & $2,483$ & $1,820$ & $977$ & $474$ & $192$ & $63$ \\ [0.5ex] 
         \hline
    \end{tabular}
    \tablecomments{In total, there are 86,688 \lya\ absorbers (16,621 quasars) at $2<\zabsorb <4$ and with $-1.0<\deltalya<-0.6$.}
 \end{table*}

\subsection{Stacked Spectra Based on \lya\ Absorbers}
\label{sec:stacked}

\begin{figure*}
\centering
\includegraphics[width=\textwidth,height=12cm]{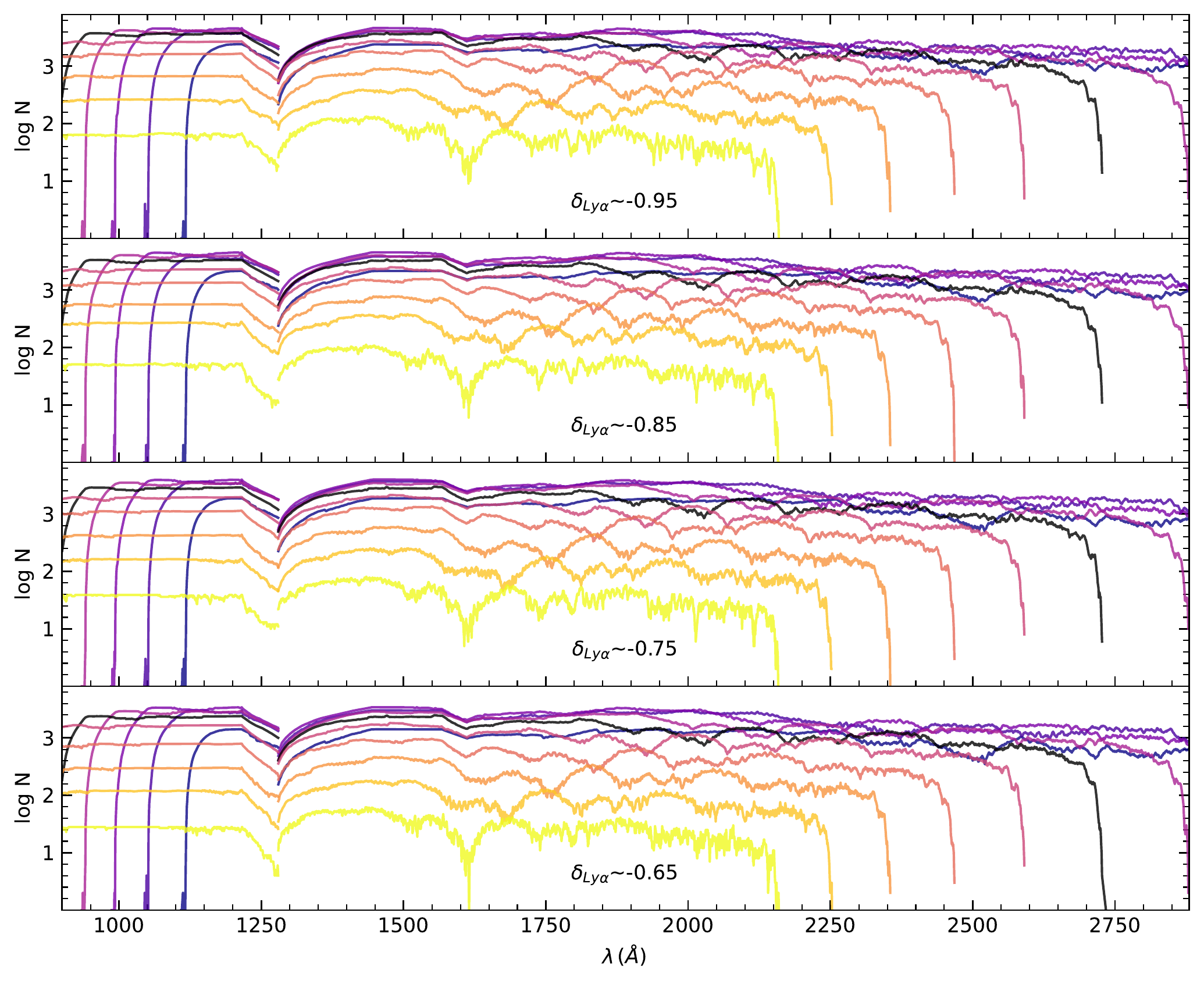} 
\caption{Number of \lya\ absorber-spectral pixel pairs (see the tables in Appendix~\ref{sec:appendix_N_metals} 
for detail). From top to bottom panels are for \lya\ absorbers with $\deltalya\sim -0.95$, $-0.85$, $-0.75$, and $-0.65$. In each panel, the color-coded curves (same as in Fig.~\ref{fig:N_metals}) correspond to 10 absorber redshift bins, from $\zabsorb=2.1$ to 3.9 with step 0.2 and roughly with lower curve for higher redshift (at rest-frame 1500\AA).}
\label{fig:N_pairs}
\end{figure*}  

The selected \lya\ absorbers here are connected to neutral hydrogen in the CGM or IGM, and they can be affected by metal enrichment caused by star formation and feedback processes. While the corresponding metal absorption features may be too weak to be identified in the eBOSS quasar spectra for individual absorbers, they can be revealed by stacking the spectra associated with a large number of \lya\ absorbers \citep[e.g.,][]{Pieri2010,Pieri2014}. 

With the catalog of \lya\ absorber pixels and the fluctuation fields computed for the Lyman series and metal forest at different wavelengths, we perform a stacking analysis to reveal metal absorption associated with \lya\ absorbers of different strengths (\dl). We follow the same technique as in \citet{Font-Ribera2014} and \citet{dMdB17} to do the stacking. That is, we compute the cross-correlation function between \lya\ absorber pixels and the fluctuation field $\delta$ of various forests, which is carried out with \picca. In detail, the cross-correlation is computed as 
\begin{equation}
\xi(\lambda)=\frac{\sum\limits_{(i,k)\in A}w_i w_k \delta_i}{\sum\limits_{(i,k)\in A} w_i w_k},
\end{equation}
where $i$ denotes the forest pixel and $k$ the \lya\ absorber pixel, $\delta_i$ is the flux fluctuation in one of the forests, and $w_i$ and $w_k$ are the weights [as defined in equations (4) and (7) of \citealt{Helion2020}]\footnote{ The weights include contributions form both instrumental noise (readout and photostatistics) and large-scale structure. Applying the weights helps increase the signal-to-noise ratio in the stacked spectrum. In the regime where the weights are dominated by photon noise, the stacking may be potentially biased toward selecting weaker absorbers. We perform tests with no weights applied and find that such a bias is negligible.}. The wavelength bin $A$ represented by $\lambda$ is determined through $\lambda=\lambda_{\rm Ly\alpha}(\lambda_i/\lambda_k)$, where $\lambda_{\rm Ly\alpha}=1215.67$\AA\ is the rest-frame wavelength of the \lya\ line and $\lambda_i$ and $\lambda_k$ are the observed wavelengths of the forest pixel and \lya\ absorber pixel, respectively. Clearly, $\lambda$ is the wavelength in the absorber's rest-frame. Later we also use $\delta(\lambda)$ to represent $\xi(\lambda)$.  The computation is performed for absorbers in each \dl\ bin and in each redshift bin. Because of systematics in continuum fitting, the upper envelope of each stacked spectrum, $1+\xi$, can slightly deviate from unity, and we apply a low-pass pseudo continuum to rectify this \citep[e.g.,][]{Pieri2014}. 
{
Specifically, for illustrating the stacked spectrum, we perform a spline fit to the $1+\delta(\lambda)$ in regions excluding absorption lines and normalize the spectrum by the fit. For analyzing each absorption line, we perform a local linear fit to the continuum and normalize the absorption line profile by the fit.
}

The covariance matrix of the cross-correlation is calculated by subsampling the data, similar to the approach in \cite{dMdB17}. We divide the sky into HEALPix pixels \citep{Gorski2005} and compute the cross-correlation using data in each HEALPix pixel (subsample). Using a division of the sky with nside=16, we obtain the number of subsamples to be between 27 and 790, depending on \dl\ and \zabs\ bins. The covariance is then given by the following, neglecting the small correlations between subsamples:
\begin{equation}
C_{\alpha\beta} = \frac{1}{W_\alpha W_\beta} \sum_{s} W^s_\alpha W^s_\beta [\xi^s_\alpha \xi^s_\beta-\xi_\alpha \xi_\beta].
\end{equation}
Here $\xi_\alpha=\xi(\lambda_\alpha)$ is the mean measured correlation for wavelength $\lambda_\alpha$, $s$ denotes a subsample with summed weight $W^s_\alpha$ and measured correlation $\xi^s_\alpha$, and $W_\alpha=\sum_s{W^s_\alpha}$.

The signal-to-noise ratio of the stacked spectra depends on the number of \lya\ absorber--forest pixel pairs. In Figure~\ref{fig:N_pairs}, we show the dependence of the number of such pairs on wavelength, \dl, and \zabs. Given our selection, the number has almost no dependence on \dl\ and only a weak dependence on wavelength. The redshift dependence has two parts. First, the coverage shifts towards shorter wavelength at higher \zabs, a reflection of redshift effect with the fixed observed wavelength range. At the lowest absorber redshift, $\zabsorb\sim 2.1$, we are not able to see Lyman series beyond \lya\, while at the long wavelength end it covers up to $\sim$2900\AA. In contrast, at the highest absorber redshift, $\zabsorb\sim 3.9$, at the low wavelength end the Lyman limit can be covered, but at the long wavelength end we cannot go above $\sim$2150\AA. Second, at a given wavelength, the redshift distribution of quasars leads to a strong redshift dependence of absorber-forest pairs, varying from a few times $10^3$ ($\zabsorb\sim 2.1$) to a few times $10^1$ ($\zabsorb\sim 3.9$). Therefore, we expect to have less constrained and fewer metal lines at higher absorber redshift. Since \dl\ is already selected to have signal-to-noise ratio higher than 10, the number of pairs also indicates that the stacked spectra for each \dl\ and at each \zabs\ have a much higher signal-to-noise ratio, 
making it possible to identify weak metal absorption lines. 
{We note that there is a break around $1280$\AA, since for the absorber-forest pixel pairs we use the \lya\ and Lyman series forest (the metal forest) below (above) this wavelength for the forest pixels, which is also adopted in our stacked spectra.}

\begin{figure*}
\centering 
     \includegraphics[width=\textwidth,height=0.94\textheight]{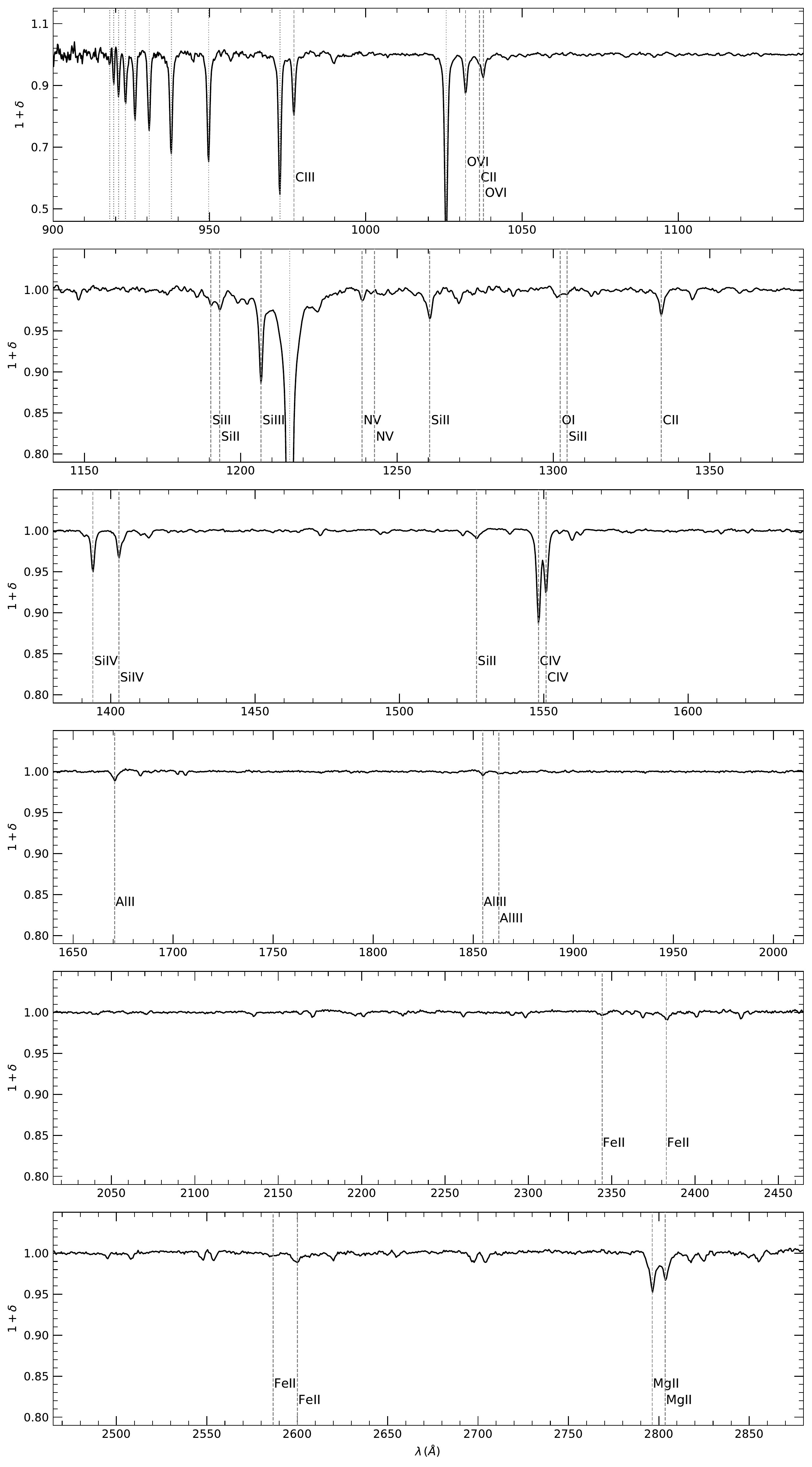}
    \caption{Illustration of the stacked spectrum by selecting  $-1.0<\deltalya<-0.9$ in the whole redshift range of $2<z_{\rm obs}<4$. Vertical dashed lines indicate metal lines identified and vertical dotted lines denote the locations of the Lyman series.
    The complete stacked spectrum is available as data behind the figure.
}
    \label{fig:stacked_spectra}
\end{figure*}

As an illustration, in Figure~\ref{fig:stacked_spectra}, we show the stacked spectra for \lya\ absorbers in the whole redshift range of $2<\zabsorb<4$ with $-1.0<\deltalya<-0.9$. The Lyman series, from \lya\ to Ly11, can be easily identified. We leave the discussion of Lyman series (including those in under-dense regions) to a separate paper and focus our discussion here to metal lines that show up in relatively strong \lya\ absorption systems. 

The high signal-to-noise ratio of such stacked spectra clearly helps reveal an array of metal lines in the wavelength range of 900--2880\AA, including low-ionization lines (e.g., \ion{Mg}{2} and \ion{Fe}{2}) and high-ionization lines (e.g., \ion{O}{6} and \ion{C}{4}). In particular, the lines with strong oscillator strengths are very prominent (e.g., \ion{Si}{4} 1394/1403\AA, \ion{C}{4} 1548/1551\AA, and \ion{Mg}{2} 2796/2804\AA\ doublets). There also appears to have a series of shadow lines. For example, the doublets at $\lambda\sim 1560$\AA\ are the shadow of the \ion{C}{4} 1548/1551\AA\ doublets caused by the contamination of \ion{Si}{3} 1206\AA\ in \lya. Similarly, the doublets at $\lambda\sim 2700$\AA\ are the shadow of the \ion{Mg}{2} 2796/2804\AA\ doublets from the contamination of \ion{Si}{2} 1260\AA\ in \lya. To avoid confusion, these shadow lines are not labeled in the figure, 
{
and a similar plot with major shadow lines identified and labelled can be found in the Appendix \ref{sec:appendix_shadow_lines} (Figure~\ref{fig:stacked_spectra_withshadowlines}).
}
The lines  contaminating \lya\ are weak, and they play no significant role in our \dl\ selection.

For each identified Lyman series line and metal line, we perform Voigt profile fitting to obtain the column density of each species. In doing the fitting, for a set of model parameters (column density $N$ and Doppler/broadening parameter $b$) the Voigt model is convolved with a Gaussian kernel of full-width-at-half-maximum $\sim 150\, {\rm km\,s^{-1}}$ to account for the eBOSS spectral resolution. 
{
For a given line, we typically choose more than 5 pixels on each side to determine the continuum, and we visually inspect the pixels to remove those belonging to shadow lines. For example, for \ion{Si}{4} 1394\AA, there is a shadow line of \ion{Si}{2} 1527\AA\ at $\lambda \sim 1391$\AA\ caused by the contamination of \ion{C}{2} 1335\AA\ in \lya;  for \ion{Si}{4} 1403\AA, there is a shadow line of \ion{Si}{4} 1394\AA\ at $\lambda\sim 1404$\AA\ caused by the contamination of \ion{Si}{3} 1207\AA\ in \lya.}
In fitting the lines, we also exclude the pixels near the shadow lines associated with the contamination of transitions around the \lya\ line (such as \ion{Si}{2} 1190\AA, \ion{Si}{2} 1193\AA, \ion{Si}{3} 1207\AA, \ion{N}{5} 1239\AA, \ion{N}{5} 1243\AA, \ion{Si}{2} 1260\AA, \ion{O}{1} 1302\AA, \ion{Si}{2} 1304\AA, and \ion{C}{2} 1335\AA). For example, the absorption feature around 1404.5\AA\ in the third panel of Figure~\ref{fig:stacked_metal2} is the shadow of the \ion{Si}{4} 1393\AA\ associated with the contamination of \ion{Si}{3} 1207\AA\ in \lya. The pixels around this feature are not included when we perform Voigt fitting for the \ion{Si}{4} 1402\AA\ line. Following previous work \citep[e.g.,][]{Pieri2014}, for doublet, we choose to fit each line separately, and the difference in the constraints can provide an assessment of possible systematics.
{
For strongly blended lines that are hard to be measured independently, such as the blend of \ion{C}{2} and \ion{O}{6} around 1037\AA\ and that of \ion{O}{1} and \ion{Si}{2} around 1303\AA, we perform a joint fit.
}

In general, we find that consistent with previous work \citep[e.g.,][]{Pieri2014} the broadening parameter $b$ appears to be high, typically at the level of a couple of hundred of ${\rm km\, s^{-1}}$. This is related to the resolution of the eBOSS spectra. 
The relatively low resolution means that each identified \lya\ absorber can consist of multiple absorbers, and the $b$ parameter includes the contributions of thermal motion in each individual absorber and the relative motion among these absorbers. This work focuses on the column densities of various species. In Tables \ref{table_Zabs[2.0,2.4]} in the Appendix \ref{sec:appendix_N_metals}, the parameter constraints from Voigt fitting for each metal line as a function of \dl\ and \zabs\ are provided.

\section{Results}
\label{sec:results}

With the stacked spectra, we first derive the \ion{H}{1} column density of each set of \lya\ absorbers by analyzing the Lyman series lines. Then we study how the metal lines depend on the strength of \lya\ absorbers and their redshift evolution. Finally, we present an illustrative model of metal column densities to provide insights on the physical properties of \lya\ absorbers selected in this work.

\subsection{Stacked Lyman Series and \ion{H}{1} Column Density}

\begin{figure*}
\centering
     \includegraphics[width=.98\textwidth]{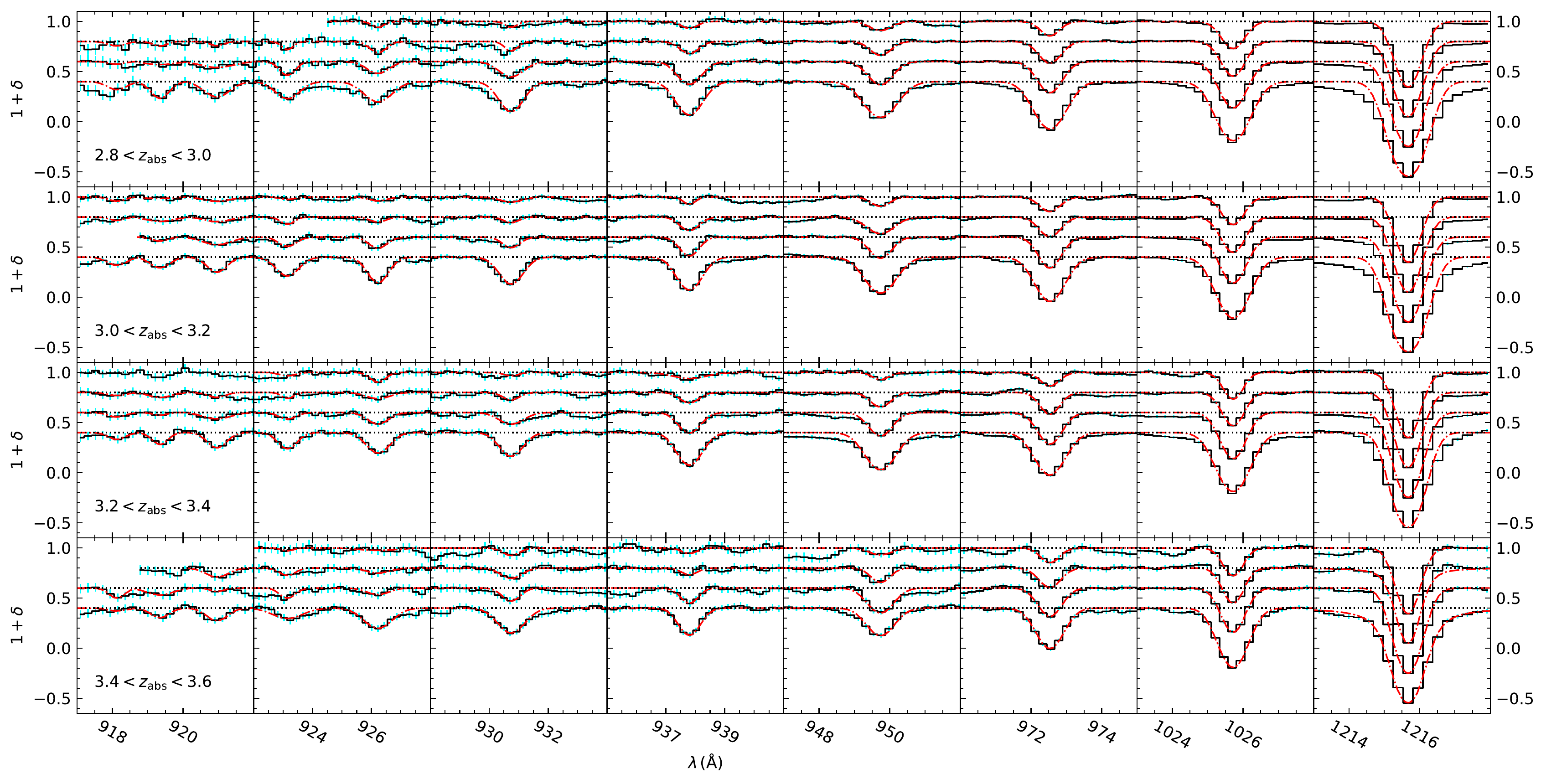}
    \caption{
    Stacked \ion{H}{1} Lyman series absorption line profiles as a function of \dl\ and absorber redshift. The absorber redshift is labelled in each panel in the first column. At each redshift, the four set of profiles are for $\deltalya\sim -0.65$, $-0.75$, $-0.85$, and $-0.95$ (from top to bottom) with offset of $0$, $-0.2$, $-0.4$, and $-0.6$, respectively, for clarity. The dotted lines show the continuum level and the dot-dashed curves are the Voigt profile fits. Seen from right to left panels in each row are line profiles of Ly$\alpha$, Ly$\beta$, Ly$\gamma$, ..., and Ly$\lambda$ (i.e., Ly11). For this figure, the absorption redshifts are chosen so that the Ly11 line can be detected for $\deltalya\sim -0.95$.
    }
    \label{fig:stacked_LySeries}
\end{figure*}

Figure~\ref{fig:stacked_LySeries} shows the stacked \ion{H}{1} Lyman series absorption profiles as a function of \dl. Compared with previous work \citep[e.g.,][]{Pieri2014}, the substantial increase in the signal-to-noise ratio allows us to reach Lyman series of much higher order. Four absorber redshifts are chosen for the illustration, which have clear detection up to Ly$\lambda$ (i.e., Ly11). For each set of a given \dl, the absorption depth decreases with decreasing line wavelength, resulting from the lower oscillator strength $f$ for higher-order Lyman lines. 

\begin{figure}
\centering
    \includegraphics[angle=0, width=.45\textwidth]{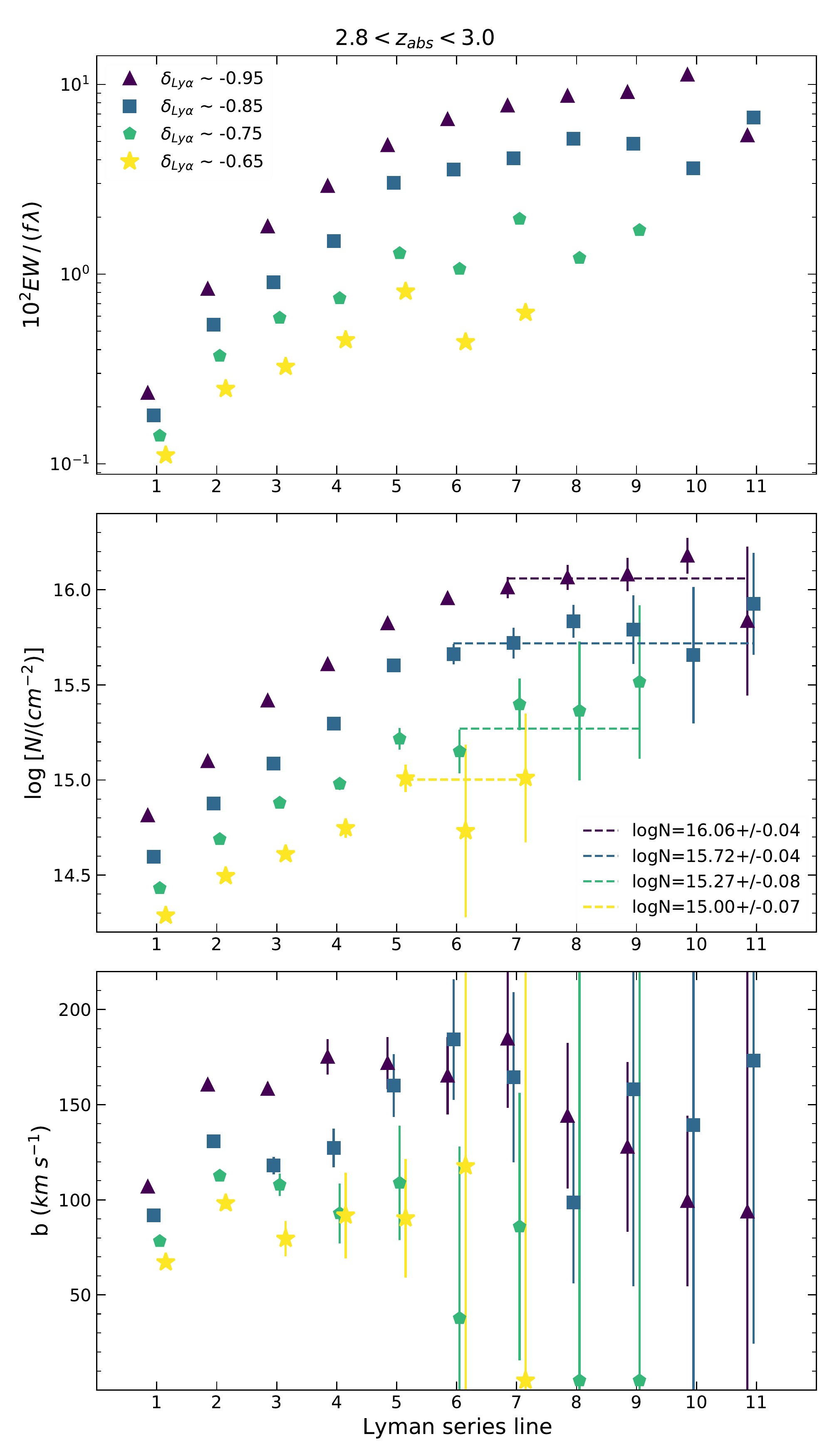}
    \caption{Measurements of the \ion{H}{1} column density with Lyman series lines at $\zabsorb\sim 2.9$. In each panel, four sets of symbols correspond to absorbers with different values of \dl.
    The bottom two panels show the Voigt profile fitting constraints on column density $N$ and Doppler parameter $b$. The top panel shows the derived equivalent width. In the middle panel, each dashed horizontal line represents the average column density in the plateau part from higher-order Lyman series, taken as the inferred \ion{H}{1} column density. See details in the text.
    }
    \label{fig:Ly_NHI}
\end{figure}

We perform Voigt profile fitting for each Lyman series line to obtain the column density $N_{\rm HI}$ and the $b$ parameter (and hence the equivalent width). An example is shown in Figure~\ref{fig:Ly_NHI} for $\zabsorb\sim 2.9$. As pointed out in \citet{Pieri2014}, lower-order Lyman series lines (e.g., Ly$\alpha$ and Ly$\beta$) are challenging to be used to derive precise measurements of the equivalent width. Higher-order, unsaturated lines are more suited for such a task, and the equivalent width (normalized by $f\lambda$) is expected to approach a constant value. As we are able to reach Lyman series lines of much higher order, we indeed find that the equivalent width reaches a plateau for each set of lines of a given \dl\ (as shown in Fig.~\ref{fig:Ly_NHI}). {This can be seen more clearly in the middle panel, where the column densities from higher-order Lyman series lines are consistent with being at a constant value within the error bars. 
}
We take the average column density from Ly5/Ly5/Ly5/Ly6 and above as the $N_{\rm HI}$ measurement for $\deltalya\sim -0.65/-0.75/-0.85/-0.95$, respectively. We find that $N_{\rm HI}$ drops from $\sim 10^{16}{\rm cm^{-2}}$ to $\sim 10^{15}{\rm cm^{-2}}$ as \dl\ varies from $\sim -0.95$ to $\sim -0.65$ and that there is almost no dependence on absorber redshift. 

In what follows, the \ion{H}{1} column density is presented together with those of metal lines and adopted in modelling the absorber systems.

\subsection{Dependence of Metal Column Densities on \dl\ and Absorber Redshift}

\begin{figure*}
\centering
\raggedleft

\includegraphics[angle=0,width=.49\textwidth,height=0.09\textheight]{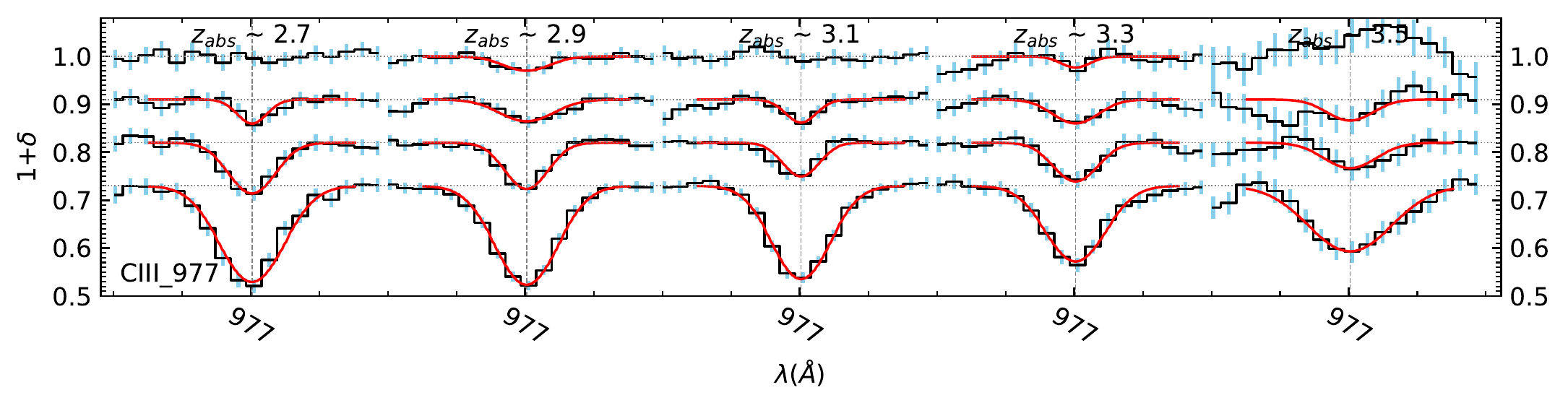}
\includegraphics[angle=0,width=.2\textwidth,height=0.09\textheight]{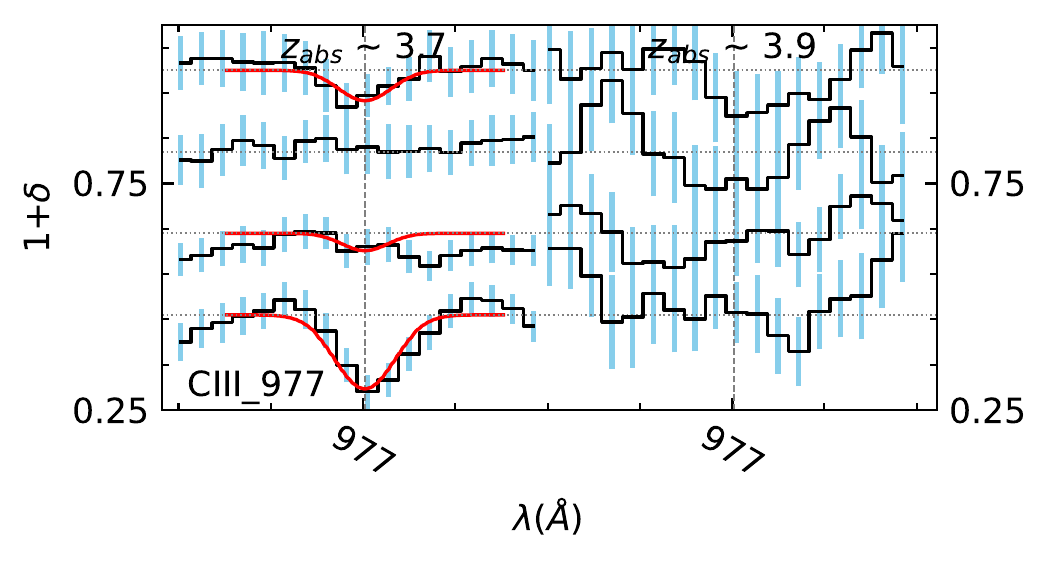} 

\includegraphics[angle=0,width=.59\textwidth,height=0.09\textheight]{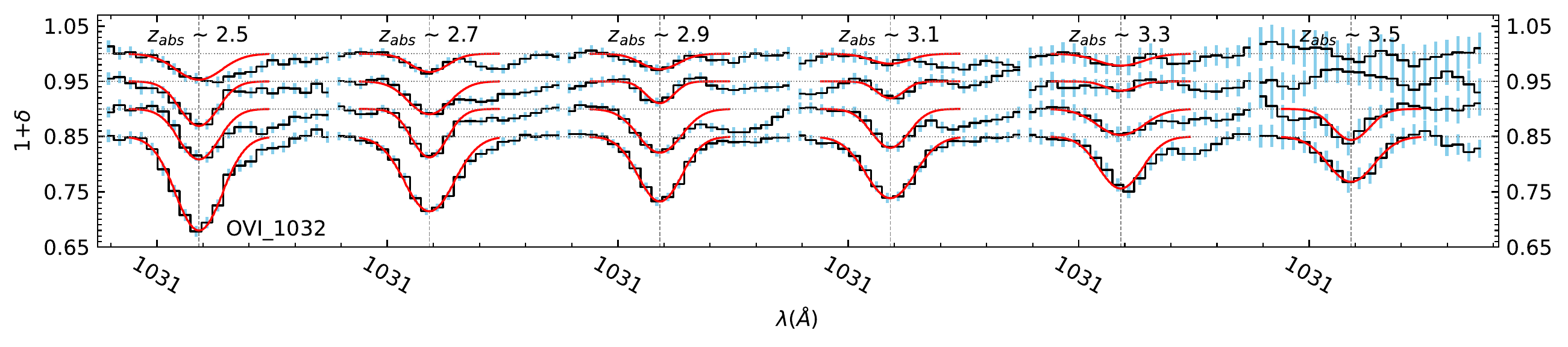}
\includegraphics[angle=0,width=.2\textwidth,height=0.09\textheight]{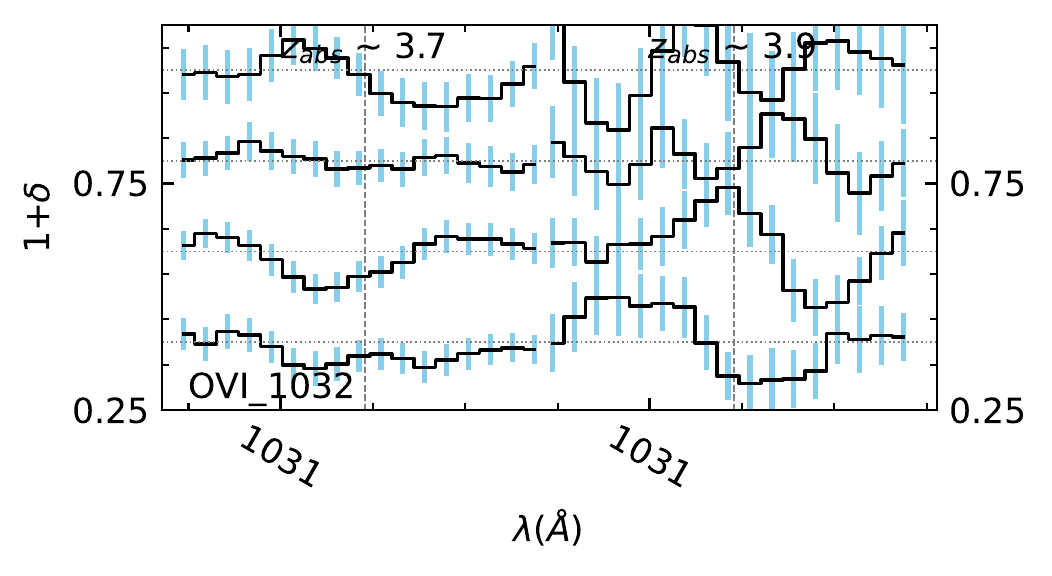}

\includegraphics[angle=0,width=.59\textwidth,height=0.09\textheight]{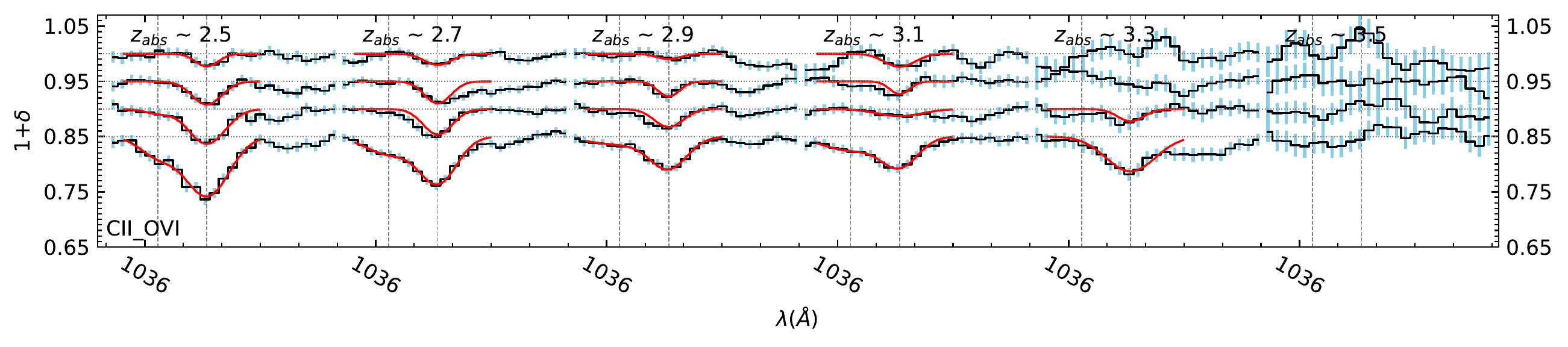} 
\includegraphics[angle=0,width=.2\textwidth,height=0.09\textheight]{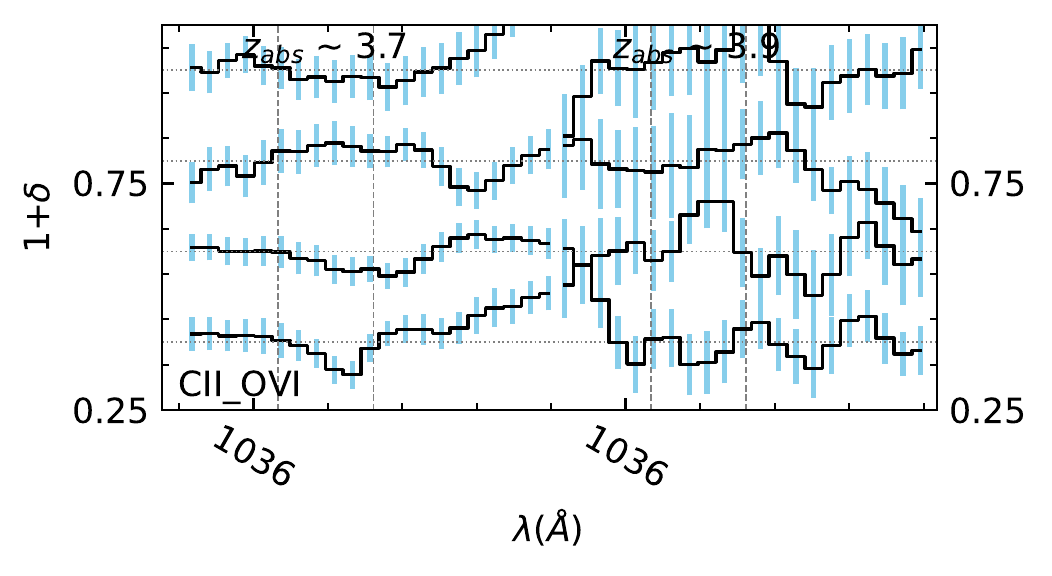}

\includegraphics[angle=0,width=.79\textwidth,height=0.09\textheight]{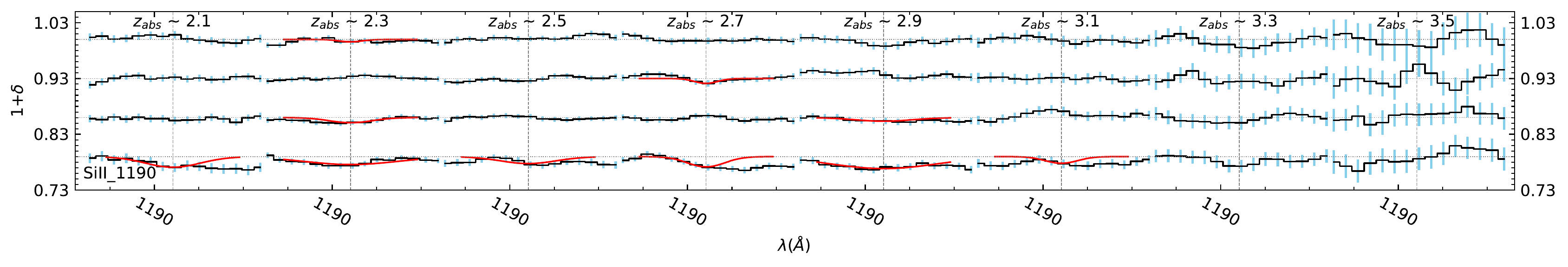} 
\includegraphics[angle=0,width=.2\textwidth,height=0.09\textheight]{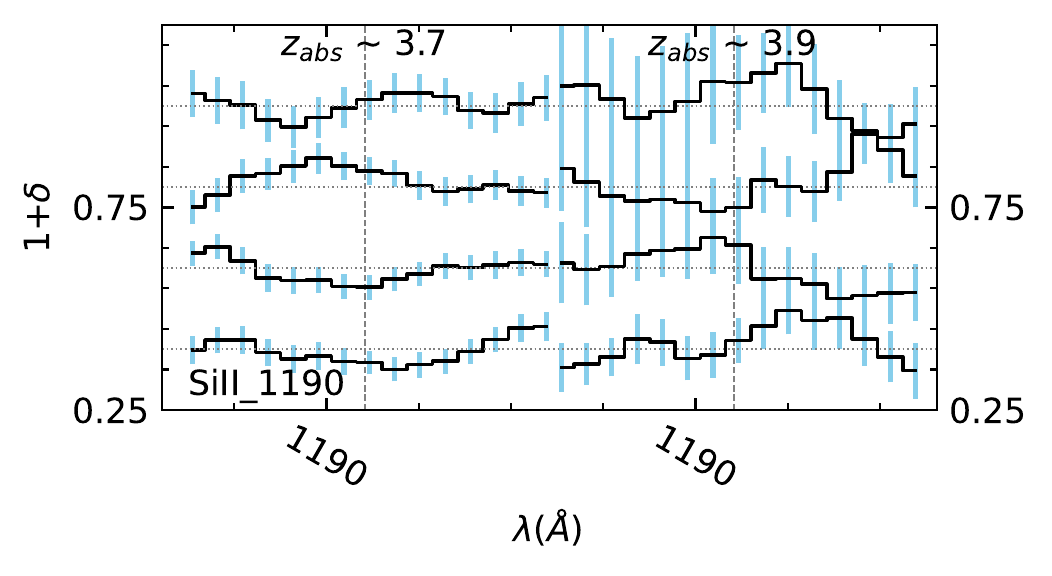}

\includegraphics[angle=0,width=.79\textwidth,height=0.09\textheight]{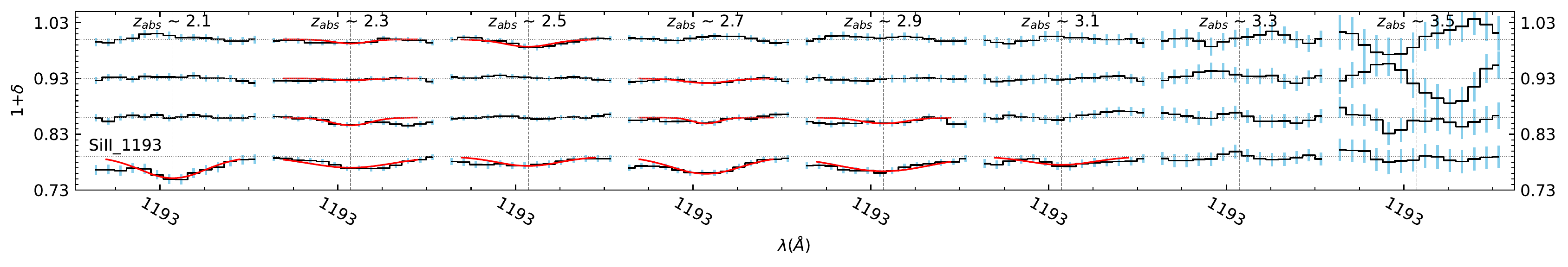} 
\includegraphics[angle=0,width=.2\textwidth,height=0.09\textheight]{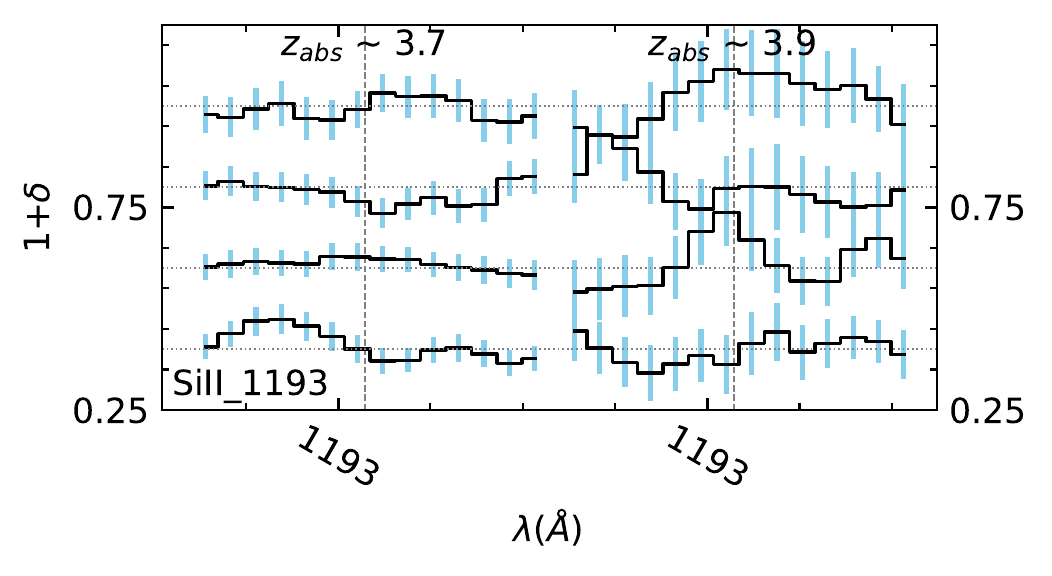}

\includegraphics[angle=0,width=.79\textwidth,height=0.09\textheight]{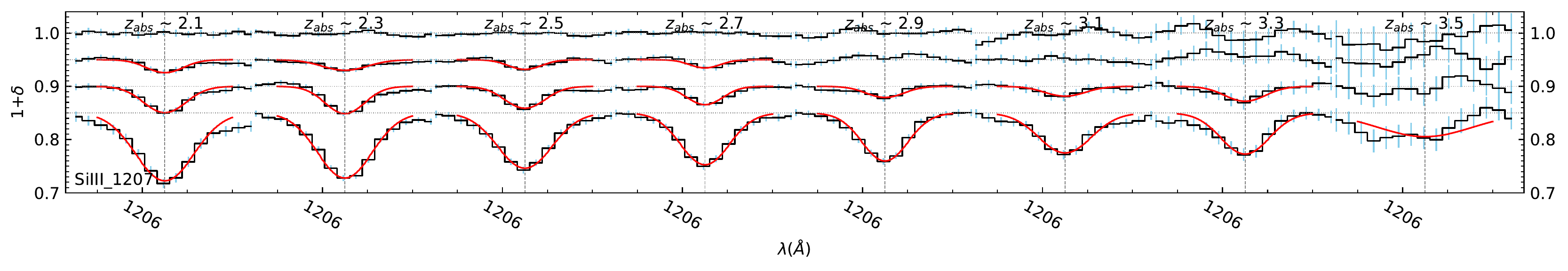} 
\includegraphics[angle=0,width=.2\textwidth,height=0.09\textheight]{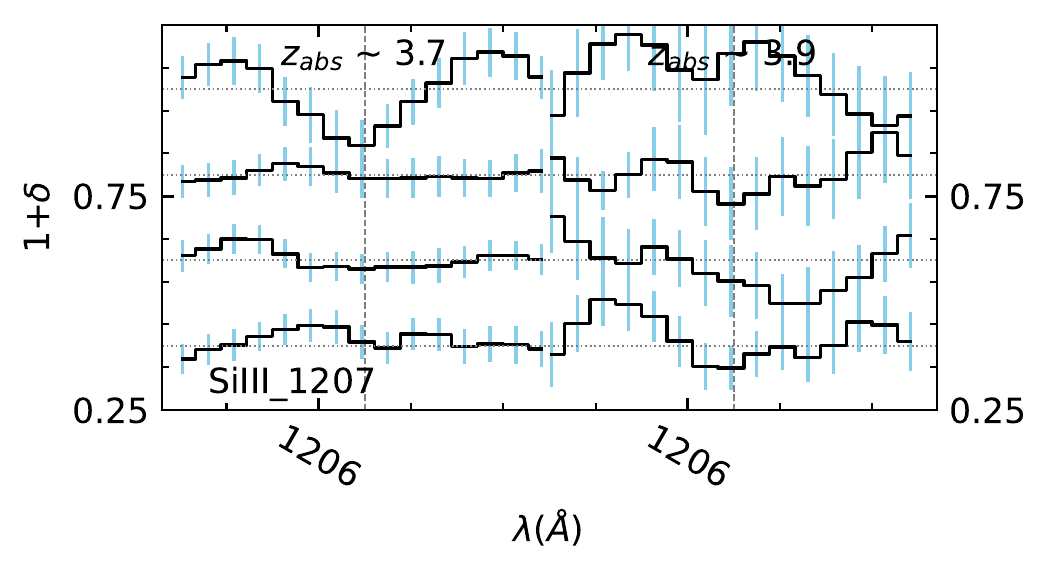}

\includegraphics[angle=0,width=.79\textwidth,height=0.09\textheight]{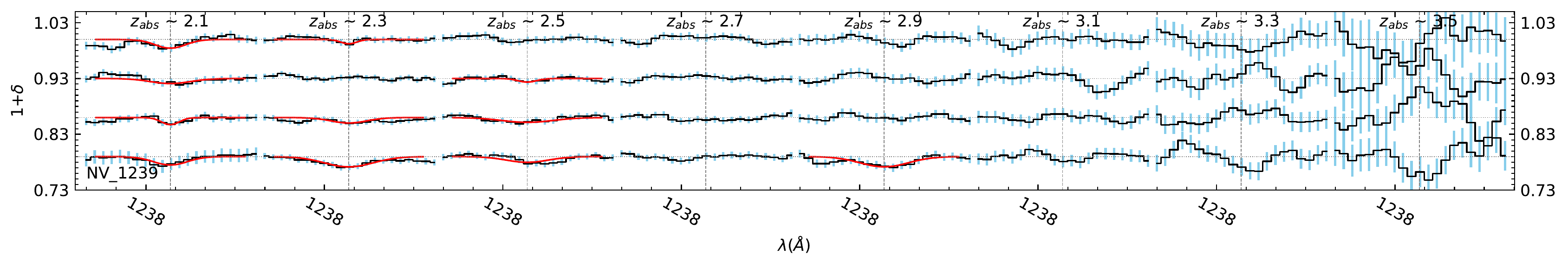} 
\includegraphics[angle=0,width=.2\textwidth,height=0.09\textheight]{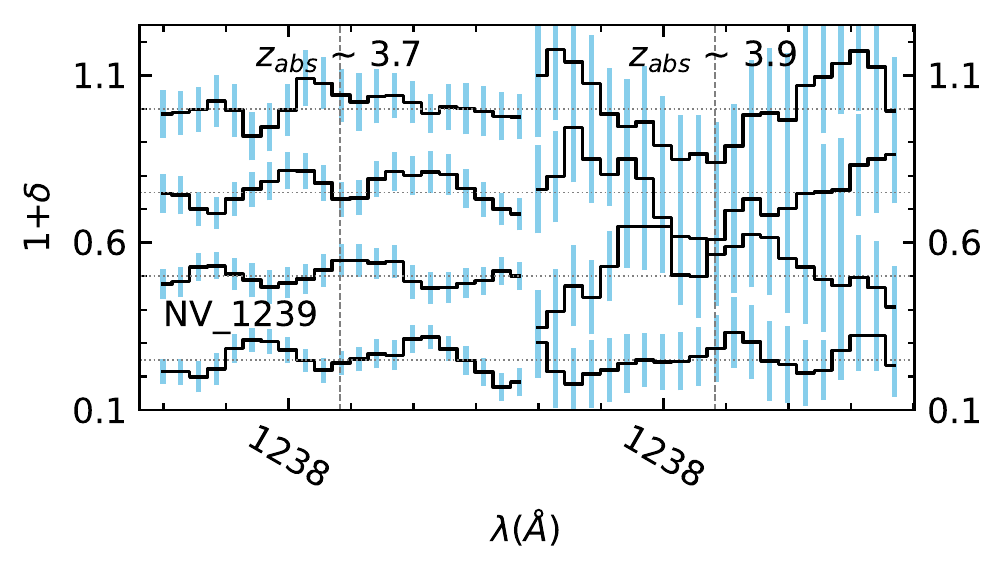}

\includegraphics[angle=0,width=.79\textwidth,height=0.09\textheight]{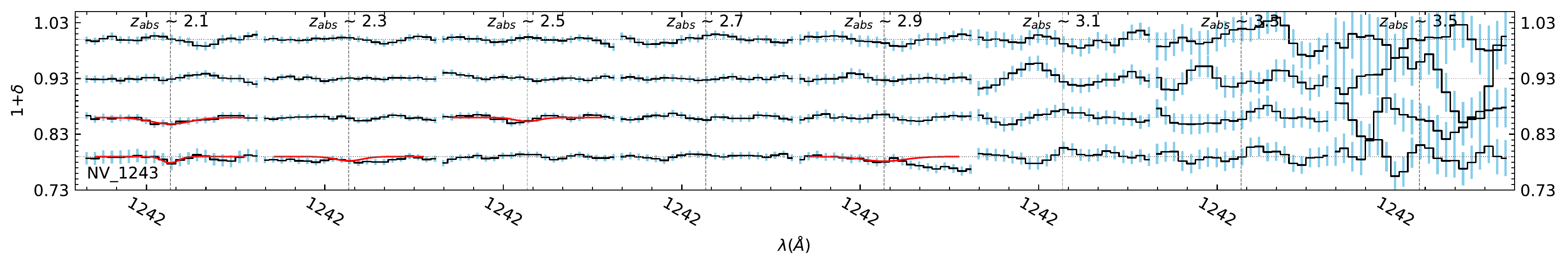} 
\includegraphics[angle=0,width=.2\textwidth,height=0.09\textheight]{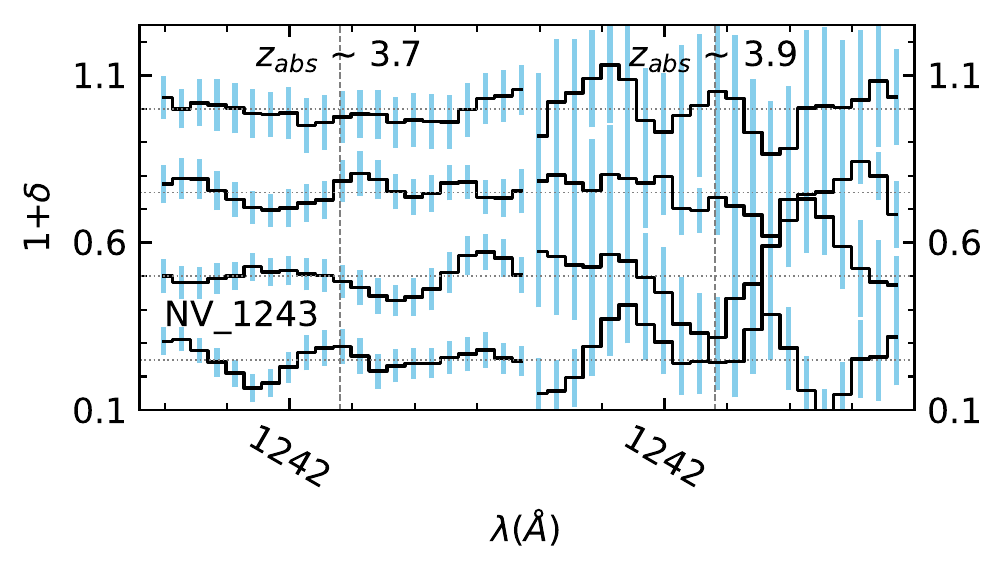} 

\includegraphics[angle=0,width=.79\textwidth,height=0.09\textheight]{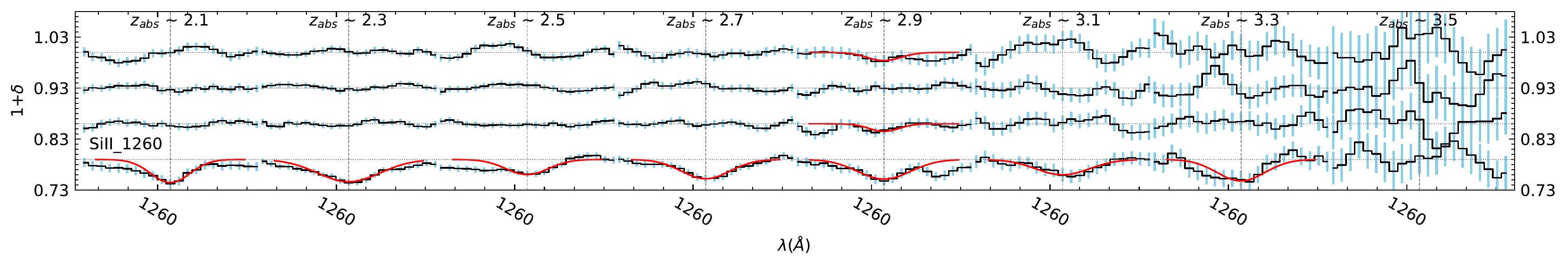} 
\includegraphics[angle=0,width=.2\textwidth,height=0.09\textheight]{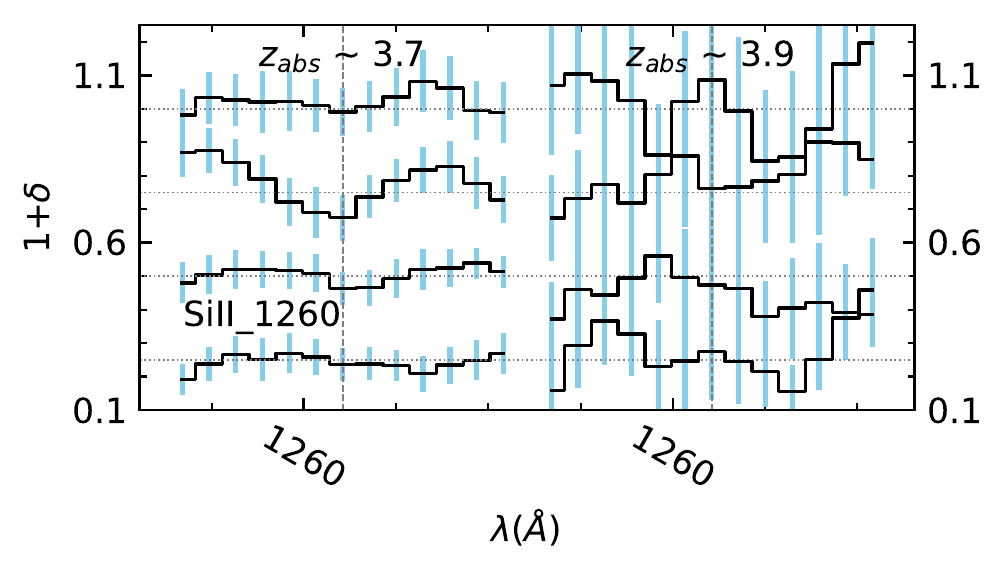}

\includegraphics[angle=0,width=.79\textwidth,height=0.09\textheight]{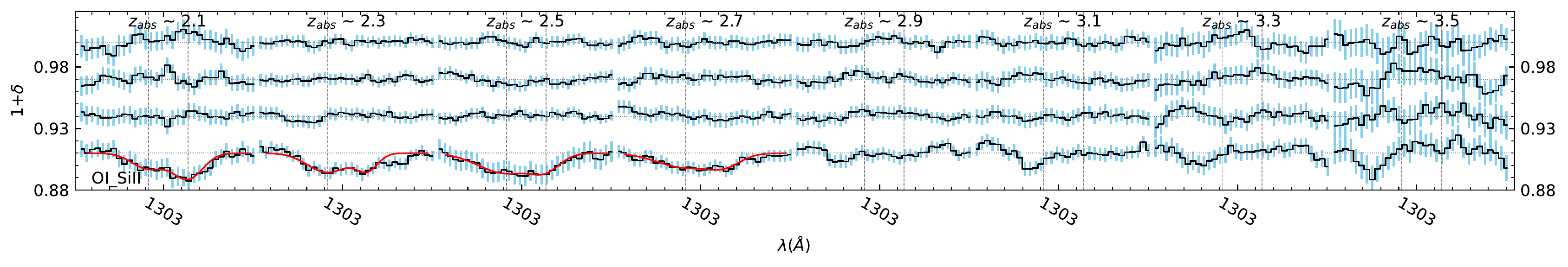} 
\includegraphics[angle=0,width=.2\textwidth,height=0.09\textheight]{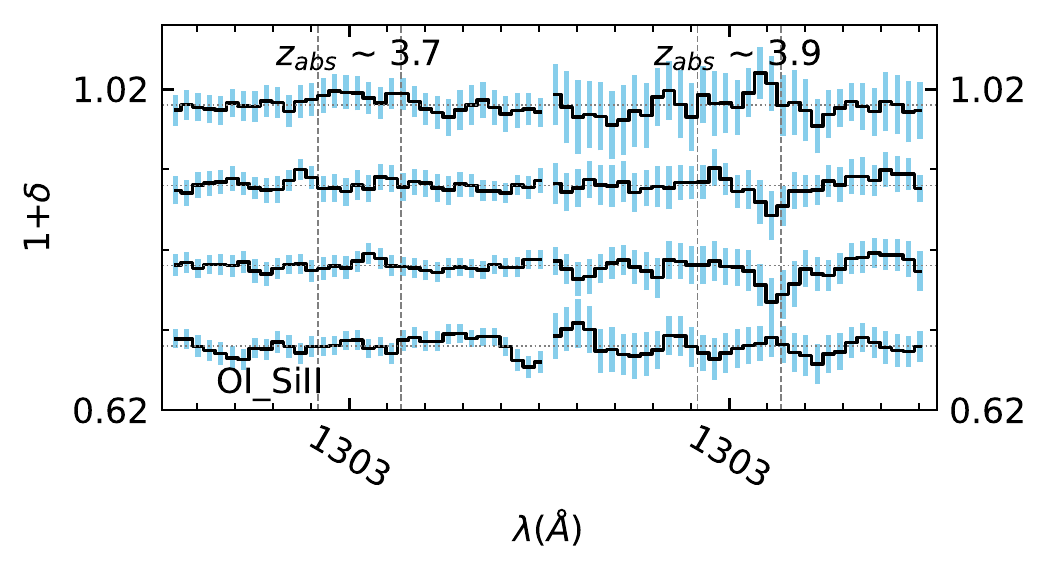}
\caption{
Dependence of line profiles in the stacked spectra on \dl\ and \zabs\ for 13 metal species. In each panel, the relevant species are labelled at the lower-left corner, and the horizontal tick mark interval is 1\AA. The \dl\ and \zabs\ dependences are shown to be along the vertical and horizontal direction, respectively. Offsets are added for different values of \dl. The vertical lines mark the wavelengths of the relevant lines, and the red curves show the Voigt profile fits. 
{
In each row, the cases for $\zabsorb<3.6$ and $3.6<\zabsorb<4.0$ absorbers are plotted in the left and right panel, respectively, and the division is mainly driven by the potentially different dynamical range in the $y$--axis of the right panel to accommodate the larger uncertainties in the profiles at higher redshifts.
}
}

\label{fig:stacked_metal1}
\end{figure*}

\begin{figure*}
\centering
\includegraphics[angle=0,width=.79\textwidth,height=0.09\textheight]{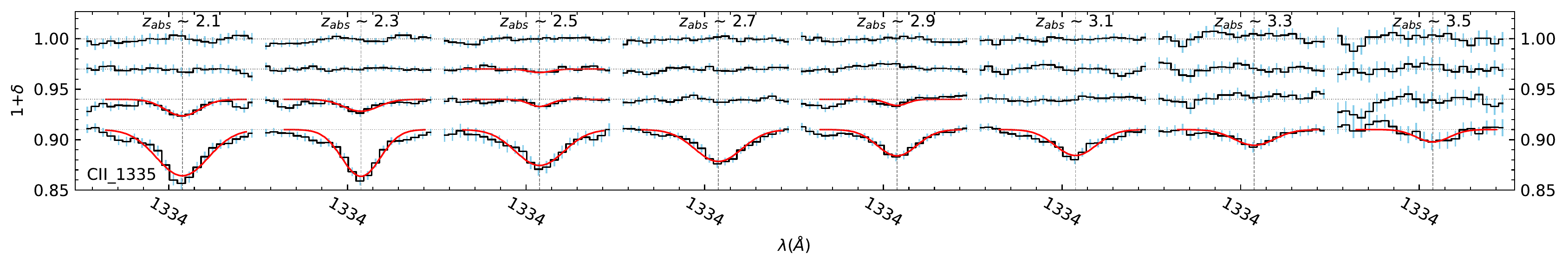} 
\includegraphics[angle=0,width=.2\textwidth,height=0.09\textheight]{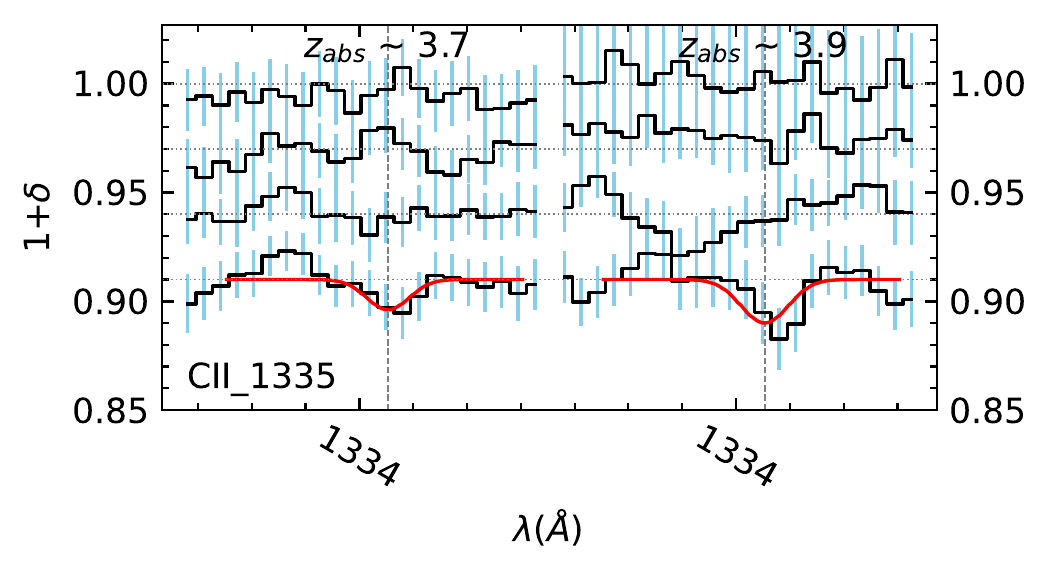}

\includegraphics[angle=0,width=.79\textwidth,height=0.09\textheight]{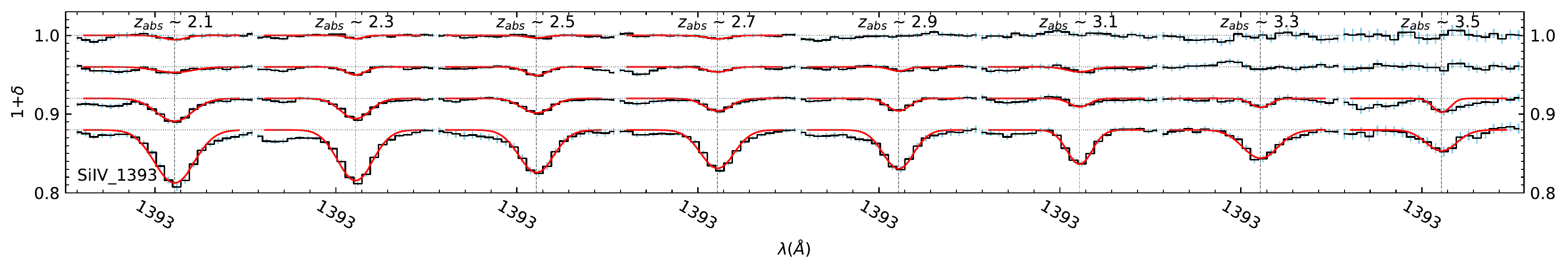} 
\includegraphics[angle=0,width=.2\textwidth,height=0.09\textheight]{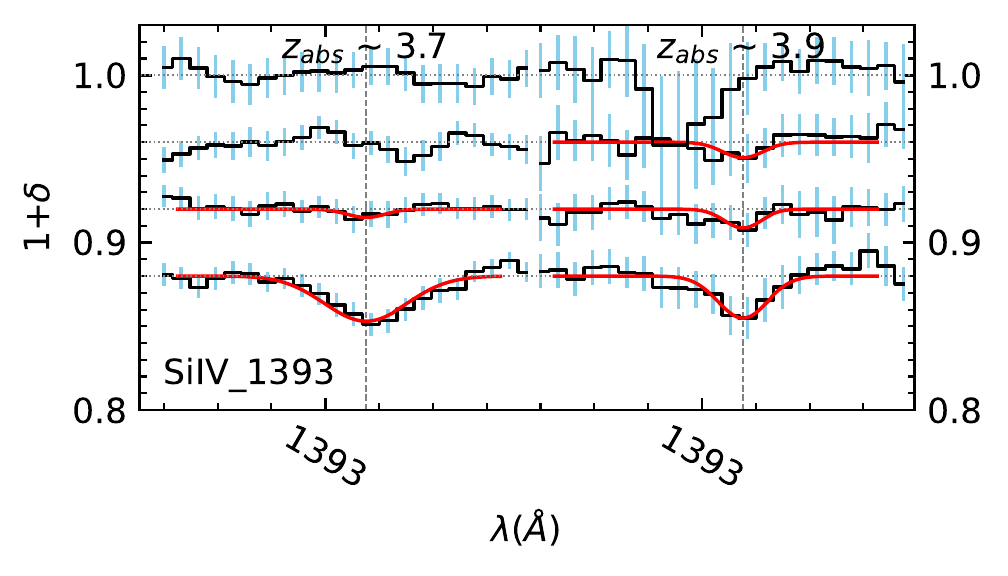}

\includegraphics[angle=0,width=.79\textwidth,height=0.09\textheight]{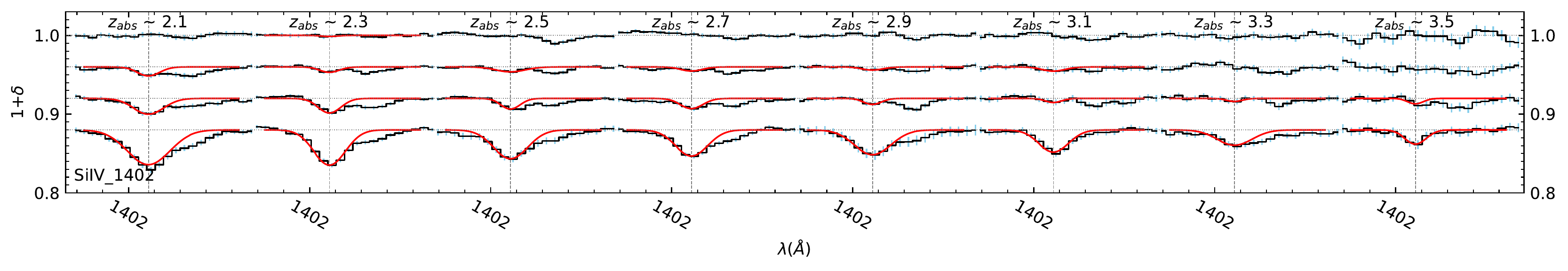} 
\includegraphics[angle=0,width=.2\textwidth,height=0.09\textheight]{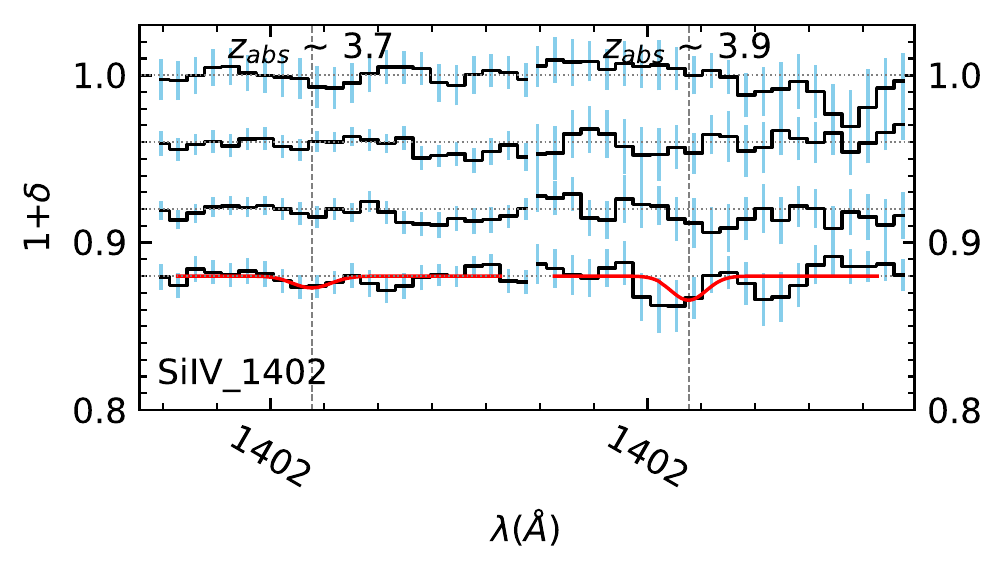}

\includegraphics[angle=0,width=.79\textwidth,height=0.09\textheight]{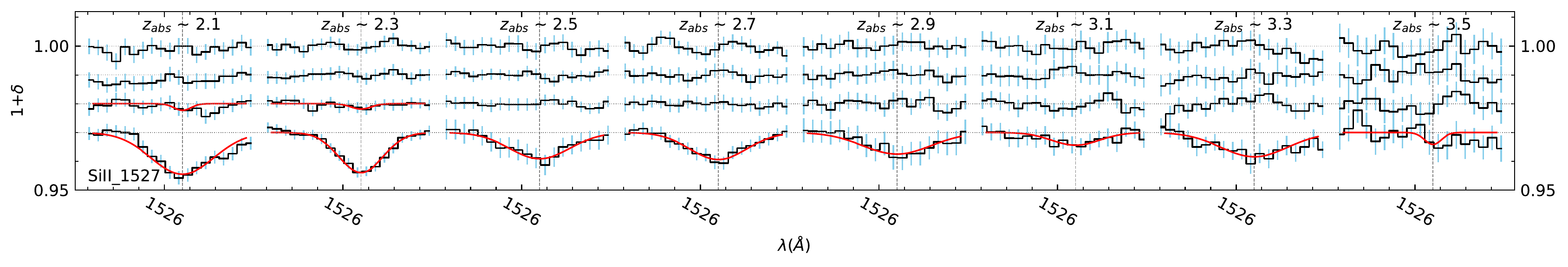} 
\includegraphics[angle=0,width=.2\textwidth,height=0.09\textheight]{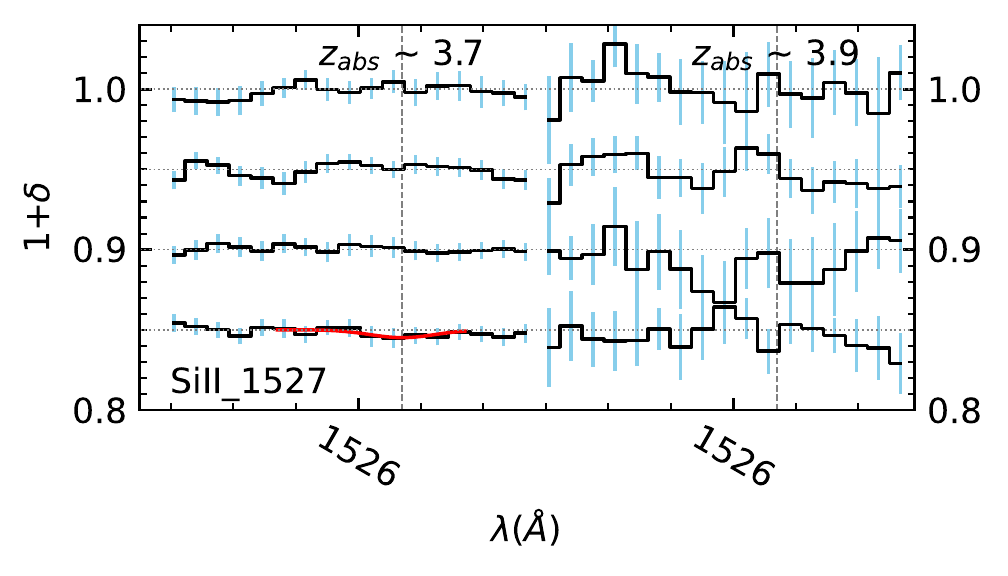} 

\includegraphics[angle=0,width=.79\textwidth,height=0.09\textheight]{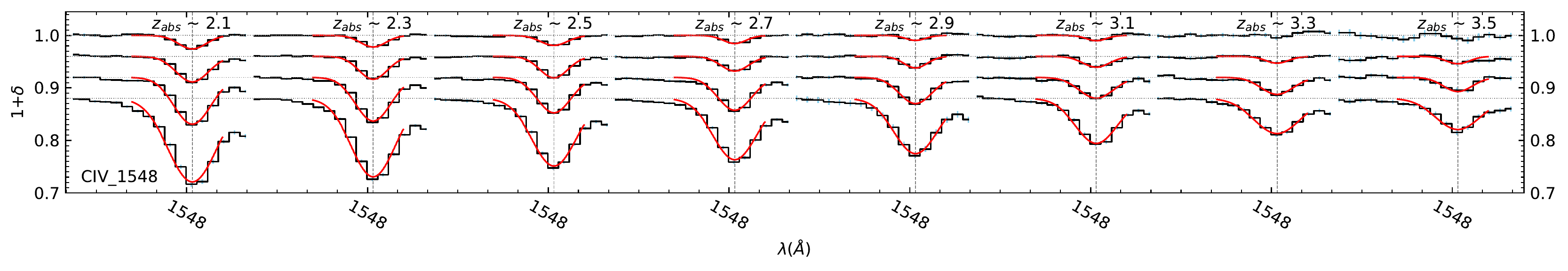} 
\includegraphics[angle=0,width=.2\textwidth,height=0.09\textheight]{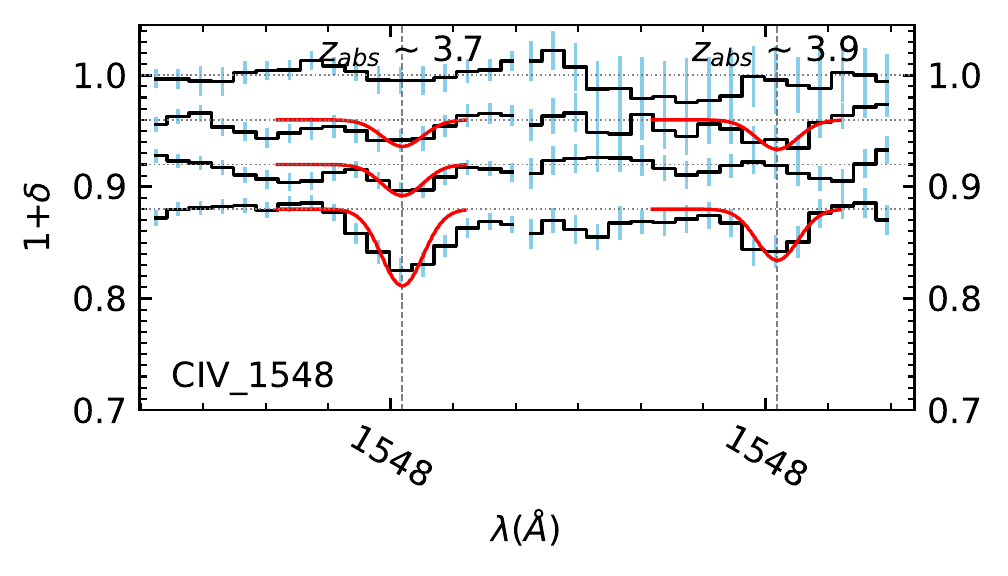}

\includegraphics[angle=0,width=.79\textwidth,height=0.09\textheight]{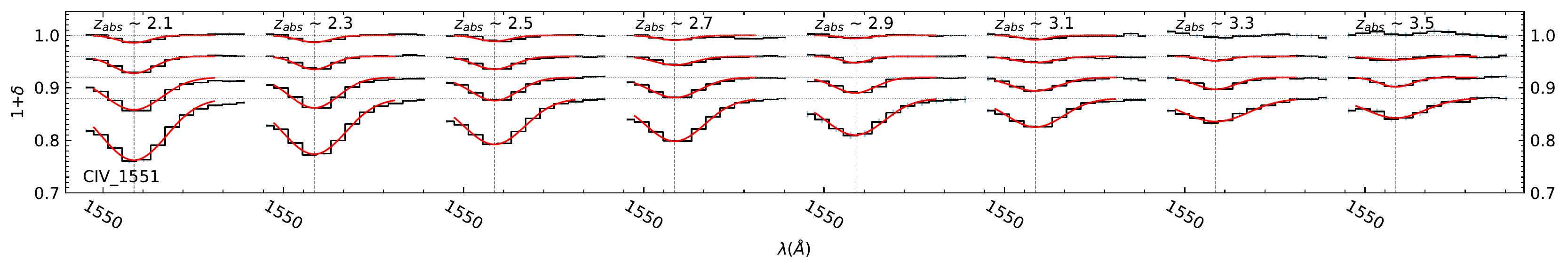} 
\includegraphics[angle=0,width=.2\textwidth,height=0.09\textheight]{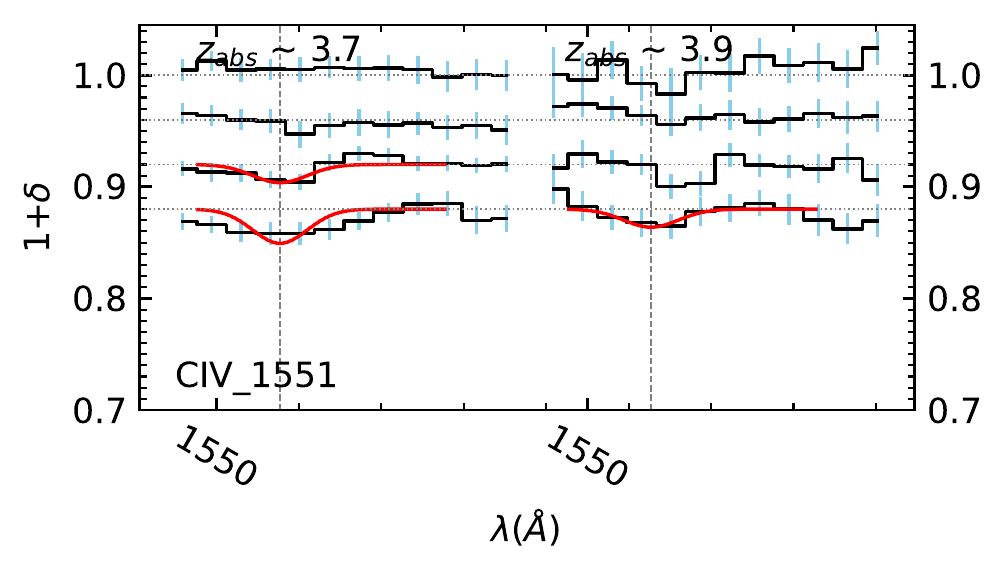}

\includegraphics[angle=0,width=.79\textwidth,height=0.09\textheight]{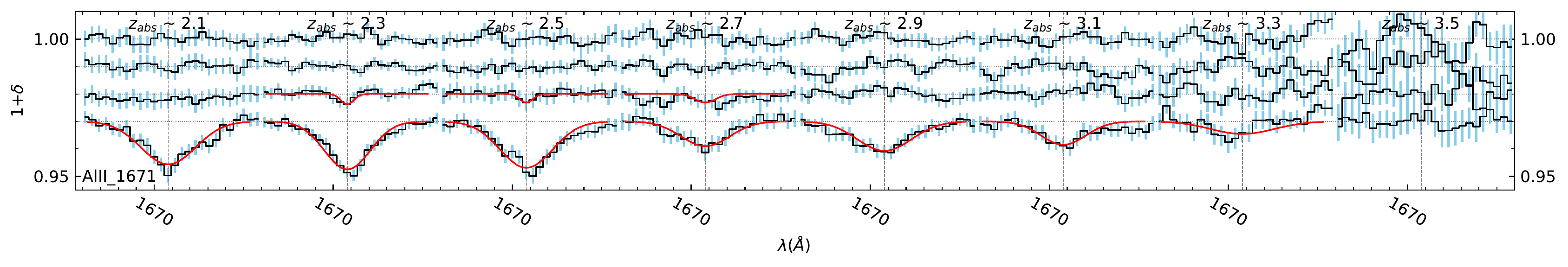} 
\includegraphics[angle=0,width=.2\textwidth,height=0.09\textheight]{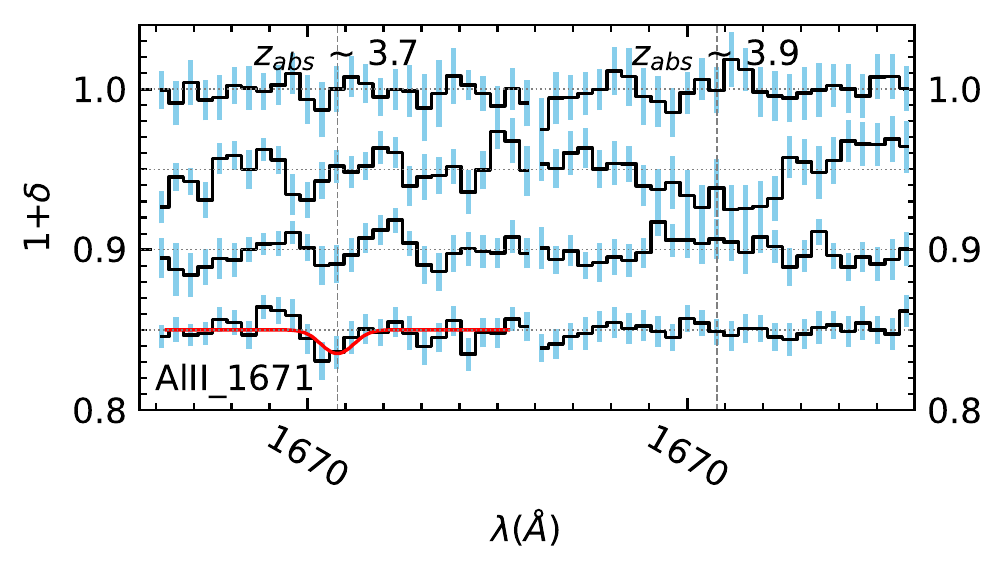}

\includegraphics[angle=0,width=.79\textwidth,height=0.09\textheight]{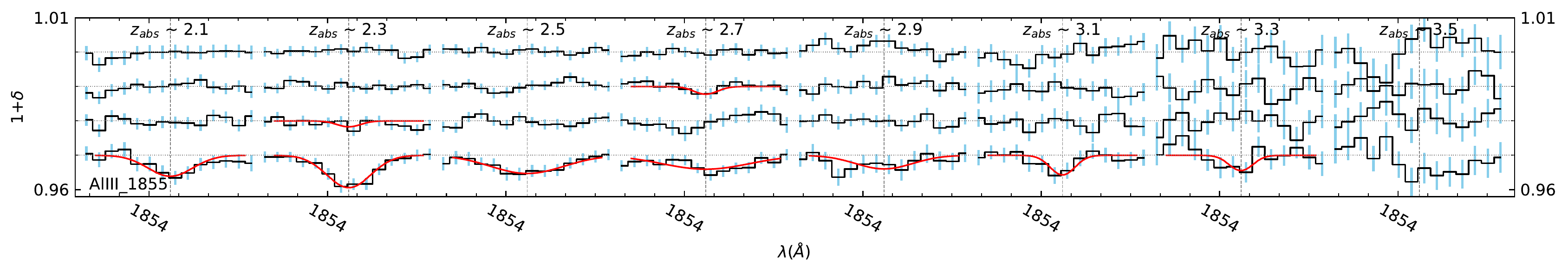} 
\includegraphics[angle=0,width=.2\textwidth,height=0.09\textheight]{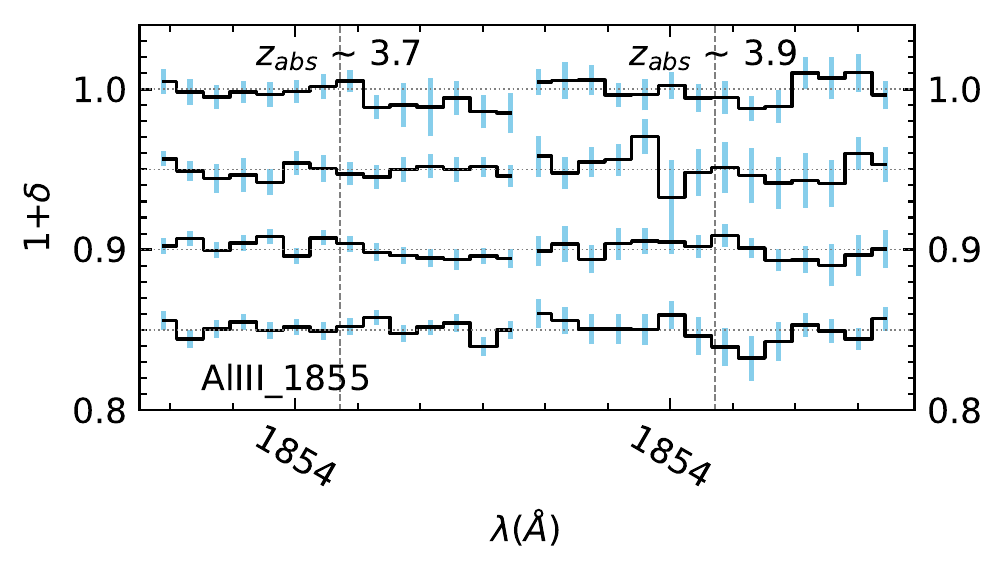}

\includegraphics[angle=0,width=.79\textwidth,height=0.09\textheight]{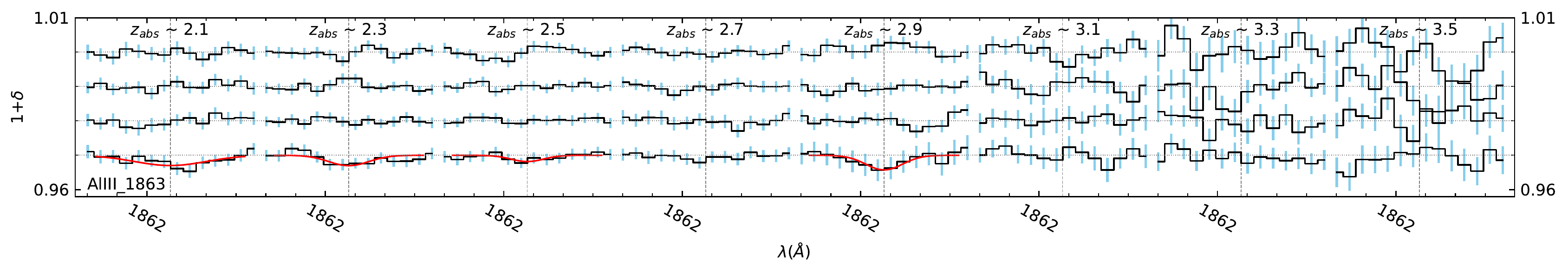}
\includegraphics[angle=0,width=.2\textwidth,height=0.09\textheight]{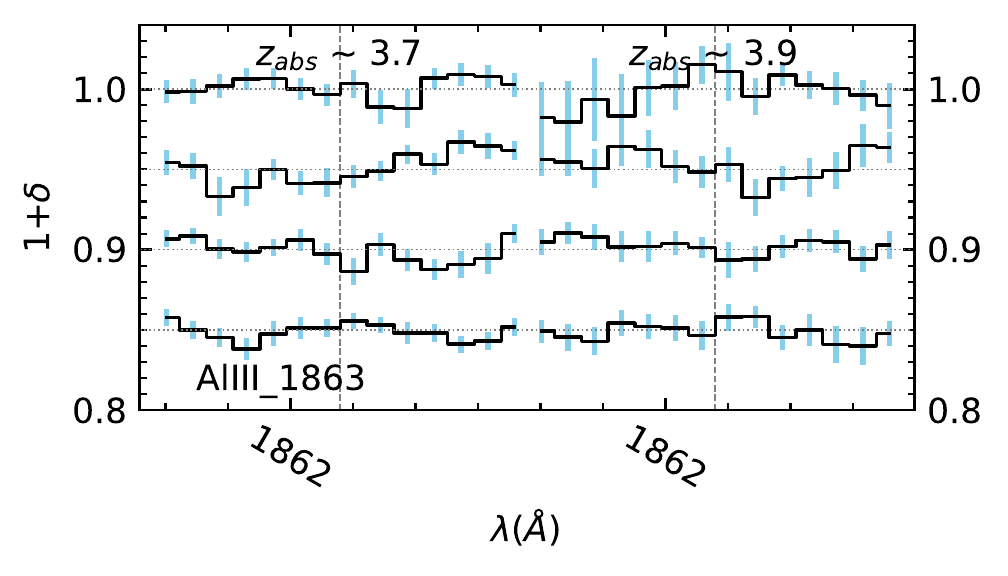} 

\caption{Continued: Dependence of line profiles in the stacked spectra on \dl\ and \zabs\ for 13 metal species.
}
\label{fig:stacked_metal2}
\end{figure*}

\begin{figure}
\includegraphics[angle=0,width=.47\textwidth,height=0.09\textheight]{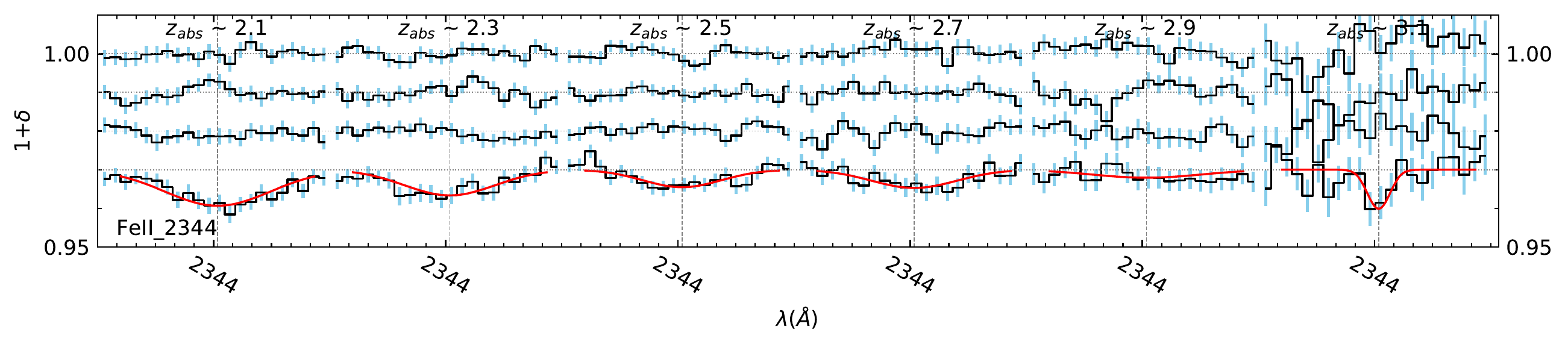} 

\includegraphics[angle=0,width=.47\textwidth,height=0.09\textheight]{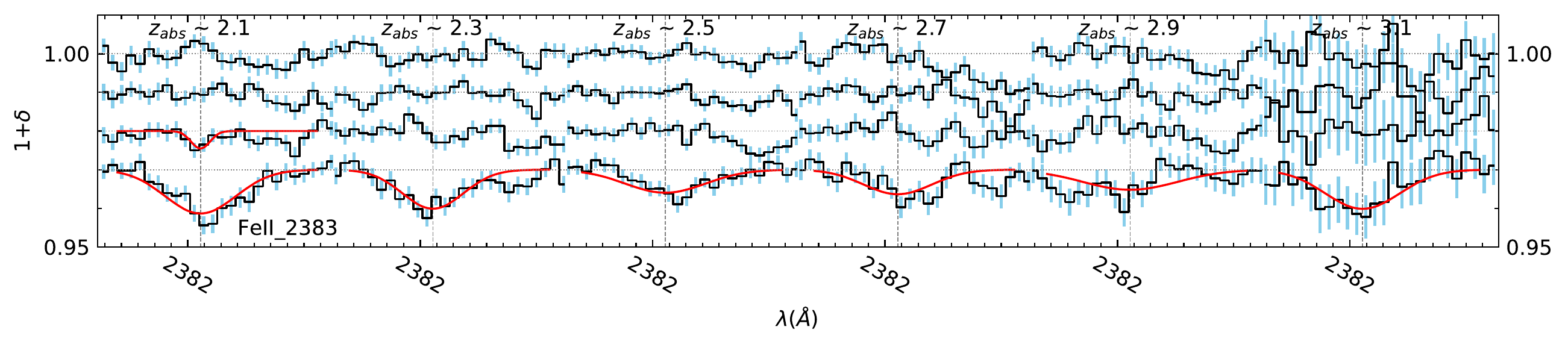} 

\includegraphics[angle=0,width=.33\textwidth,height=0.09\textheight]{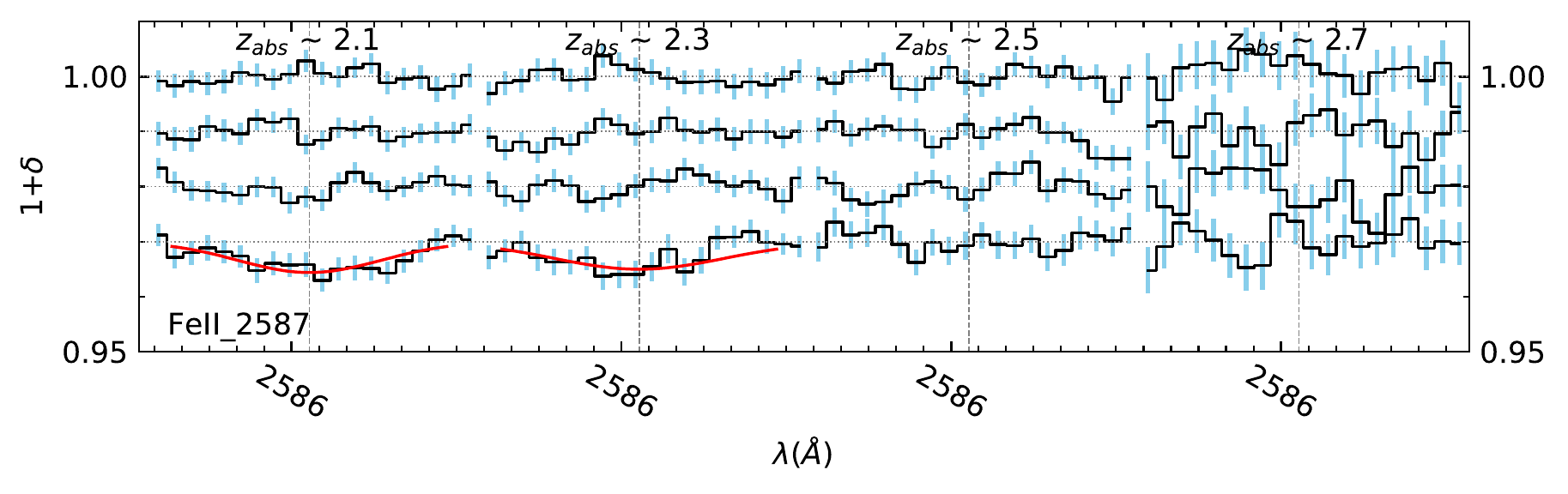}

\includegraphics[angle=0,width=.33\textwidth,height=0.09\textheight]{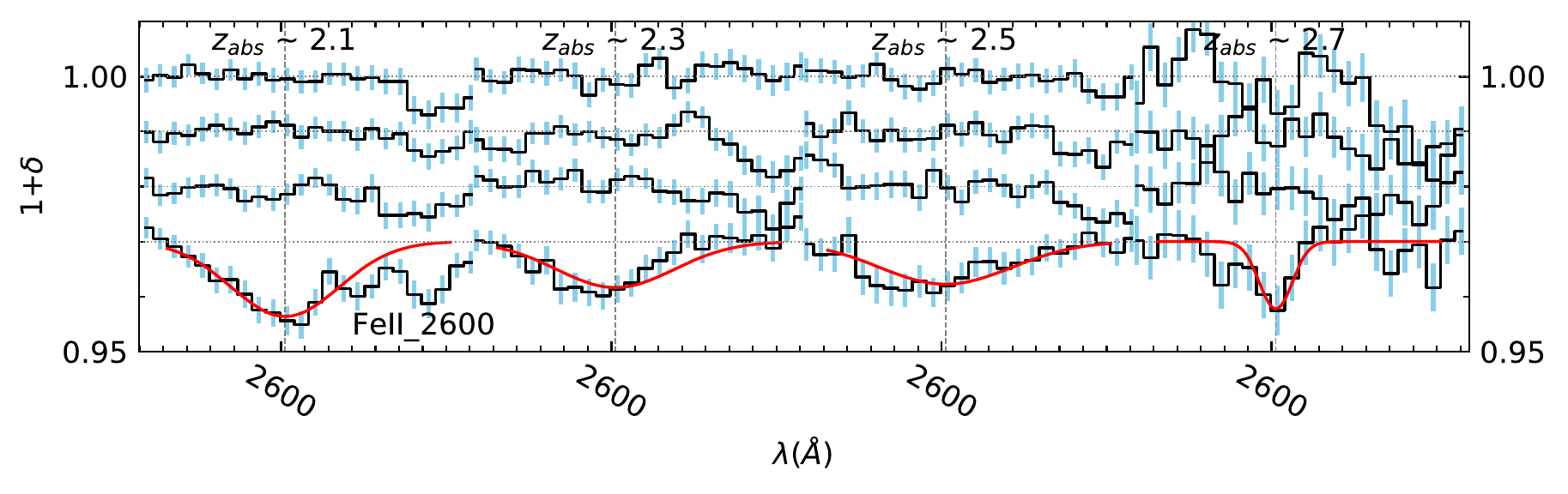}

\includegraphics[angle=0,width=.33\textwidth,height=0.09\textheight]{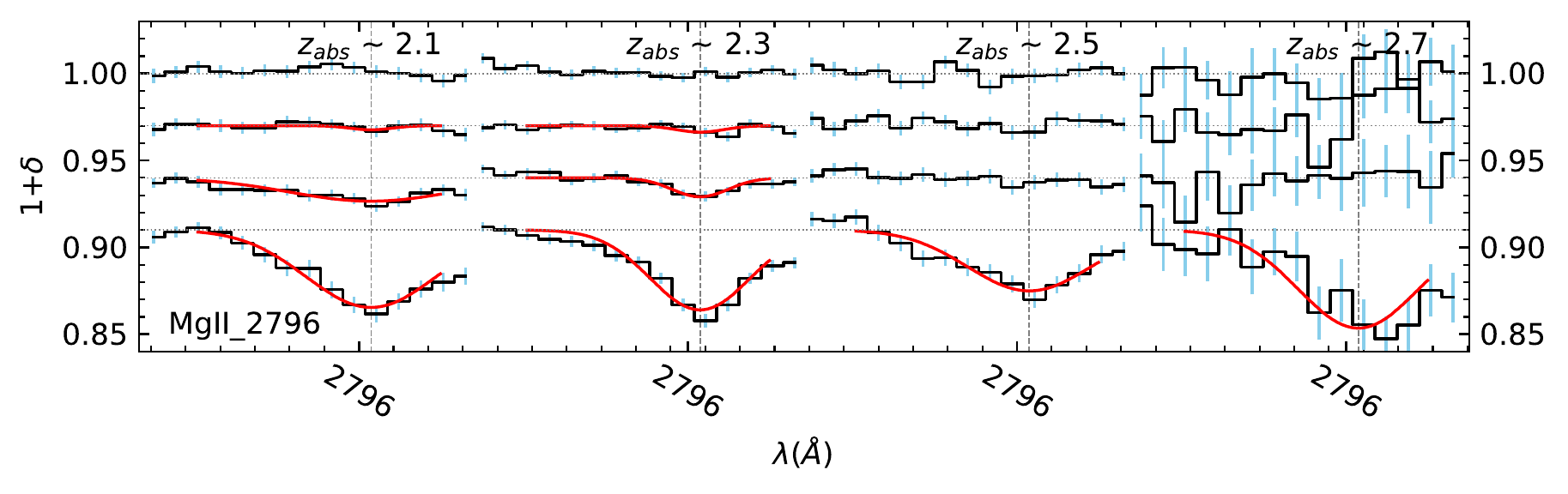}

\includegraphics[angle=0,width=.33\textwidth,height=0.09\textheight]{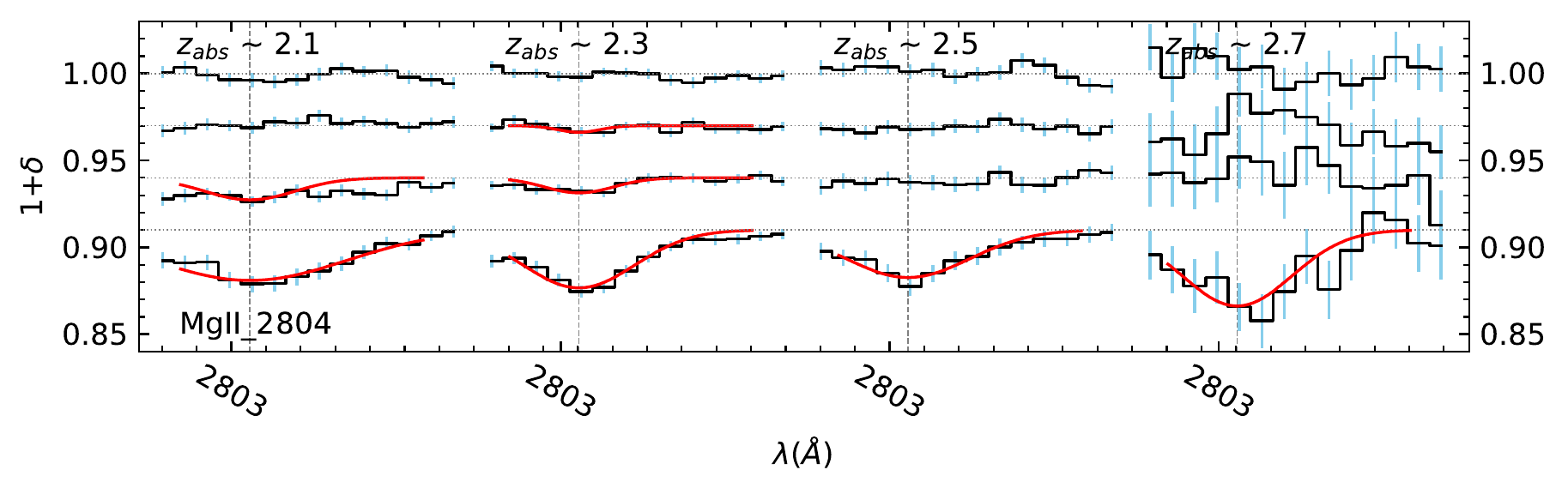}

\caption{Continued: Dependence of line profiles in the stacked spectra on \dl\ and \zabs\ for 13 metal species.
}
\label{fig:stacked_metal3}
\end{figure}

Since our focus in this work is on metal lines, we obtain the stacked profiles for 13 major metal species.  We show the profiles in Figures~\ref{fig:stacked_metal1}, \ref{fig:stacked_metal2}, and \ref{fig:stacked_metal3} for each line at all the four values of \dl\ and at different absorber redshift \zabs. We choose to present all the stacked profiles, regardless of the signal-to-noise ratio, to obtain a full view of the data. The two panels in the same row represent those profiles with high and low signal-to-noise ratio, respectively. The stacked spectra in the right panels typically do not lead to any robust results with Voigt profile fitting.  

In bins of the lowest redshift ($\zabsorb\sim 2.1$), we miss the species with the shortest wavelengths, like \ion{C}{3} 977\AA, \ion{O}{6} 1032\AA, \ion{C}{2} 1036\AA, and \ion{O}{6} 1038\AA, as they do not enter into the observed wavelength range. Similarly, in bins of the highest redshifts ($\zabsorb>3.2$), the species with the longest wavelengths (\ion{Fe}{2} and \ion{Mg}{2}) are not probed by the data. 

At each redshift, for each metal line, there is a clear trend that the line strength increases with increasing $|\deltalya|$, implying that more metals are present in denser regions. At fixed \dl, each metal line becomes weaker toward higher redshift.

At a given \dl\ and \zabs, Voigt profile fitting is carried out for each metal line, as long as the signal-to-noise ratio allows. 
{
Specifically, if the minimum pixel value of an absorption line lies below $1\sigma$ of the noise level of the ``continuum'', we perform the profile fitting.
}
The bestfit is shown as the red solid curve. The derived column density $N$ and Doppler parameter $b$ can be found in Tables~\ref{table_Zabs[2.0,2.4]} in Appendix~\ref{sec:appendix_N_metals}. Overall, if the results are averaged over the redshift range, we find broad agreements with those in \citet{Pieri2010} and \citet{Pieri2014}. 

\begin{figure*}
\centering
\includegraphics[width=.45\textwidth,height=.8\textheight]{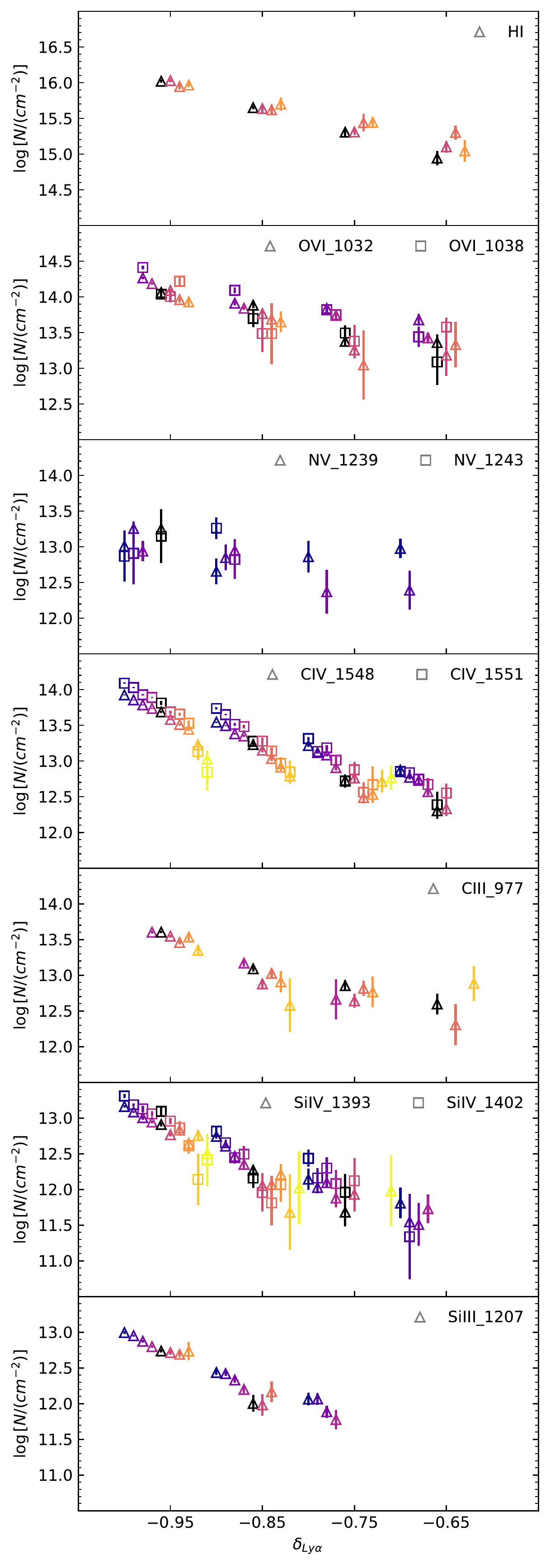} 
\includegraphics[width=.45\textwidth,height=.8\textheight]{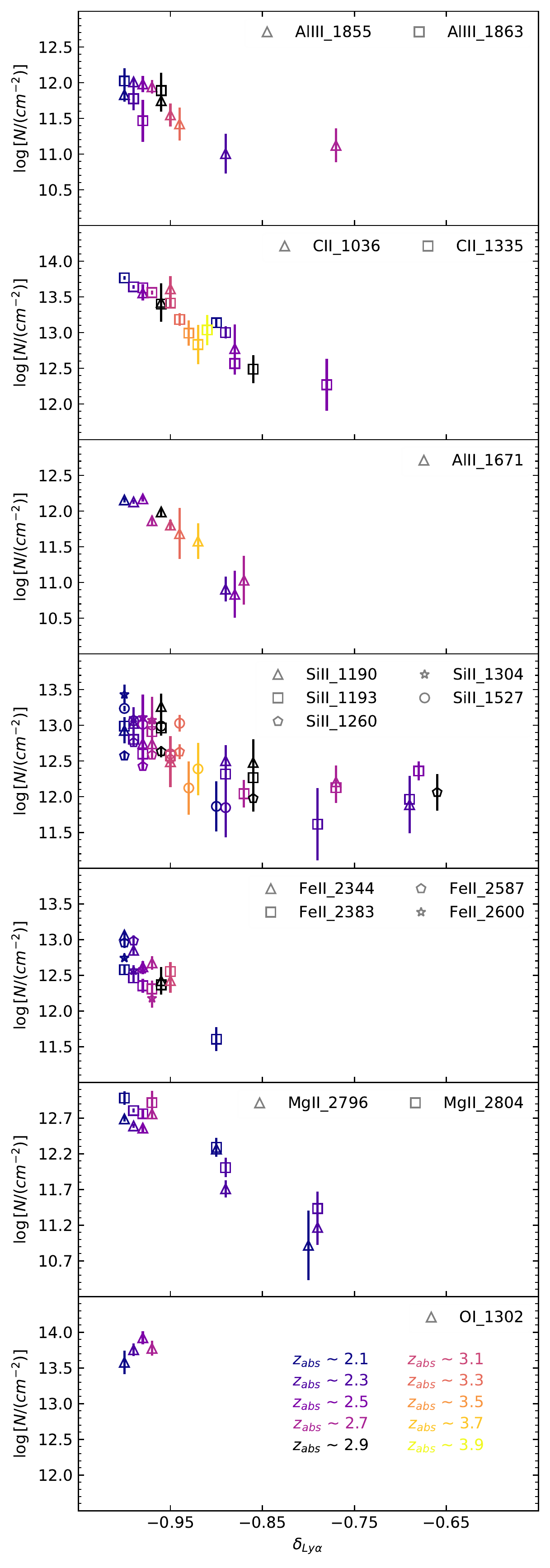} 
\caption{
Dependence of column density on \dl\ and absorption redshift for 13 metal species as well as neutral hydrogen. In each panel, there are up to four sets of data points for the four values of \dl\ ($-0.95$, $-0.85$, $-0.75$, and $-0.65$). Each set of data points at a given \dl\ are color-coded by the absorber redshift. If there are multiple lines of a species, they are distinguished by different symbols.
}
\label{fig:N_metals}
\end{figure*}

Figure~\ref{fig:N_metals} summarizes the dependence of the column density on \dl\ and \zabs\ for various metal species as well as neutral hydrogen. In each panel, up to four sets of data points are shown, corresponding to the four values of \dl\ ($-0.95$, $-0.85$, $-0.75$, and $-0.65$). Each set of data points are color-coded by the absorber redshift \zabs. The column density ranges of the panels are all chosen to across 3 dex for an easy comparison between the strengths of the column density dependence on \dl\ and \zabs\ of different species.

At each redshift, the column density of each metal species drops as \dl\ increases from $-0.95$ to $-0.65$. The column density of neutral hydrogen (top-left panel) drops by about one dex, from $\log (N_{\rm HI}/{\rm cm}^{-2})\sim$16 to $\sim$15. The change of column densities of high-ionization metal species (such as \ion{O}{6}, \ion{N}{5}, \ion{C}{3}, and \ion{C}{4}) seems to track that of neutral hydrogen, all with a $\sim$1 dex decrease over the \dl\ range. In contrast, the column densities of low-ionization metal species show a stronger dependence on \dl. For example, that of \ion{Mg}{2} drops by almost 2 dex over the \dl\ range. Such a difference implies that high- and low-ionization species arise from different environments, as will be shown with a simple model in the next subsection.

At a given \dl, the column density of neutral hydrogen show almost no dependence on redshift. In comparison, those for metal species at fixed \dl\ show a clear dependence on redshift, typically more than 0.5 dex from $\zabsorb\sim 2.1$ to $\zabsorb\sim 3.5$, which may result from the evolution in temperature and ionizing background.

To help understand these results, we turn to a simple model of \lya\ absorbers.

\subsection{Illustrative Model of Metals in the \lya\ Forest of High Neutral Hydrogen Column Density}

The column densities of various species encode information about the physical properties of the absorption system. Similar to \citet{Pieri2014}, we apply a simple, illustrative photoionization model to interpret the column density measurements.

\subsubsection{Parameter Dependence}

\begin{figure}
\centering
\includegraphics[angle=0, width=.45\textwidth]{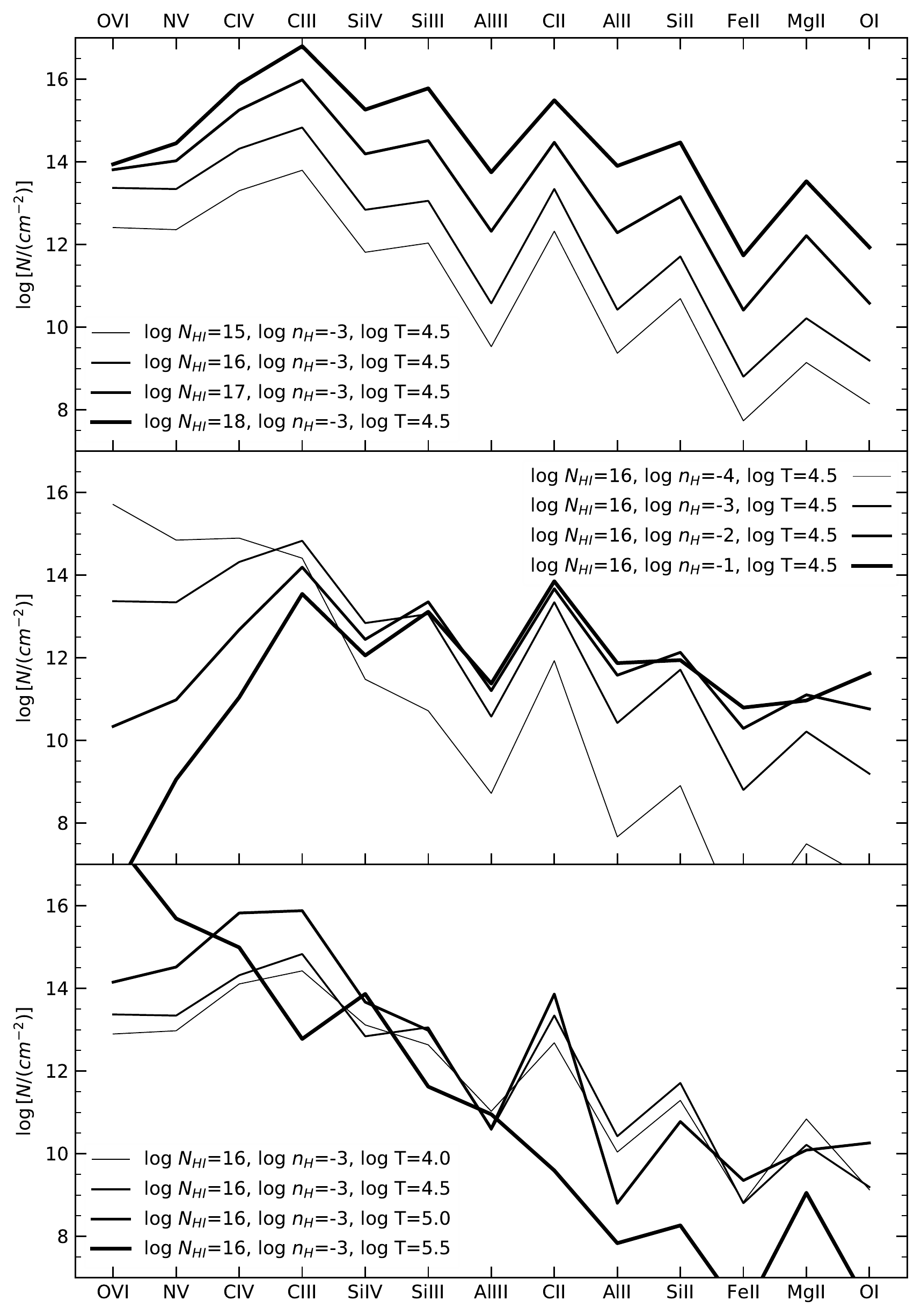}
\caption{
Dependence of column densities of various species on neutral hydrogen column density (top), total hydrogen number density (middle), and temperature (bottom), based on Cloudy models. The species are ordered from high-ionization to low-ionization potential from left to right. The metallicity is fixed to ${\rm [X/H]}=-1$ for all the cases. 
}
\label{fig:N_element_par}
\end{figure}

To guide our choice of models to compare to the measurements, in Figure~\ref{fig:N_element_par}, we show the dependence of the column density of various species on the stop neutral hydrogen column density ($N_{\rm HI}$), total hydrogen number density ($\nH$), and temperature ($T$), with gas in a plane-parallel geometry photoionized by an external ionizing UV background. The nearly linear dependence of the column density of a species on metallicity (with other parameters fixed) is not shown here. Figure~\ref{fig:model_trend} in Appendix~\ref{sec:appendix_N_metals} further explores the parameter dependence. 

In Figure~\ref{fig:N_element_par}, the species are ordered such that the ionization potential decreases from left to right. Calculations are performed with version 17.02 of photoionization code Cloudy, last described by \citet{Cloudy17}. {The gas was assumed to be a uniform slab of constant density in thermal and ionization equilibrium and exposed to the external UV background.} The models explored are centered at a model with total hydrogen density $\log (\nH/{\rm cm}^{-3})=-3$ and temperature $\log(T/{\rm K})=4.5$. The metallicity is fixed at ${\rm [X/H]}=-1$. Solar abundance is assumed and the \citet{HM12} UV background at $z=2.75$ is adopted in the calculation. With the central model, we vary the three parameters, $N_{\rm HI}$, $\nH$, and $T$, one at a time. Each panel shows one set of change, and thicker lines are for higher values of the corresponding varied parameter.

As with the metallicity dependence, there is a clear trend that a species' column density increases with increasing stop neutral hydrogen column density (top panel). The dependence is stronger for low-ionization species. For example, for a 3 dex change in $N_{\rm HI}$, the column density of \ion{Mg}{2} increases by about 4.5 dex, while the change is only about 1.5 dex in that of \ion{O}{6}. This is expected in the photoionization model. At higher hydrogen column density, the ionizing background is more attenuated within the absorption system. The subsequent lower photoionization rate causes a given species to stay at a low-ionization state than a high-ionization state.

As for the dependence on total hydrogen density $\nH$, a higher $\nH$ results in a higher recombination rate, making it hard for a species to maintain a state of high-level ionization. As a consequence the column densities of high-ionization species drop by a large factor (middle panel). For example, as $\nH$ increases from $\log (\nH/{\rm cm}^{-3})=-3$ by 1 dex, the column densities of \ion{O}{6}, \ion{N}{5}, and \ion{C}{4} decrease by about 3 dex, 2.5 dex, and 2 dex, respectively. The recombination from higher-level ionization, on the other hand, increases the column densities of low-ionization species by a mild amount, e.g., $\sim$1 dex for \ion{Mg}{2}. In contrast, lowering $\nH$ by 1 dex leads to a substantial drop in the column densities of low-ionization species (e.g., by $\sim$2.5 dex for \ion{Mg}{2}) and a clear increase in column densities of high-ionization species (e.g., by $\sim$2.5 dex for \ion{O}{6}). As we will see, such an increase for high-ionization lines at low density provides a possible mechanism to explain the measurements.

Increasing the temperature by 0.5 dex from $\log(T/{\rm K})=4.5$ (hence reducing the recombination rate) has the effect of enhancing the high-ionization species relative to low-ionization ones (bottom panel). Compared to the lower density case (middle panel), the relative abundances of high-ionization lines do not change much. For example, the column density of \ion{O}{6} still appears to be lower than that of \ion{C}{4}. The column density change in the low-ionization species is not as drastic as the lower density case. Compared with the case of increasing temperature by 0.5 dex, decreasing the temperature by 0.5 dex only has a small effect, shifting the column densities by at most 0.5 dex.

\subsubsection{A Two-Component Model}

\begin{figure}
    \centering
    \includegraphics[width=0.45\textwidth]{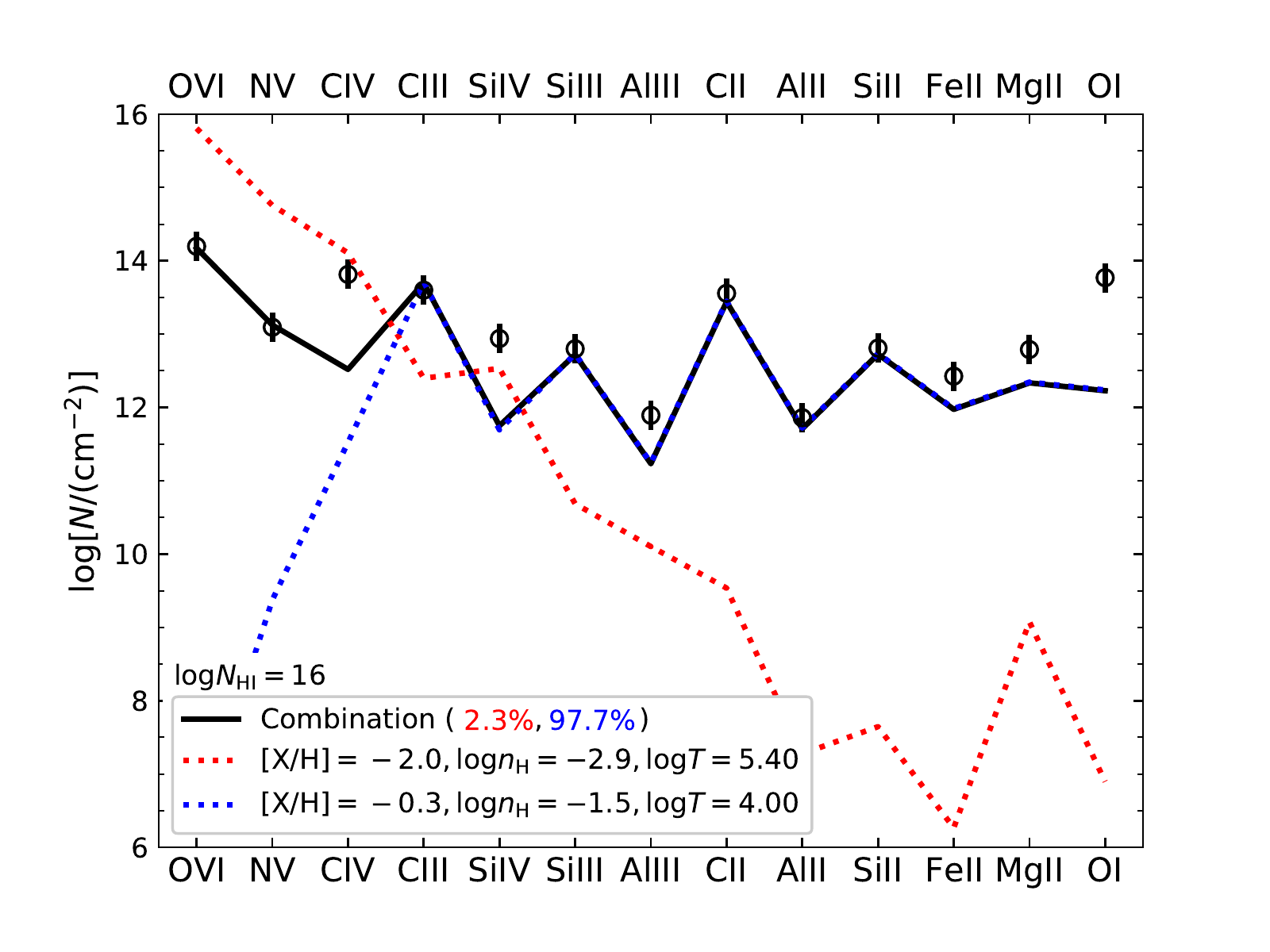}
    \includegraphics[width=0.45\textwidth]{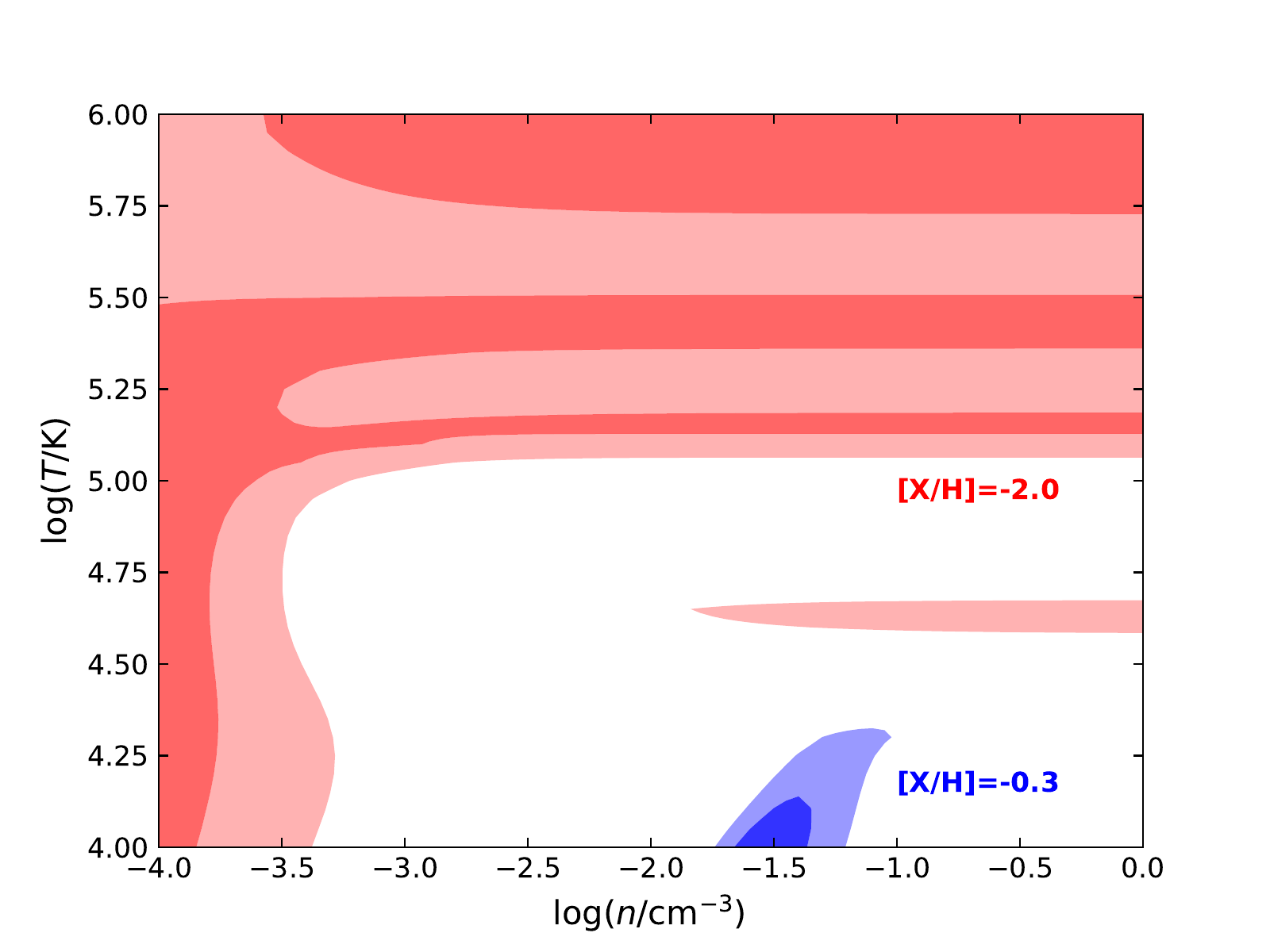}
    \caption{Illustration of the combination of two absorption systems to compare to the observed column densities of various species with $\deltalya\sim-0.95$ and $2.6<\zabsorb <2.8$. Top: the best fit (black) from a combination of a low-metallicity system (red) and a high-metallicity system (blue). Bottom: the 1$\sigma$ and 2$\sigma$ constraints in $n_{\rm H}$ and $T$ for the combination of a low-metallicity and a high-metallicity system. 
    }
    \label{fig:cloudy_fits}
\end{figure}

To see how the measured column densities of various species compare with Cloudy models, we use the absorption system selected by $\deltalya=-0.95$ and $2.6<\zabsorb<2.8$ as an example. The data points in the left panel of Figure~\ref{fig:cloudy_fits} show the inferred column densities (see Table~\ref{table_Zabs[2.0,2.4]} {for $2.6<\zabsorb<2.8$ absorbers}). For those lines with column density measurements barely missed at this redshift but with measurements at the neighboring redshifts, such as \ion{C}{2} 1036 \AA\ and \ion{O}{6} 1038 \AA, we apply interpolations to obtain the $z\sim 2.7$ values. For the same species with multiple lines, we find that the (logarithmic) column density measurements from different lines can have a systematic difference up to 0.10--0.25 dex (e.g., \ion{C}{4} or \ion{Fe}{2}), and we take the
maximum column density as the measurement\footnote{We find that the discrepancies between column densities from the different lines (e.g., doublets) of an ion are related to the difference in their cross-sections (oscillator strengths). The line with the larger cross-section is more easily saturated. With the low-resolution spectra, the Voigt fit has the tendency of pushing the model toward the linear regime by fitting it with a large $b$ value. This would lead to a smaller fitted column density. Therefore, the column density from the line with the minimum oscillator strength is a better estimate. We perform tests of using the inverse variance weighted mean column densities in doing the fitting with the Cloudy model, and the results remain similar.}.

Overall, the column densities of high- and low-ionization species are within a relatively narrow range of about 2 dex. In contrast, the single component models presented in Figure~\ref{fig:N_element_par} all show a large range in column density. In particular, these models are unable to make the column densities of high-ionization species to be at a comparable level as those of low-ionization species, as seen in the data. Motivated by \citet{Pieri2014}, where the high-ionization and low-ionization species are compared to different models, we explore the scenario that the inferred column density pattern results from a superposition of the contributions from a high-metallicity and a low-metallicity component. 

We build two sets of Cloudy models with gas in a plane-parallel geometry, one with metallicity $[{\rm X/H}]=-2$ and the other with $[{\rm X/H}]=-0.3$. For each set, we compute models on a grid, $\log(n_{\rm H}/{\rm cm^{-3}})=-4$ to 0 with step 0.05 and $\log(T/{\rm K})=$ 4 to 6 with step 0.05, all with stop neutral hydrogen column density $\log(N_{\rm HI}/{\rm cm^{-2}})=16$ assuming the solar abundance and the \citet{HM12} UV background at $z=2.75$. Note that the mean hydrogen density at this redshift is about $10^{-5}{\rm cm}^{-3}$ and the hydrogen density in the set of models covers 10 to $10^5$ times the cosmic mean. For each combination of the low-metallicity and high-metallicity models, the relatively fraction is solved to best fit the measurements. Our purpose here is to look for combinations that can provide a reasonable explanation to the overall trend of the column densities of various species. If we use the measurement uncertainties, the results would be completely driven by the few species with small error bars (e.g., 0.02 dex for \ion{C}{3} and \ion{Si}{4} in Table~\ref{table_Zabs[2.0,2.4]} for $2.6<\zabsorb<2.8$). At such an uncertainty level, the model assumptions may also matter (e.g., the adopted solar abundance pattern). To focus on the overall trend and given the possible systematics in deriving the column densities (e.g., seen in the differences from multiple lines), we adopt a uniform error bar of 0.2 dex in this exercise. 

The solid line in the top panel of Figure~\ref{fig:cloudy_fits} is the bestfit model, consisting of a small contribution (2.3\%) of low-metallicity systems with $\log(n_{\rm H}/{\rm cm}^{-3})=-2.9$ and $\log(T/{\rm K})=5.4$ (red dotted line) and a large contribution (97.7\%) of high-metallicity systems with $\log(n_{\rm H}/{\rm cm}^{-3})=-1.5$ and $\log(T/{\rm K})=4$ (blue dotted line). To have a better idea of the possible combinations, we show the constraints on $\nH$ and $T$ in the bottom panel. The high-metallicity component is well constrained to be around $\log(\nH/{\rm cm}^{-3})\sim -1.5$ and $T\sim 10^4$ K. The low-metallicity component, on the other hand, is only loosely constrained. It can come from systems with either low density ($\sim 10^{-4} {\rm cm^{-3}}$) or high temperature ($\gtrsim  10^5$ K), both driven by the high column densities of high-ionization species (e.g., \ion{O}{6}, \ion{N}{5}, and \ion{C}{4}; see the parameter dependence in Fig.~\ref{fig:N_element_par}). Such physical conditions appear to be in line with the prediction for highly ionized metals from hydrodynamic simulations \citep[e.g.,][]{Rahmati16}. The high-density component likely comes from cool gas in the CGM, while the other one is contributed by gas in the IGM. The two-component model is consistent with the finding based on high-resolution observation that metal absorbers show bimodal physical properties with low- and high-metallicity branches \citep{Kim16}.

{
We note that we have performed a test with Cloudy by switching to the HM05 UV background model. We find the overall trend, e.g., as seen in Figure~\ref{fig:cloudy_fits}, remains almost the same, with a slight shift in the number density direction. For example, the bestfit number density for the low-metallicity component shifts by about 0.2 dex, from $\log (n/{\rm cm^{-3}}) \sim -1.5$ to -1.3.
}

\subsubsection{High- and Low-Ionization Species and Multi-Component Model}
\begin{figure}
    \centering
    \includegraphics[width=0.45\textwidth]{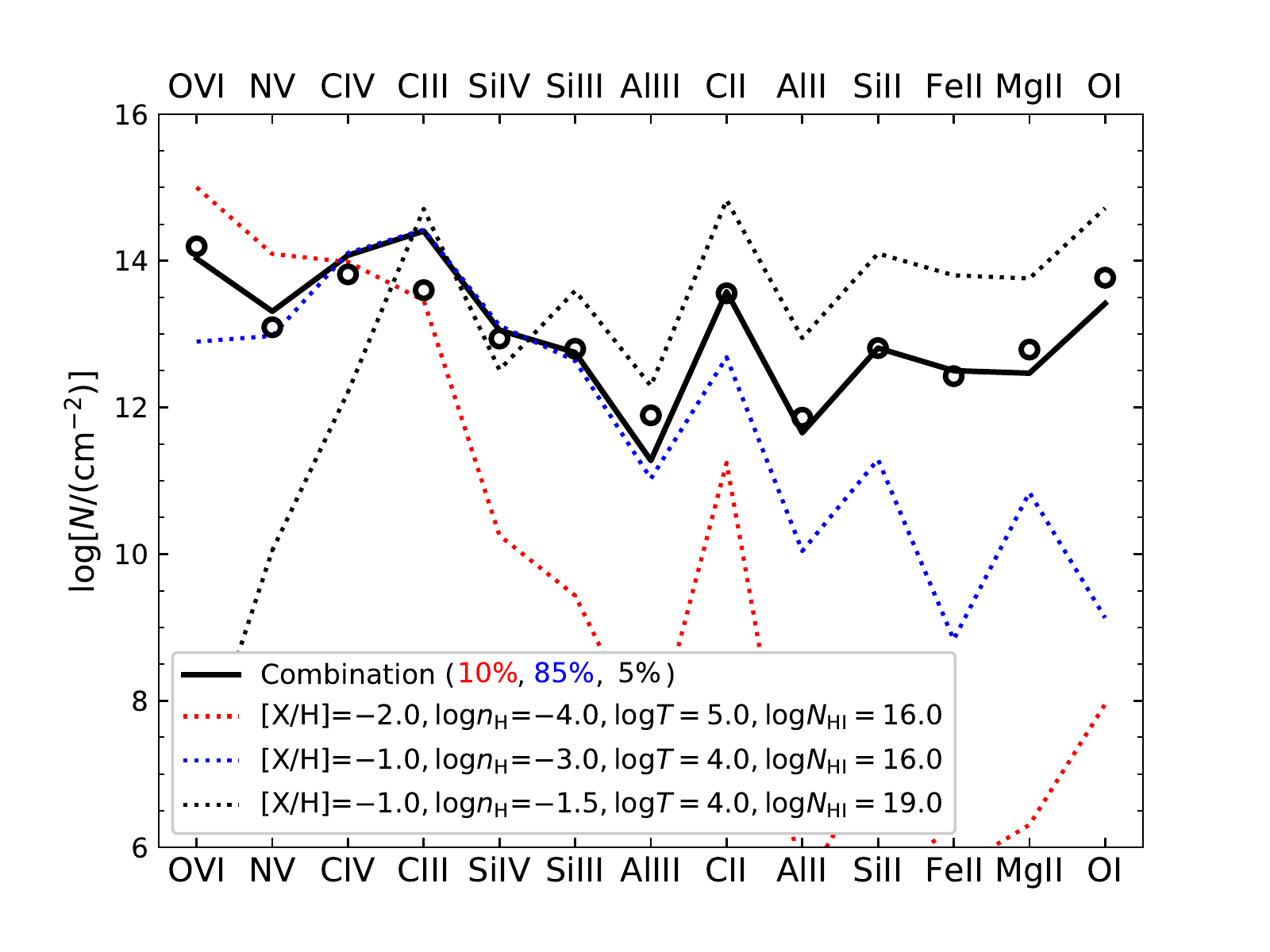}
    \caption{Illustration of a three-component model toward explaining the high measured column density of \ion{O}{1} (or more generally the species with low ionization potential). Open circles show the observed column densities of various species with $\deltalya\sim-0.95$ and $2.6<\zabsorb <2.8$. A high-column-density system (black dotted line) is added, and the combination with the other two systems (red and blue dotted line) is shown as the solid line, with the combination fractions labelled in the legend (not a model fit). This is only for illustration purpose, as there is degeneracy among model parameters (e.g., column density and metallicity) and the system is unlikely to be limited to three components. See the text for more detail.}
    \label{fig:three_component}
\end{figure}

It should be emphasized that the exercise shown in Figure~\ref{fig:cloudy_fits} indicates that the selected absorption systems cannot be described by a simple single component model. However, it does not necessarily mean that the right model is the two-component model, which is merely used here to illustrate the possible complexity of such systems. In fact, the fit shown in the top panel of Figure~\ref{fig:cloudy_fits} misses a few data points, both in the high- and low-ionization species. 

For example, the pattern in the high-ionization species is not fully reproduced, with the predicted \ion{C}{4} and \ion{Si}{4} lower than measured. While the column densities of most low-ionization species can be reasonably described by the model, the model underpredicts the column densities of \ion{Fe}{2} and \ion{Mg}{2} by about 0.5 dex and that of \ion{O}{1} by about 1.5 dex. 

An adjustment of metallicity can bring the model to better match the column densities of low-ionization species. 
{
With $N_{\rm HI}=10^{16}{\rm cm^{-2}}$, we find that  a model with super-solar metallicity (${\rm [X/H]}=0.3$) 
and $\log(n_{\rm H}/{\rm cm}^{-3})=-1$ can lead to a better match to the column densities of \ion{Fe}{2} and \ion{Mg}{2}, but the predicted \ion{O}{1} column density still falls short of the measurement by $\sim$0.75 dex.
}
We note that in \citet{Pieri2014} a model with ${\rm [X/H]}=0.3$ is able to provide a match to their \ion{O}{1} column density measurement (see their Fig.12). The difference may have two causes. First, we have a different selection of hydrogen absorption systems and the measured \ion{O}{1} column density, $\log (N_{\rm OI}/{\rm cm^{-2}})=13.77$, from our selection is higher than theirs (by about 0.37 dex). Second, the solar abundance pattern used in our work appears to be different from theirs. As it is unlikely to have metallicity much higher than ${\rm [X/H]}=0.3$, within our adopted model we find it is hard to explain the high \ion{O}{1} column density measurement solely by metallicity effect.

The mismatch between the measured and model \ion{O}{1} column density could suggest that O is more abundant in real systems, i.e., the abundance is different from the solar abundance adopted in the calculation.  This may also help resolve the difference seen in the high-ionization species, which relates to the CNO enrichment history. The high O/N ratio can be caused by a series of major star formation episodes in galaxies with short interval between them \citep[e.g.,][]{Kobulnicky98,Pettini02,Pettini04}, given that O is produced in core-collapse supernovae ($>10$ Myr after star burst) and N comes from intermediate-mass stars (4--8$M_\odot$; $>100$ Myr after star burst). Galactic winds may bring such an abundance pattern into the CGM/IGM, where the absorption systems likely reside.

There are certainly other possibilities. Given the resolution of the eBOSS spectra, the strong absorption systems we select, in this case, $\deltalya\sim -0.95$, may have a small fraction of contamination from high column density systems. In high-column-density systems (e.g., Lyman limit systems), the column density of \ion{O}{1}, the one with the lowest ionization potential for species considered here, can become enhanced, owing to the low photo-ionization rate inside the systems from the attenuated ionizing background
{
(see the bottom two panels of Fig.~\ref{fig:model_trend} in Appendix~\ref{sec:appendix_N_metals}). 
}
Such an enhancement in \ion{O}{1} column density is seen in DLAs (see e.g., \citealt{Cooke11a,Cooke11b}). 
{
Figure~\ref{fig:three_component} illustrates a three-component model (not from model fitting). Loosely motivated by Figure~\ref{fig:cloudy_fits}, the components with (${\rm [X/H]}$, $\log\nH$, $\log T$)=($-2.0$, $-4.0$, $5.0$) and ($-1.0$, $-3.0$, $4.0$), both with $N_{\rm HI}=10^{16}{\rm cm}^{-2}$, are adopted to  $\log\nH=-4$ and $\log\nH=-1.5$ components (with $[{\rm X/H}]=-0.3$ and $N_{\rm HI}=10^{16}{\rm cm}^{-2}$) are adopted to reasonably produce the trend seen in about half of the species with higher ionization potentials. The third component represents a contamination from $N_{\rm HI}=10^{19}{\rm cm}^{-2}$ systems ($[{\rm X/H}]=-1.0$ and $\log \nH=-1.5$), which is able to produce the high column density of \ion{O}{1} and improves the match to those of \ion{Fe}{2} and \ion{Mg}{2}. In combination, the three components lead to a trend following the measurements reasonably well.
}
Increasing the metallicity, $\nH$, and $N_{\rm HI}$ all have the effect of boosting the column density of \ion{O}{1}, the species with the lowest ionization potential, and the latter two have the additional effect of causing a steeper jump between \ion{Fe}{2}/\ion{Mg}{2}  and \ion{O}{1} column densities. The contamination from the third component is assumed to be {5\%} in the illustration. 
{
While the metallicity ${\rm [X/H]}=-1$ is motivated by the measurements in Lyman-limit systems \citep[e.g.,][]{Fumagalli2016,Lehner2016}, given the degeneracy among ${\rm [X/H]}$, $\nH$, and $N_{\rm HI}$, the values used here are only for the purpose of illustrating the possible contribution from high-column density systems. In addition, there is hardly any reason to limit to three components for the selected absorbers.
}

{If we broadly take the contamination from Lyman limit systems to be at a level of a few percent, is it reasonable?} To answer this we must understand the systems that meet the selection function. Similar to the case of flux transmission $F<0.25$ in \citet{Pieri2014}, isolated lines without damping wings do not reach $-1<\deltalya<-0.9$ in moderate resolution spectra (such as those observed in eBOSS). The selection function requires lines that are both saturated in high resolution and clustered with one another on scales within the full width half maximum of the SDSS resolution element (see Fig.2 of \citealt{Pieri2014}). For this reason we cannot simply integrate the column density distribution function in order to assess the proportion of Lyman limit systems present in the selected absorbers. Put in logic terms, the presence of $N_{\rm HI}\gtrsim 10^{14}{\rm cm}^{-2}$ is a necessary but not sufficient condition for selection, since clustering is also required; however, $N_{\rm HI}>10^{17.5}{\rm cm}^{-2}$ is neither necessary nor sufficient, since such systems are not necessarily sufficiently clustered with other $N_{\rm HI}>10^{14} {\rm cm}^{-2}$ lines on $\sim 150\, {\rm km\, s}^{-1}$ scales. Hence the column density distribution function is largely orthogonal to our selection function. 

In order to understand the populations of individual lines selected here we must compare with results from line fits of high resolution data and verify what the selection function provides. This was done in \citet{Pieri2014} with the conclusion that $N_{\rm HI} > 10^{17.5}{\rm cm}^{-2}$ systems can contribute no more than 3.7\% of the systems with a true flux transmission of $F<0.25$ with SDSS resolution. While our selection function is for stronger features than those in \citet{Pieri2014}, the selection function is otherwise the same. We have no basis to conclude that the somewhat stronger blended absorption here differs from the line model test in \citet{Pieri2014}. We therefore also assume their conclusion that systems with $N_{\rm HI} > 10^{17.5}{\rm cm}^{-2}$ should be no more than 3.7\% of our sample.

The discussion here focuses on $\deltalya \sim -0.95$ systems. The \ion{O}{1} line becomes too weak to be measured robustly for $\deltalya \gtrsim -0.85$ systems (bottom panel of Fig.~\ref{fig:stacked_metal1}), and this is nearly so for \ion{Fe}{2} and \ion{Mg}{2} as well. Such a behavior lends support to the scenario of contamination from high-column-density systems to the strongest absorption systems ($\deltalya\sim -0.95$) in this study with eBOSS spectra. 
{
In Appendix~\ref{sec:appendix_high_N}, we perform further tests through selecting presumably stronger absorbers within the $\deltalya \sim -0.95$ systems. The change in the metal column densities and the feature at the corresponding Lyman limit also provide further evidence on the contribution from high-column-density systems.
}

\bigskip

To summarize, the column densities of various low-ionization species in $\deltalya\sim -0.95$ absorption systems at $2.7<\zabsorb<2.9$ can be reasonably explained by a model cloud with $\nH\sim 10^{-1.5}{\rm cm}^{-3}$ and $T\sim 10^4$ K, photoionized by a UV background \citep{HM12} to have a neutral hydrogen column density of $N_{\rm HI}\sim 10^{16}{\rm cm}^{-2}$. The relatively high column densities of high-ionization species suggests an additional component, with either low $\nH$ or high $T$, or a deviation from solar abundance. The high column density of \ion{O}{1}, the one with the lowest ionization potential for species considered in this work, may signal contamination from high column density systems in view of the spectral resolution of eBOSS. 

\begin{figure}
\begin{center}
\includegraphics[angle=0, width=.47\textwidth]{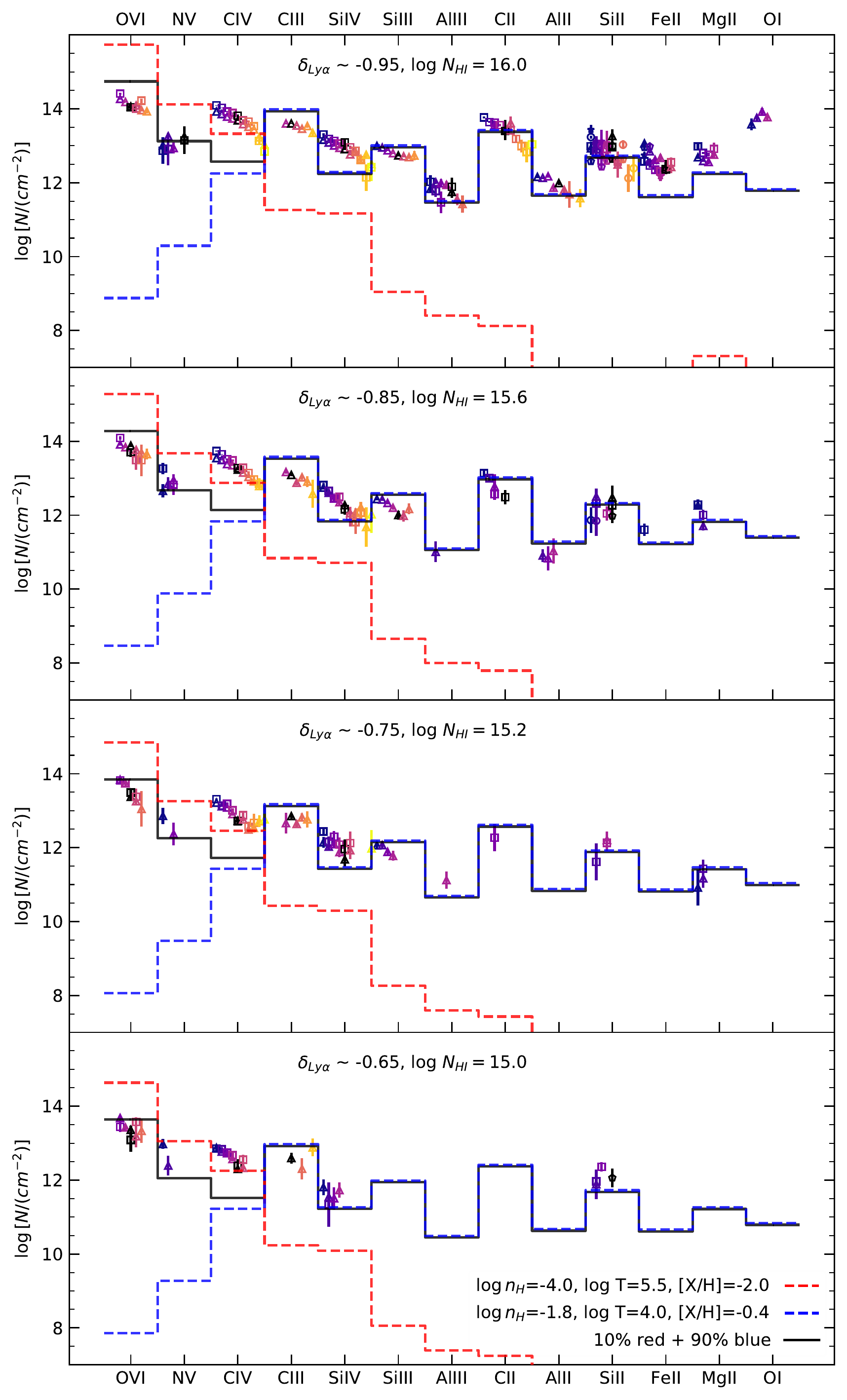}
\end{center}
\caption{
Column densities of 13 metal species as a function of \dl\ and \zabs. 
The four panels are for different values of \dl. In each panel, the column density of each species is color-coded according to absorber redshift, low \zabs\ to high \zabs\ from left to right within each species. The solid black line illustrates the model combination of a low-metallicity component (low $\nH$ and high $T$; red-dashed line) and a high metallicity component (high $\nH$ and low $T$; blue dashed line), and the difference in the models across the four panels is in the neutral hydrogen column density, as labelled.
}
\label{fig:N_element_model}
\end{figure}

Multiple components appear to also apply to other \dl\ systems and to systems at different redshifts. In Figure~\ref{fig:N_element_model}, the column densities of various species at different \dl\ values are shown in different panels, and in each panel the redshift dependence is color-coded. The Cloudy model (not fit) with a high-metallicity and a low-metallicity component of different values of $\nH$ and $T$ is used to illustrate the possible multiple components. Histogram, instead of lines connecting column density of each species, is used here for the model and its components to avoid the possible confusion with the redshift dependence. The models across the four values of \dl\ are the same, except for the stop neutral hydrogen column density.  The \ion{O}{1} line is not robustly measured, except for the strongest absorption system with $\deltalya\sim -0.95$, consistent with it being contaminated by systems of high column density. Given the uncertainties discussed above, we make no attempt to further model the redshift dependence, which can be related to metallicity and density evolution of the selected systems.

\section{Summary and Discussion}
\label{sec:summary}

We study metal absorption lines associated with relatively strong \lya\ absorbers in the \LAF\ using quasar spectra from the eBOSS survey. Various metal species are revealed in the stacked spectra, and their column densities are derived as a function of \lya\ absorber strength and redshift. We find that multiple components, including low- and high-metallicity gas of different densities and temperatures, are needed to interpret the inferred column density pattern of the metal species. The results can be used to study the chemical enrichment of CGM and IGM. They also provide inputs to model the metal contamination in the \LAF\ BAO measurement.

We construct a catalog of \lya\ absorbers in 10 redshift bins over $2<\zabsorb<4$, selected according to the \lya\ flux fluctuation \dl. We focus on the relatively strong absorbers, with $\deltalya \sim -0.95$, $-0.85$, $-0.75$, and $-0.65$. The \lya\ absorbers in each redshift bin are cross-correlated with the flux fluctuation in quasar spectra in the full wavelength range (shifted in accordance with the \lya\ absborber redshift) to obtain the stacked spectrum. The neutral hydrogen column density of these absorbers, from analyzing higher-order Lyman series lines, ranges from $\sim 10^{15}{\rm cm}^{-2}$ (for $\deltalya\sim -0.65$) to $\sim 10^{16}{\rm cm}^{-2}$ (for $\deltalya\sim -0.95$), which has only weak redshift evolution.

In the stacked spectrum, up to 13 metal species can be identified, including high-ionization ones (e.g., \ion{O}{6}, \ion{N}{5}, and \ion{C}{4}) and low-ionization ones (e.g., \ion{Al}{3}, \ion{Al}{3}, \ion{C}{2}, \ion{Si}{2}, \ion{Fe}{2}, \ion{Mg}{2}, and \ion{O}{1}). The column densities of these species drop as \dl\ increases and toward higher redshift.

We use a Cloudy model with gas cloud photoionized by an external UV background to help understand the results. In agreement with previous work \citep{Pieri2010,Pieri2014}, we find that the column density distribution of the various metal species can not be explained by a single gas component. In the model we consider, the low-ionization species point to absorption systems of metallicity $[{\rm X/H}]\sim -0.3$ with high density ($\nH\sim 10^{-1.5}{\rm cm}^{-3}$) and low temperature ($T\sim 10^4$ K). The high-ionization species are likely low-metallicity ($[{\rm X/H}]\sim -2$) with either low density ($\nH\sim 10^{-4}{\rm cm}^{-3}$) or high temperature ($T>10^{4.5}$ K). The two components likely correspond to cool gas in the CGM and gas in the IGM, respectively. For absobers selected to have $\deltalya\sim -0.95$, there is also a possibility of contamination from \lya\ absorbers of higher neutral hydrogen column density.

The model explored in this work is for illustration purpose, highlighting the need for more than one component. In reality, for systems selected by \dl\, there may be more than two or three components with a distribution of gas properties ($\nH$, $T$, abundance pattern, etc). Large-volume hydrodynamic simulations of galaxy formation \citep[e.g.,][]{Oppenheimer12} with photo-ionization computation for metal enriched gas \citep[e.g.,][]{Oppenheimer13} can help guide the modelling effort by identifying \lya\ absorption systems in the synthetic spectra (matched with eBOSS resolution) and connecting them to the physical properties of the underlying gas \citep[e.g.,][]{Turner16}. Meanwhile, high-resolution observation of the \lya\ forest \citep[e.g.,][]{Cowie95,DOdorico16} can be used to study the components in the \lya\ absorbers identified under eBOSS resolution and their associated metal properties. The complementary use of the high-resolution spectra and stacked eBOSS spectra would provide an ideal approach to understanding the metal content in the \LAF.

\begin{acknowledgements}
This work is supported by National Key R\&D Program of China (grant No. 2018YFA0404503).
L.Y. gratefully acknowledges the support of China Scholarship Council (No. 201804910563) and the hospitality of the Department of Physics and Astronomy at the University of Utah during her visit. Z.Z. is supported by NSF grant AST-2007499. The support and resources from the Center for High Performance Computing at the University of Utah are gratefully acknowledged. G.R. acknowledges support from the National Research Foundation of Korea (NRF) through Grant No. 2020R1A2C1005655 funded by the Korean Ministry of Education, Science and Technology (MoEST).

Funding for the Sloan Digital Sky 
Survey IV has been provided by the 
Alfred P. Sloan Foundation, the U.S. 
Department of Energy Office of 
Science, and the Participating 
Institutions. 

SDSS-IV acknowledges support and 
resources from the Center for High 
Performance Computing  at the 
University of Utah. The SDSS 
website is www.sdss.org.

SDSS-IV is managed by the 
Astrophysical Research Consortium 
for the Participating Institutions 
of the SDSS Collaboration including 
the Brazilian Participation Group, 
the Carnegie Institution for Science, 
Carnegie Mellon University, Center for 
Astrophysics | Harvard \& 
Smithsonian, the Chilean Participation 
Group, the French Participation Group, 
Instituto de Astrof\'isica de 
Canarias, The Johns Hopkins 
University, Kavli Institute for the 
Physics and Mathematics of the 
Universe (IPMU) / University of 
Tokyo, the Korean Participation Group, 
Lawrence Berkeley National Laboratory, 
Leibniz Institut f\"ur Astrophysik 
Potsdam (AIP),  Max-Planck-Institut 
f\"ur Astronomie (MPIA Heidelberg), 
Max-Planck-Institut f\"ur 
Astrophysik (MPA Garching), 
Max-Planck-Institut f\"ur 
Extraterrestrische Physik (MPE), 
National Astronomical Observatories of 
China, New Mexico State University, 
New York University, University of 
Notre Dame, Observat\'ario 
Nacional / MCTI, The Ohio State 
University, Pennsylvania State 
University, Shanghai 
Astronomical Observatory, United 
Kingdom Participation Group, 
Universidad Nacional Aut\'onoma 
de M\'exico, University of Arizona, 
University of Colorado Boulder, 
University of Oxford, University of 
Portsmouth, University of Utah, 
University of Virginia, University 
of Washington, University of 
Wisconsin, Vanderbilt University, 
and Yale University.

\end{acknowledgements}
 
\appendix
\section{Column Density Measurements for Metal Species and Model Trend}
\label{sec:appendix_N_metals}

In Tables~\ref{table_Zabs[2.0,2.4]}, we list the column density measurements of all metal species in the ten absorber redshift bins over $2<\zabsorb<4$ and in the four \dl\ bins at each redshift. The measurements result from Voigt fits with the stacked spectra. The vacuum wavelengths and oscillator strengths used in the fits are taken from the \texttt{Atomic Line List}\footnote{\url{https://github.com/jkrogager/VoigtFit/blob/master/VoigtFit/static/linelist.dat}} \citep{Krogager2018}.

\begin{figure}[h]
\centering
\includegraphics[width=0.45\textwidth,height=0.75\textheight]{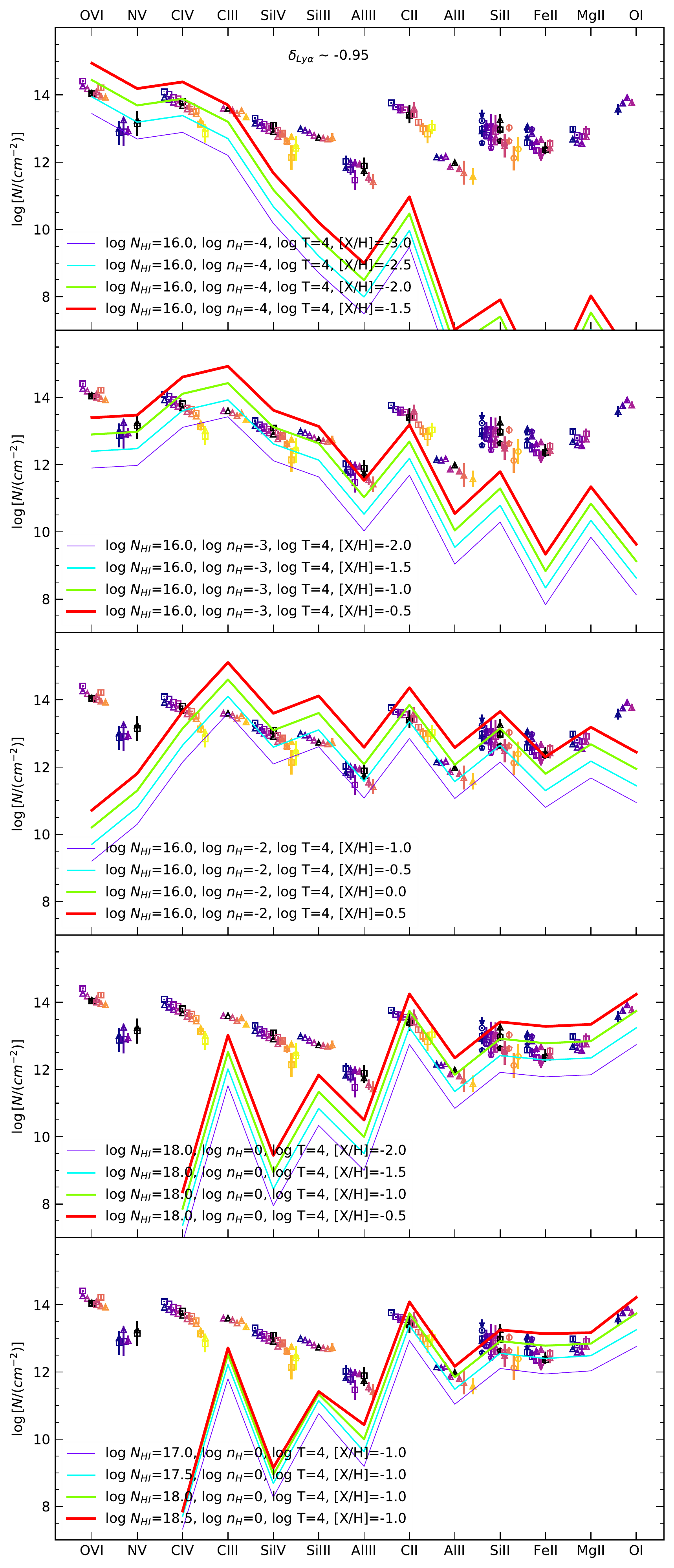}
\caption{Dependence of column densities of metal species from Cloudy model. The gas temperature is fixed at $T=10^4$ K. Across the top four panels, the total hydrogen density $\nH$ increases, and in each of these panels, different lines show the dependence on metallicity. Note that in the fourth panel the neutral hydrogen column density $N_{\rm HI}$ changes from $10^{16}\cm^{-2}$ to $10^{18}\cm^{-2}$. The bottom panel shows the dependence on $N_{\rm HI}$, with other parameters fixed. For comparison, the measured column densities for $\deltalya\sim -0.95$ absorbers at different absorber redshift (color-coded) are overlaid. 
}
\label{fig:model_trend}
\end{figure}

In Figure~\ref{fig:model_trend}, we extended the exploration of Cloudy models in the main text, overlaid with the measured column densities. The temperature is fixed at $T=10^4$ K. From top to bottom panels, the density $\nH$ increases from $10^{-4}\cm^{-3}$ to 1 $\cm^{-3}$. Note that in the fourth panel the neutral hydrogen column density $N_{\rm HI}$ changes from $10^{16}\cm^{-2}$ to $10^{18}\cm^{-2}$. From the top four panels, we see that as $\nH$ increases, the overall trend is an increase in the column density of low-ionization species and a decrease in those of high-ionization species, a result of the increase in the recombination rate. The dependence on metallicity is to change the amplitude. The bottom panel shows the dependence on neutral hydrogen column density $N_{\rm HI}$. At a higher $N_{\rm HI}$, species with the lowest ionization potential tends to have a stronger increase in column density.

\section{Shadow Lines}
\label{sec:appendix_shadow_lines}
\begin{figure*}
\centering 
     \includegraphics[width=\textwidth,height=0.94\textheight]{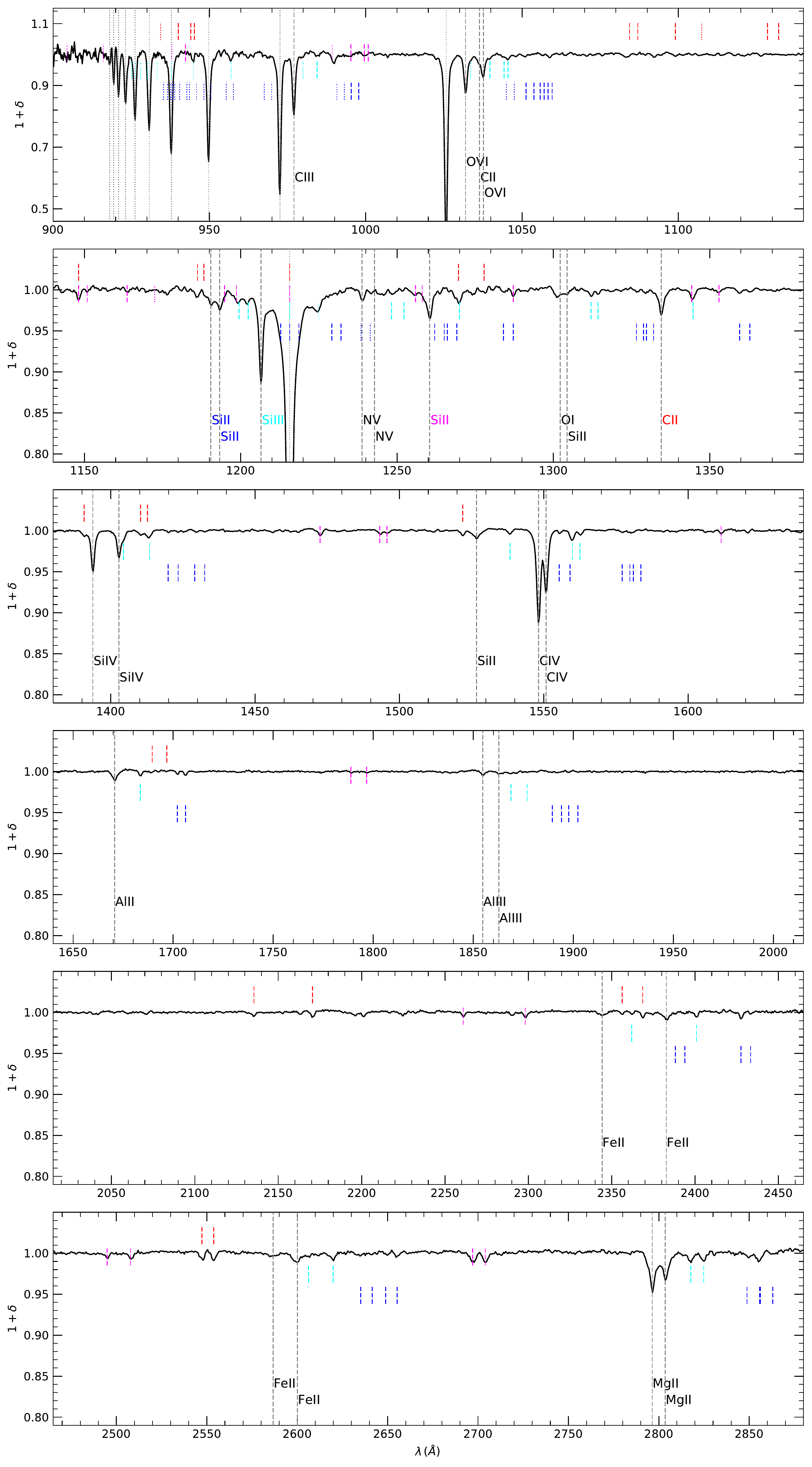}
    \caption{Similar to Figure~\ref{fig:stacked_spectra}, but with positions of major shadow lines indicated. Four species contaminating \lya\ absorption selection are considered, including \ion{Si}{2} 1190\AA\ \& 1193\AA\ (blue), \ion{Si}{3} 1207\AA\ (cyan), \ion{Si}{2} 1304\AA\ (magenta) and \ion{C}{2} 1335\AA\ (red). 
    For each contaminating line with wavelength $\lambda_{\rm contamination}$, the shadow of a given line $\lambda_{\rm line}$ (e.g., indicated by one of the long vertical lines) is shifted to a nearby position $\lambda_{\rm shadow\, line}$ at a fixed wavelength ratio, $\lambda_{\rm shadow\, line}=\lambda_{\rm Ly\alpha}/\lambda_{\rm contamination} \times \lambda_{\rm line}$. These shadow lines are indicated by short dashed (dotted) vertical lines of the corresponding color for metals (Lyman series).
    Vertical Offsets are added to the vertical lines for different contaminating species.
}
    \label{fig:stacked_spectra_withshadowlines}
\end{figure*}

{

In the stacked spectrum shown in Figure~\ref{fig:stacked_spectra}, we label the major metal lines associated with the selected \lya\ absorber. Many shadow lines can be seen in the stacked spectrum. These shadow lines are caused by metal lines in \lya\ absorbers with wavelengths close to \lya\ -- the \lya\ absorption used to select a type of absorbers of interest (e.g., with $-1.0<\deltalya<-0.9$) can be contaminated by metal absorptions from neighboring \lya\ absorbers. Then in the stacked spectrum of the specific absorbers of interest,  the metal lines in the neighboring absorbers  show up with shifted wavelengths, emerging as shadow lines.

As the shadow lines are not directly related to what we intend to study in this work, they are not labelled in Figure~\ref{fig:stacked_spectra} to avoid any confusion. In Figure~\ref{fig:stacked_spectra_withshadowlines}, we provide a version of the stacked spectrum with major shadow lines identified and labelled to aid any possible investigation of shadow lines. 

First, in the second row from the top we label the five metal lines in the \lya\ absorber with wavelengths close to \lya,  including \ion{Si}{2} 1190\ \& 1193\AA\ (blue), \ion{Si}{3} 1207\AA\ (cyan), \ion{Si}{2} 1260\AA\ (magenta), and \ion{C}{2} 1335\AA\ (red). If one of these lines with restframe wavelength $\lambda_{\rm contamination}$ contaminates the selected \lya\ absorption, a metal line with restframe wavelength $\lambda_{\rm line}$ associated with the absorber where the contaminating line originates will show up as a shadow line at the following wavelength,
\begin{equation}
    \lambda_{\rm shadow\, line}
 = \frac{\lambda_{\rm Ly\alpha}}{\lambda_{\rm contamination}}\lambda_{\rm line},
\end{equation}
where $\lambda_{\rm Ly\alpha}=1215.67$\AA\ is the \lya\ wavelength. We then use dashed (dotted) vertical lines of the corresponding color to indicate shadowed metal lines (Lyman series lines). 

As an example, consider the shadow lines caused by \ion{Si}{2} 1190\AA\ \& 1193\AA\ contaminating \lya\ absorption. Here $\lambda_{\rm contamination}=1190$\AA\ or $1193$\AA. The corresponding \ion{Si}{4} 1394\AA\ and 1403\AA\ shadow lines show up at wavelengths indicated by the four blue dashed vertical lines around 1420--1430\AA. Redward of 1550\AA, we see the corresponding shadow lines of \ion{Si}{2} 1527\AA\ and \ion{C}{4} 1548/1551\AA\ doublet. Around 2850\AA\ are the shadow lines of the \ion{Mg}{2} 2796/2804\AA\ doublet. The shadow Ly$\beta$ lines are indicated by the two blue dotted vertical lines around 1045\AA.

}

\section{Further Tests on the Contribution of High-Column Density Systems}
\label{sec:appendix_high_N}

\begin{figure}
\centering 
     \includegraphics[width=0.5\textwidth]{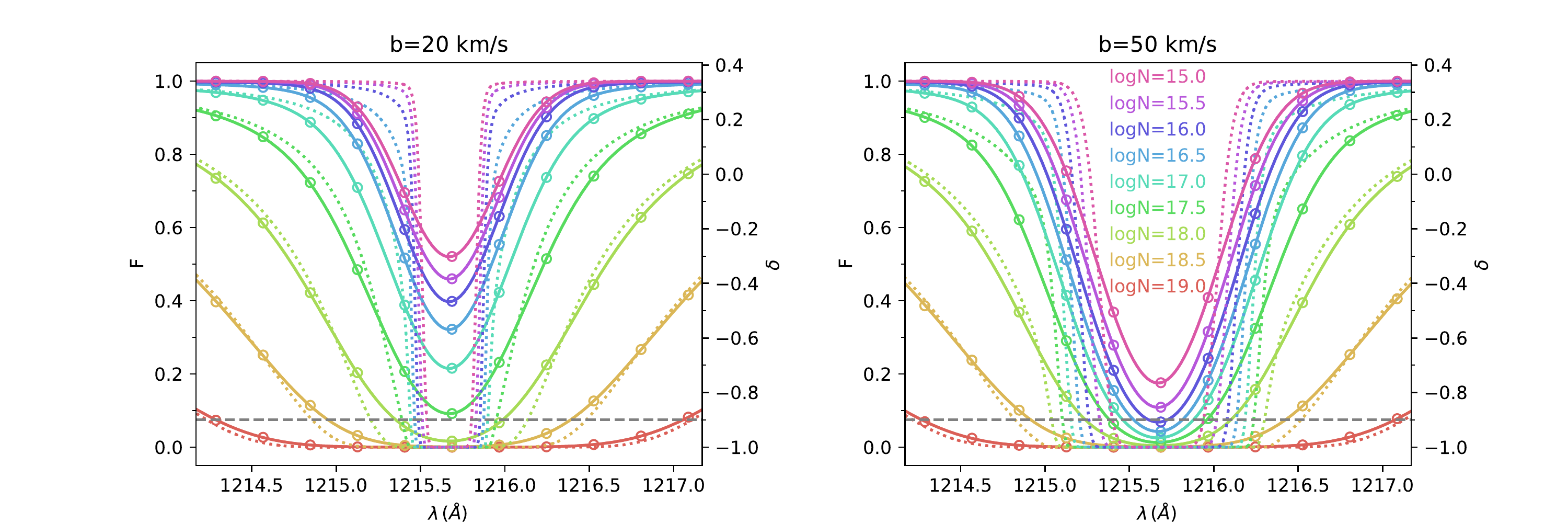}
    \caption{Dependence of \lya\ absorption profile on neutral hydrogen column density $\log (N_{\rm HI} /{\rm cm}^{-2})$ and Doppler parameter $b$ for the case of a single \lya\ absorber ($b=20$ and 50 ${\rm km\, s^{-1}}$ in the left and right panel, respectively). In each panel, the dotted curves are intrinsic profiles. The solid curves are those with eBOSS resolution ($R\sim 2000$) and open circles indicate the eBOSS pixels ($\Delta \log \lambda \sim 10^{-4}$). The left y-axis is the normalized flux, and the right y-axis is the corresponding flux fluctuation $\delta=F/F_m-1$, where the mean transmitted flux $F_m(z_{\rm abs}=2.7) \approx 0.745$ \citep{Becker2013}. The dashed horizontal line represents the $\delta=-0.9$ level for our selection of $\zabsorb \sim 2.7$ absorbers with $-1.0<\deltalya<-0.9$. 
}
    \label{fig:illustration_lya_selection}
\end{figure}

\begin{figure}
\centering 
     \includegraphics[width=0.48\textwidth]{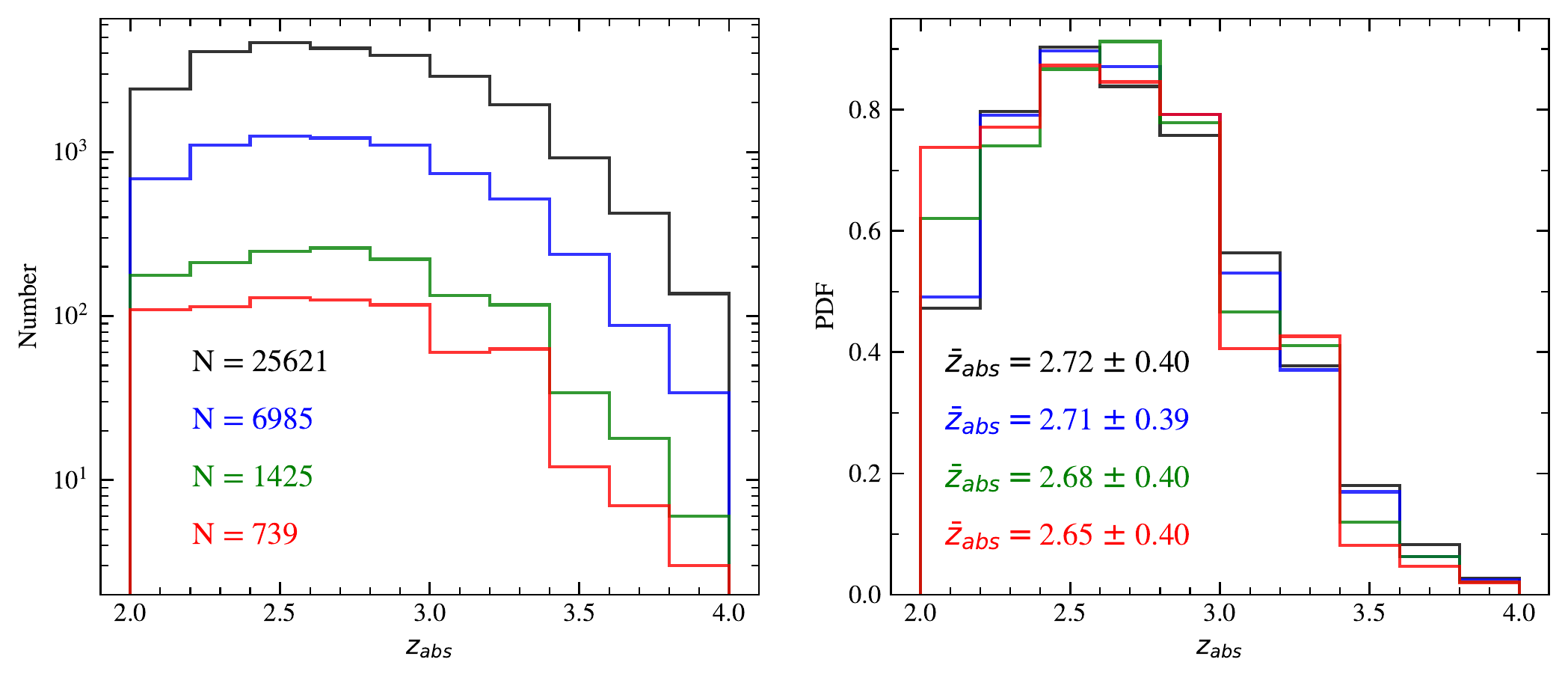}
    \caption{Redshift distributions for \lya\ absorbers in the original sample (black) with $-1.0<\deltalya<-0.9$ and in the subsamples (blue/green/red) with the 2/4/6 neighboring pixels around the selected absorbers in the original sample satisfying $\deltalya < -0.9+\sigma_{\delta}$, with $\sigma_{\delta}$ the uncertainty in $\deltalya$. The number distributions are shown in the left panel, and the normalized distributions (i.e., the probability distribution functions; PDFs) are in the right panel.
}
    \label{fig:zabshist_ss_ssorder1neigh1-2-3}
\end{figure}

\begin{figure*}
\centering 
     \includegraphics[width=\textwidth,height=0.94\textheight]
     {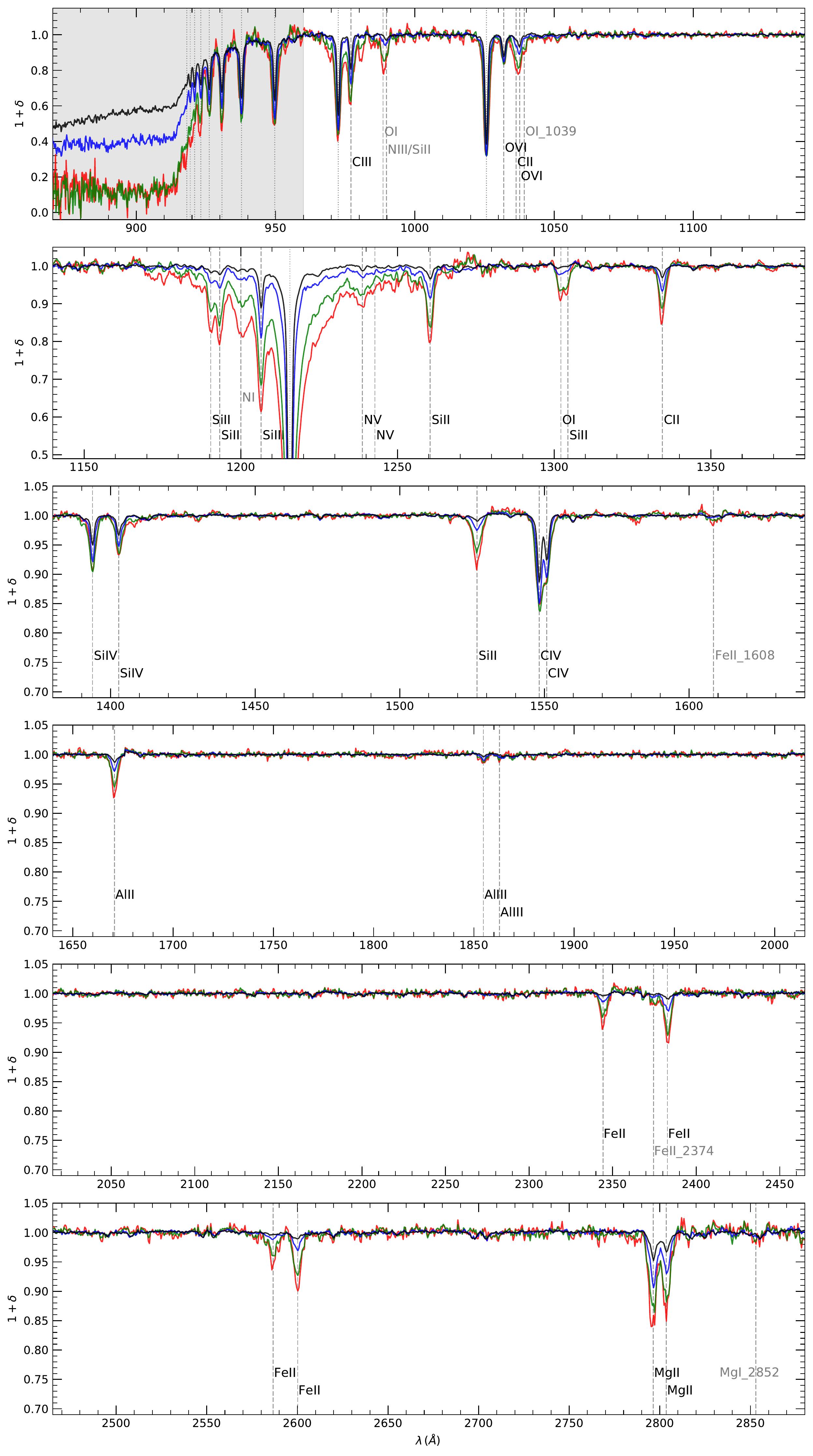}
    \caption{Similar to Figure~\ref{fig:stacked_spectra}, but with the stacked spectra of the original sample (black) and three subsamples (blue/green/red). The original sample consists of absorbers with $-1.0<\deltalya<-0.9$ in the redshift range of $2<z_{\rm abs}<4$. The subsamples (blue/green/red) are from selecting stronger absorbers by requiring the $\deltalya$ values of the neighboring 2/4/6 pixels around the absorbers in the original sample to lie below $-0.9+\sigma_{\delta}$, where $\sigma_{\delta}$ is the uncertainty in $\deltalya$. In the shaded region, the flux is normalized to the mean continuum level within $960\pm 1$\AA\ to show the absorption near the Lyman break (912\AA).}
    \label{fig:stacked_spectra_ss_ssorder1neigh1-2-3}
\end{figure*}

\begin{figure}[h!tp]
\centering 
     \includegraphics[width=0.45\textwidth]{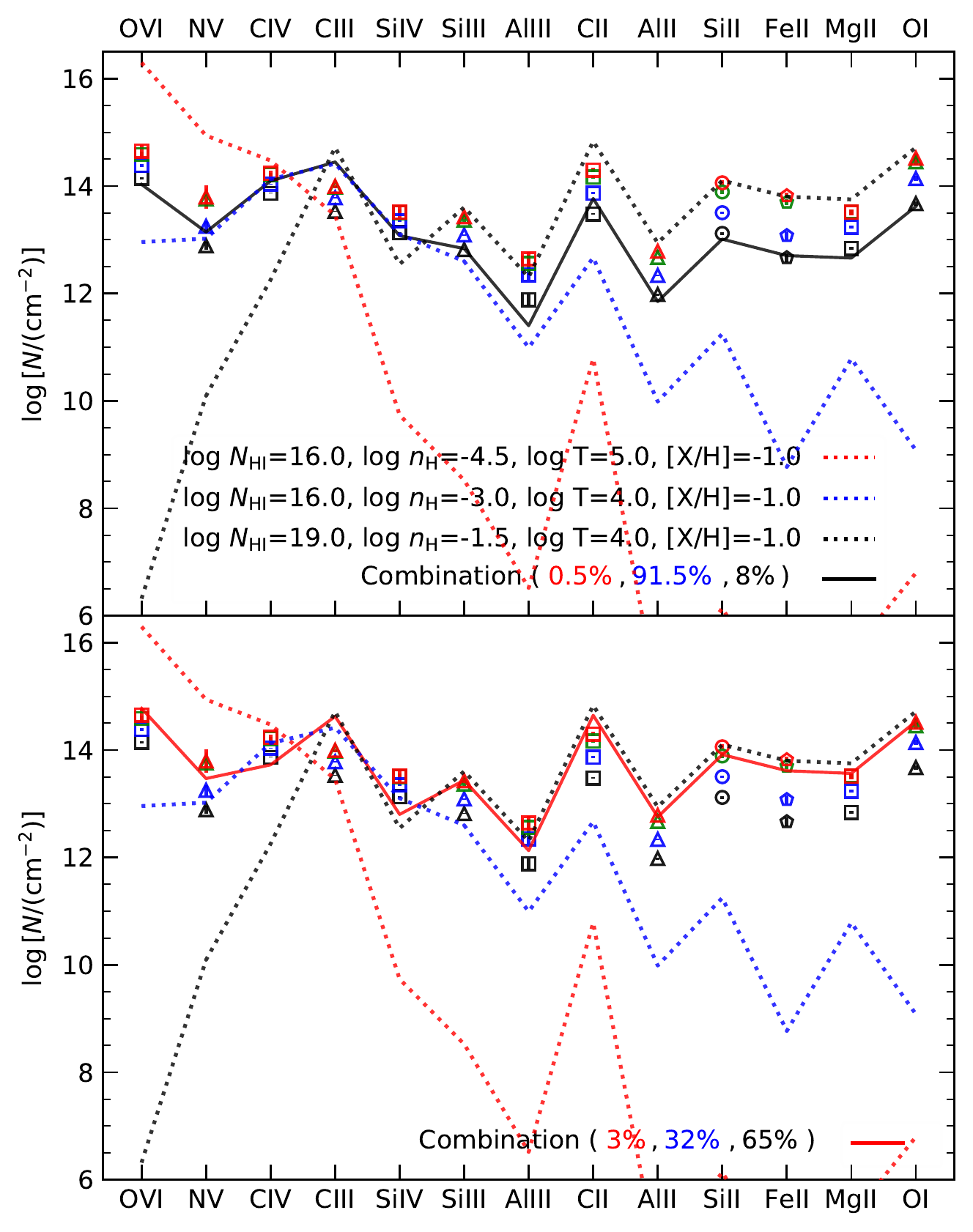}
    \caption{Column densities of metal species in the original sample (black symbols) and in the subsamples (blue/green/red symbols). The color of symbols follow the same notation as in Figure~\ref{fig:stacked_spectra_ss_ssorder1neigh1-2-3}. In each panel, three Cloudy model curves are shown -- a component with a low hydrogen number density and high temperature (red dotted line), a component with a medium hydrogen number density (blue dotted line), and a component with a high hydrogen number density and neutral hydrogen column density (black dotted line). Different combinations (not model fits) of the three components are shown as the solid black (top panel) and the solid red (bottom panel) lines to illustrate the difference between the absorbers in the original sample and those in the subsample of the strongest absorption. See the text for detail.
}
    \label{fig:cloudy_ss_ssorder1neigh1-2-3}
\end{figure}

{

We perform further tests to investigate the contribution of high-column density systems in the $-1<\deltalya<-0.9$ absorbers. 

By default, an absorber is selected to be in the sample by requiring the corresponding pixel with $-1<\deltalya<-0.9$ to be the local minimum. With such a selection, the $\deltalya$ values of its neighbouring pixels can be either below or above $-0.9$. If the absorber has a higher column density, its neighbouring pixels more likely have \dl\ lie below $-0.9$. We therefore impose an additional selection criterion by requiring the \dl\ values of the pixels to the left and that to the right of the above selected pixel to be less than $-0.9$. More exactly,  we require $\delta_{{\rm Ly\alpha}, i}<-0.9+\sigma_\delta$, where $\sigma_\delta$ is the uncertainty in \dl\ and $i$ indicates the $i$-th pixel next to the selected absorber, negative (positive) $i$ for pixels to the left (right). 

We create three subsamples, subsample-1, subsample-2, and subsample-3, by requiring 1, 2, and 3 left/right neighbouring pixels to satisfy the above additional selection, respectively. These subsamples potentially have increasingly higher column densities. As illustrated in Figure~\ref{fig:illustration_lya_selection}, in the case of a single absorber under the SDSS resolution, the subsample-1 selection prefers to have systems of $\log(N_{\rm HI}/{\rm cm}^{-2}) > 18$ or those of $\log N_{\rm HI}>17$ with large $b$ values, and subsample-2 and subsample-3 selections would select absorbers of $\log N_{\rm HI} > 18.5$. The absorbers in the three subsamples follow the same redshift distribution as the original sample, as shown in 
Figure~\ref{fig:zabshist_ss_ssorder1neigh1-2-3}.

Figure~\ref{fig:stacked_spectra_ss_ssorder1neigh1-2-3} compares the stacked spectra of the original $-1<\deltalya<-0.9$ sample and the three subsamples. It is clear that the stacked \lya\ line becomes increasingly broader from the original sample to the subsample-1/2/3, and the sequence also displays an increasingly stronger wing-like feature. The Lyman series lines appear to become broader and deeper. All of these are consistent with the expectation that the subsamples consist of absorbers with an increasing contribution from systems of high neutral hydrogen column density. 

To further test the column density, in the gray shaded region in Figure~\ref{fig:stacked_spectra_ss_ssorder1neigh1-2-3}, we normalize the stacked spectra to the mean continuum level within $960\pm 1$\AA\ in the absorber's rest frame so that the flux decrements in the Lyman continuum ($<912$\AA) can be revealed. This part of the stacked spectra comes from absorbers with $\zabsorb \gtrsim 2.95$. Similar to that seen in the spectra of Lyman-limit systems (e.g., Fig.8 in \citealt{Fumagalli2020}), we see marked flux decrements blueward of 912\AA. If the decrement were caused by a single absorber, the implied $\log N_{\rm HI}$ would be about 16.9, 17.2, 17.5, and 17.5 for the original sample and the subsample-1/2/3, respectively, estimated using the decrement around 912\AA. We note that a similar exercise in \citet{Pieri2014} leads to an estimate of $\log N_{\rm HI}\sim 16.7$ (16.4) for absorbers with flux transmission $-0.05<F<0.05$ ($0.05<F<0.15$), a similar level to that inferred for our original sample. This is argued in \citet{Pieri2014} as an upper limit because of effects like the large-scale excess in forest absorption associated with the selected absorbers. Anyway, the increasing decrement over these samples leads support to the scenario that high-column-density systems contribute to the $-1<\deltalya<-0.9$ absorbers selected in this work. A more realistic model, which is beyond the scope of this work, should consist of absorbers with a distribution of column densities, and the decrement is related to the average absorption $\langle \exp(-\tau)\rangle$ over such a distribution of absorbers. Without such a model, we could not assess the fraction of Lyman-limit systems and the relative importance of the high-column-density systems and the strongly-clustered low-column-density systems.

Next we compare the metal species. Following the original sample and the subsample-1/2/3 sequence, we see that metal absorptions become increasingly stronger. In addition, a few species not clearly detected in the original sample start to emerge in the subsamples, which are labelled in Figure~\ref{fig:stacked_spectra_ss_ssorder1neigh1-2-3} in gray fonts. For example, around 989\AA\, there are blended \ion{O}{1} (988.58/988.65/988.77\AA), \ion{N}{3} (989.80\AA), and \ion{Si}{2} (989.87\AA), and around 1200\AA\, we see \ion{N}{1}. These metal species are similar to those commonly seen in DLAs \citep[e.g.,][]{Lluis2017}. 

The measured column densities of the metal species for the subsamples are shown in Figure~\ref{fig:cloudy_ss_ssorder1neigh1-2-3}, along with those for the original sample. From the original sample to subsample-1/2/3, the column densities of low-ionization species (like \ion{O}{1} and \ion{Fe}{2}) in general increase by more than one order of magnitude, a much stronger change than high-ionization species (e.g., \ion{O}{6}, \ion{C}{4} and \ion{Si}{4}; with \ion{N}{5} being an exception caused by blending with the \lya\ shadow lines from the contamination of \ion{Si}{2} 1190/1193\AA\ in \lya\ as shown in Fig.~\ref{fig:stacked_spectra_withshadowlines}). Such a relative change between column densities of low- and high-ionization species is a typical feature in high-column-density systems (see the bottom two panels of Fig.~\ref{fig:model_trend}) and is seen in the observational study with high-resolution quasar spectra \citep[e.g.,][]{Lehner2021}. Similar to Figure~\ref{fig:three_component}, in the top panel of Figure~\ref{fig:cloudy_ss_ssorder1neigh1-2-3}, we show a three-component Cloudy model (not a fit) to illustrate the role of adding a high-column-density component in reproducing the trend in the low-ionization species. In the bottom panel, we boost the contribution of the high-column-density system, which provides a reasonable match to the column densities of the low-ionization species in subsample-3, the one presumably with the highest contribution from high-column-density systems. While the fractions in the illustration are not to be taken seriously, the numbers are self-consistent: the $\sim$65\% high-column-density systems in subsample-3 would make $\sim$2\% of the original sample, based on the ratio of the numbers of absorbers in the two samples (Fig.~\ref{fig:zabshist_ss_ssorder1neigh1-2-3}), and this is treated as a lower limit of the contribution to the original sample.

To summarize, the relative change in the column densities of low- and high-ionization metal species clearly favor the scenario that from the original sample to the subsample-1/2/3 the average neutral hydrogen column density of the absorbers increases (with possible modulations by metallicity), instead of a pure effect of absorber clustering. The change in the Lyman series absorption profiles and the decrement at the Lyman limit also support such a scenario. Future investigations based on high-resolution observations and hydrodynamic simulations will help verify and quantify the finding here.

}

\begin{table*}
\setcounter{table}{1}
\caption{Measurements of metal lines in the $2.0<\zabsorb<4.0$ stacked spectrum. Column density $N$ and Doppler parameter $b$ from Voigt profile fitting are in units of $\cm^{-2}$ and $\kms$, respectively. The complete set of metal line measurements is available in the online Journal in a machine-readable format.}
\label{table_Zabs[2.0,2.4]}
\resizebox{\linewidth}{!}{
\begin{tabular}{@{}l*{14}{r}}
\hline
\hline
    \noalign{\smallskip}
Species & $\lambda$ (\AA) &  
\multicolumn{3}{c}{$-1.0\le \,$\dl$\, < -0.9$} &  
\multicolumn{3}{c}{$-0.9\le \,$\dl$\, < -0.8$}&  \multicolumn{3}{c}{$-0.8\le \,$\dl$\, < -0.7$}&  \multicolumn{3}{c}{$-0.7\le \,$\dl$\, < -0.6$} \\
& & \multicolumn{1}{c}{$\log N$}  & \multicolumn{1}{c}{$b$} & \multicolumn{1}{c}{$N_{pair}$}  & \multicolumn{1}{c}{$\log N$}  & \multicolumn{1}{c}{$b$} & \multicolumn{1}{c}{$N_{pair}$} & \multicolumn{1}{c}{$\log N$}  & \multicolumn{1}{c}{$b$} & \multicolumn{1}{c}{$N_{pair}$} & \multicolumn{1}{c}{$\log N$}  & \multicolumn{1}{c}{$b$} & \multicolumn{1}{c}{$N_{pair}$} \\
\cutinhead{$2.0<\zabsorb<2.2$}
\CIII\ & 977& - &  - & -& - &  - & -& - &  - & -& - &  - & - \\ 
\OVI\ & 1032& - &  - & -& - &  - & -& - &  - & -& - &  - & - \\ 
\CII\ & 1036& - &  - & -& - &  - & -& - &  - & -& - &  - & - \\ 
\OVI\ & 1038& - &  - & -& - &  - & -& - &  - & -& - &  - & - \\ 
\SiII\ & 1190& $12.93 \pm 0.18$ & $204 \pm 129 $ & $1,203$ & - &  - & -& - &  - & -& - &  - & - \\ 
\SiII\ & 1193& $12.99 \pm 0.08$ & $244 \pm  63 $ & $1,183$ & - &  - & -& - &  - & -& - &  - & - \\ 
\SiIII\ & 1207& $13.00 \pm 0.02$ & $203 \pm  15 $ & $1,084$ & $12.44 \pm 0.04$ & $137 \pm  21 $ & $946$ & $12.06 \pm 0.09$ & $112 \pm  47 $ & $817$ & - &  - & - \\ 
\NV\ & 1239& $13.01 \pm 0.22$ & $174 \pm 118 $ & $828$ & $12.66 \pm 0.18$ & $ 49 \pm  88 $ & $718$ & $12.86 \pm 0.22$ & $214 \pm 148 $ & $618$ & $12.98 \pm 0.13$ & $150 \pm  75 $ & $482$  \\ 
\NV\ & 1243& $12.87 \pm 0.36$ & $ 15 \pm 243 $ & $804$ & $13.26 \pm 0.15$ & $210 \pm 103 $ & $696$ & - &  - & -& - &  - & - \\ 
\SiII\ & 1260& $12.57 \pm 0.07$ & $201 \pm  48 $ & $697$ & - &  - & -& - &  - & -& - &  - & - \\ 
\OI\ & 1302& $13.58 \pm 0.16$ & $287 \pm 103 $ & $1,617$ & - &  - & -& - &  - & -& - &  - & - \\ 
\SiII\ & 1304& $13.43 \pm 0.14$ & $251 \pm  97 $ & $1,619$ & - &  - & -& - &  - & -& - &  - & - \\ 
\CII\ & 1335& $13.77 \pm 0.03$ & $307 \pm  28 $ & $1,636$ & $13.14 \pm 0.07$ & $194 \pm  47 $ & $1,462$ & - &  - & -& - &  - & - \\ 
\SiIV\ & 1394& $13.16 \pm 0.02$ & $214 \pm  14 $ & $1,643$ & $12.74 \pm 0.03$ & $189 \pm  23 $ & $1,467$ & $12.14 \pm 0.15$ & $203 \pm 109 $ & $1,276$ & $11.81 \pm 0.22$ & $ 99 \pm  97 $ & $958$  \\ 
\SiIV\ & 1403& $13.31 \pm 0.03$ & $230 \pm  25 $ & $1,641$ & $12.81 \pm 0.06$ & $151 \pm  35 $ & $1,465$ & $12.44 \pm 0.12$ & $ 94 \pm  55 $ & $1,274$ & - &  - & - \\ 
\SiII\ & 1527& $13.24 \pm 0.04$ & $330 \pm  40 $ & $1,564$ & $11.86 \pm 0.35$ & $ 15 \pm 457 $ & $1,395$ & - &  - & -& - &  - & - \\ 
\CIV\ & 1548& $13.93 \pm 0.01$ & $195 \pm   5 $ & $1,548$ & $13.54 \pm 0.01$ & $139 \pm   5 $ & $1,380$ & $13.22 \pm 0.02$ & $115 \pm   8 $ & $1,200$ & $12.87 \pm 0.04$ & $ 88 \pm  18 $ & $901$  \\ 
\CIV\ & 1551& $14.09 \pm 0.01$ & $197 \pm   6 $ & $1,546$ & $13.74 \pm 0.01$ & $166 \pm   9 $ & $1,379$ & $13.31 \pm 0.02$ & $106 \pm  12 $ & $1,199$ & $12.85 \pm 0.08$ & $ 63 \pm  37 $ & $900$  \\ 
\AlII\ & 1671& $12.16 \pm 0.03$ & $375 \pm  38 $ & $1,496$ & - &  - & -& - &  - & -& - &  - & - \\ 
\AlIII\ & 1855& $11.83 \pm 0.10$ & $147 \pm  55 $ & $1,406$ & - &  - & -& - &  - & -& - &  - & - \\ 
\AlIII\ & 1863& $12.02 \pm 0.18$ & $260 \pm 143 $ & $1,400$ & - &  - & -& - &  - & -& - &  - & - \\ 
\FeII\ & 2344& $13.06 \pm 0.04$ & $485 \pm  58 $ & $873$ & - &  - & -& - &  - & -& - &  - & - \\ 
\FeII\ & 2383& $12.58 \pm 0.06$ & $367 \pm  76 $ & $845$ & $11.60 \pm 0.17$ & $ 46 \pm  89 $ & $751$ & - &  - & -& - &  - & - \\ 
\FeII\ & 2587& $12.95 \pm 0.07$ & $417 \pm  76 $ & $728$ & - &  - & -& - &  - & -& - &  - & - \\ 
\FeII\ & 2600& $12.74 \pm 0.04$ & $363 \pm  42 $ & $711$ & - &  - & -& - &  - & -& - &  - & - \\ 
\MgII\ & 2796& $12.69 \pm 0.04$ & $266 \pm  32 $ & $516$ & $12.26 \pm 0.11$ & $347 \pm 110 $ & $452$ & $10.92 \pm 0.49$ & $ 15 \pm 471 $ & $390$ & - &  - & - \\ 
\MgII\ & 2804& $12.98 \pm 0.10$ & $409 \pm 109 $ & $511$ & $12.29 \pm 0.13$ & $176 \pm  86 $ & $447$ & - &  - & -& - &  - & - \\ 
\hline
\end{tabular}
}
\end{table*}

\clearpage
\bibliographystyle{aasjournal}
\bibliography{bibtex_ly}{}
\label{lastpage}
\end{document}